\documentclass[aps, twocolumn, superscriptaddress, nofootinbib, longbibliography, notitlepage, floatfix]{revtex4-2}
\usepackage[T1]{fontenc}
\usepackage{graphicx}
\usepackage{xcolor}
\usepackage{tikz}
\usepackage{xfrac}
\usetikzlibrary{arrows.meta}
\usetikzlibrary{positioning}
\usetikzlibrary{svg.path}
\usepackage{amsmath,mathtools}
\usepackage{amssymb}
\usepackage{amstext}
\usepackage{amsbsy}
\usepackage{amsthm}
\usepackage{mathrsfs}
\usepackage{esint}
\usepackage{aligned-overset}
\usepackage{dsfont}
\usepackage{tabularx}
\usepackage{longtable,booktabs}
\usepackage{dcolumn}
\usepackage{multirow}
\usepackage{makecell}
\usepackage{outlines}
\usepackage{url}
\usepackage{bm}
\usepackage{extarrows}
\usepackage[super]{nth}
\usepackage{scalerel}

\definecolor{darkblue}{RGB}{0,0,150}
\definecolor{darkred}{RGB}{150,0,0}

\usepackage[
  colorlinks=true,
  linkcolor=darkblue,
  citecolor=darkblue,
  urlcolor=darkred,
  linktocpage=true
]{hyperref}
\usepackage{cleveref}

\tikzset{
    wcirc/.style={draw, circle, minimum size=1ex, inner sep=0pt},
    bcirc/.style={fill, circle, minimum size=1ex, inner sep=0pt},
	wcircL/.style={draw, circle, minimum size=1.5ex, inner sep=0pt},
    bcircL/.style={fill, circle, minimum size=1.5ex, inner sep=0pt}
}

\definecolor{orcidlogocol}{HTML}{A6CE39}
\tikzset{
	orcidlogo/.pic={
		\fill[orcidlogocol] svg{M256,128c0,70.7-57.3,128-128,128C57.3,256,0,198.7,0,128C0,57.3,57.3,0,128,0C198.7,0,256,57.3,256,128z};
		\fill[white] svg{M86.3,186.2H70.9V79.1h15.4v48.4V186.2z}
		svg{M108.9,79.1h41.6c39.6,0,57,28.3,57,53.6c0,27.5-21.5,53.6-56.8,53.6h-41.8V79.1z M124.3,172.4h24.5c34.9,0,42.9-26.5,42.9-39.7c0-21.5-13.7-39.7-43.7-39.7h-23.7V172.4z}
		svg{M88.7,56.8c0,5.5-4.5,10.1-10.1,10.1c-5.6,0-10.1-4.6-10.1-10.1c0-5.6,4.5-10.1,10.1-10.1C84.2,46.7,88.7,51.3,88.7,56.8z};
	}
}

\allowdisplaybreaks[4]
\newcommand\orcidicon[1]{\href{https://orcid.org/#1}{\mbox{\scalerel*{
				\begin{tikzpicture}[yscale=-1,transform shape]
					\pic{orcidlogo};
				\end{tikzpicture}
			}{|}}}}
\tikzset{bold/.style={color=blue, line width=2pt}}
\tikzset{redop/.style={circle,fill=red}}
\tikzset{blueop/.style={circle,fill=blue}}
\graphicspath{{pic/}{Pictures/}}

\makeatletter

\newcommand{\Rmnum}[1]{\expandafter\@slowromancap\romannumeral #1@}

\makeatother

\begin{document}
\title{Kosterlitz--Thouless Criticality in a Dipole-Conserving XY Model}

\author{Han-Xie Wang\orcidicon{0009-0002-6414-0830}}\thanks{These authors contributed equally.}
\affiliation{Guangdong Provincial Key Laboratory of Magnetoelectric Physics and Devices,  State Key Laboratory of Optoelectronic Materials and Technologies,
    and School of Physics,  Sun Yat-sen University,  Guangzhou,  510275,  China}
\author{Shuai A. Chen}\thanks{These authors contributed equally.}
\affiliation{Max Planck Institute for the Physics of Complex Systems,  N\"othnitzer Stra{\ss}e~38,  Dresden~01187,  Germany}
\author{Zheng Yan}
\affiliation{Department of Physics, School of Science and Research Center for
    Industries of the Future, Westlake University, Hangzhou 310030, China}
\affiliation{Institute of Natural Sciences, Westlake Institute for Advanced Study, Hangzhou 310024, China}
\author{Peng Ye\orcidicon{0000-0002-6251-677X}}
\email{yepeng5@mail.sysu.edu.cn}
\affiliation{Guangdong Provincial Key Laboratory of Magnetoelectric Physics and Devices,  State Key Laboratory of Optoelectronic Materials and Technologies,
    and School of Physics,  Sun Yat-sen University,  Guangzhou,  510275,  China}

\date{\today}

\begin{abstract}
In this Letter, we study finite-temperature phase transitions in a two-dimensional classical statistical model called ``dipole-conserving XY model''. Unlike the conventional XY model, the original phase field $\theta$ has no quasi-long-range order, conventional phase vortices have finite self-energy, and the standard helicity modulus vanishes identically. Analytic vortex energetics and Gaussian continuum theory show that Kosterlitz--Thouless (KT) criticality is instead controlled by phase-gradient (PG) vortices defined in the two compact dipole fields $\chi_\alpha=a\partial_\alpha\theta$, whose self-energies grow logarithmically with system size. We determine the phase diagram using parallel-tempering Metropolis Monte Carlo simulations and three diagnostics tailored to dipole conservation. KT finite-size scaling of dipole-field correlation-ratio crossings locates the critical temperatures; generalized helicity moduli defined through quadratic phase twists measure the stiffness of individual dipole channels; and PG-vortex densities obtained from plaquette winding numbers identify the proliferating vortex species. At isotropic couplings, the two PG-vortex species unbind simultaneously at a single KT transition. Spatial anisotropy separates their unbinding temperatures, yielding two KT transitions and an intermediate phase with quasi-long-range order in only one dipole channel. The transitions merge again when the mixed-derivative channel is removed. Our results establish the finite-temperature phase structure of the dipole-conserving XY model and identify thermal PG-vortex unbinding as a novel route to KT criticality in systems with higher-moment conservation.

\end{abstract}

\maketitle

\textit{Introduction}.--- Understanding how symmetry and conservation laws shape phase transitions and critical phenomena is a central goal of statistical physics.  Topological phase transitions, driven by the binding and unbinding of topological defects rather than by the onset of a local order parameter, represent a fundamental departure from the Landau--Ginzburg paradigm of symmetry breaking. The Kosterlitz--Thouless (KT) transition~\cite{kosterlitzOrderingMetastabilityPhase1973,kosterlitzCriticalPropertiesTwodimensional1974,kosterlitzKosterlitzThoulessPhysics2016} is the paradigmatic example of topological phase transitions. 
Its canonical realization is the two-dimensional (2D) XY model~\cite{nagaosaQuantumFieldTheory1999}, which has a charge $U(1)$ symmetry under uniform phase rotations $\theta\to\theta+\lambda_0$ of a phase field $\theta$. A phase vortex $\Theta_0$ with a $2\pi$ winding of $\theta$ carries both energy and entropy $\propto\ln(L/a)$, so heating drives the unbinding of vortex-antivortex pairs. 
This KT mechanism has been realized in superfluid helium films~\cite{bishopStudySuperfluidTransition1978}, thin superconducting films~\cite{hebardEvidenceKosterlitzThouless1980}, Josephson-junction arrays~\cite{resnickKosterlitzThoulessTransition1981}, and cuprate superconductors~\cite{leeDopingMottInsulator2006}. 
In 2D magnetic materials~\cite{bedoyaPintoIntrinsic2DXY2021}, KT phases have recently been identified in the frustrated magnet TmMgGaO$_4$~\cite{liEvidenceBerezinskiiKosterlitz2020} and in the triangular-lattice antiferromagnet K$_2$Co(SeO$_3$)$_2$~\cite{zhuWannierStatesSpin2025}. These developments cement KT physics as a robust organizing principle for finite-temperature order in 2D. 
Theoretically and numerically, the KT paradigm remains an active frontier: systems exhibit nontrivial thermodynamics from the interplay of spin waves and vortices~\cite{maccariInterplaySpinWaves2020,maccariVortexSupersolidXY2023}, and long-range couplings can alter the transition~\cite{giachettiBerezinskiiKosterlitzThoulessPhaseTransitions2021,yaoNonclassicalRegimeTwodimensional2025,waltherPersistenceBerezinskiiKosterlitzThoulessTransition2026}.
In all these examples, KT criticality derives from a single charge $U(1)$ symmetry. What happens when additional dipole $U(1)$ symmetries are imposed?

In this Letter, we address this question by studying the minimal dipole-conserving XY model, whose Hamiltonian [Eq.~\eqref{eq:H_lattice}] is defined on a square lattice with couplings $J_1, J_2, J_3\ge 0$. The model possesses three compact $U(1)$ symmetries---a global charge rotation $\theta\to\theta+\lambda_0$ and two dipole shifts $\theta\to\theta+\lambda_1^{(1)}x_1/a$, $\theta\to\theta+\lambda_1^{(2)}x_2/a$, where $x_1,x_2$ are spatial coordinates and $\lambda_0,\lambda_1^{(1)},\lambda_1^{(2)}$ are arbitrary constants. These minimal charge- and dipole-conserving terms also emerge from the dipolar Bose-Hubbard model~\cite{lakeDipolarBoseHubbardModel2022}. In the continuum, the model reduces to the Lifshitz theory~\cite{gorantlaGlobalDipoleSymmetry2022a}, which plays a critical role in the context of fractonic superfluids~\cite{yuanFractonicSuperfluids2020b,chenFractonicSuperfluidsII2021a,wangFractonicSuperfluidsIII2025a,liRenormalizationGroupAnalysis2021a,yuanQuantumHydrodynamicsFractonic2022a,yuanHierarchicalProliferationHigherrank2023a} and the dual gauge-theory description~\cite{pretkoFractonElasticityDuality2018,pretkoCrystaltofractonTensorGauge2019,radzihovskyLifshitzGaugeDuality2022,gorantla2+1dimensionalCompactLifshitz2023a,nelsonDislocationmediatedMeltingTwo1979,manojFractonicViewFolding2021}. Yet the finite-temperature statistical
physics of such generalized XY model has not been systematically explored. The $q^{-4}$ propagator destroys quasi-long-range order in $\theta$ via a modified Mermin--Wagner mechanism~\cite{kapustinHohenbergMerminWagnertypeTheoremsDipole2022,stahlSpontaneousBreakingMultipole2022}, and a conventional vortex no longer carries logarithmically divergent self-energy. Thus, both pillars of standard KT physics identified in the standard XY model---power-law correlations of the phase field and logarithmically costly vortices---seem to disappear. Does finite-temperature KT physics survive once dipole symmetry removes the logarithmic energy from the phase vortex?

\begin{figure}[t]
    \centering
    \begin{tabular}{@{}c@{\hspace{3pt}}c@{\hspace{3pt}}c@{}}
        \includegraphics[width=0.32\columnwidth]{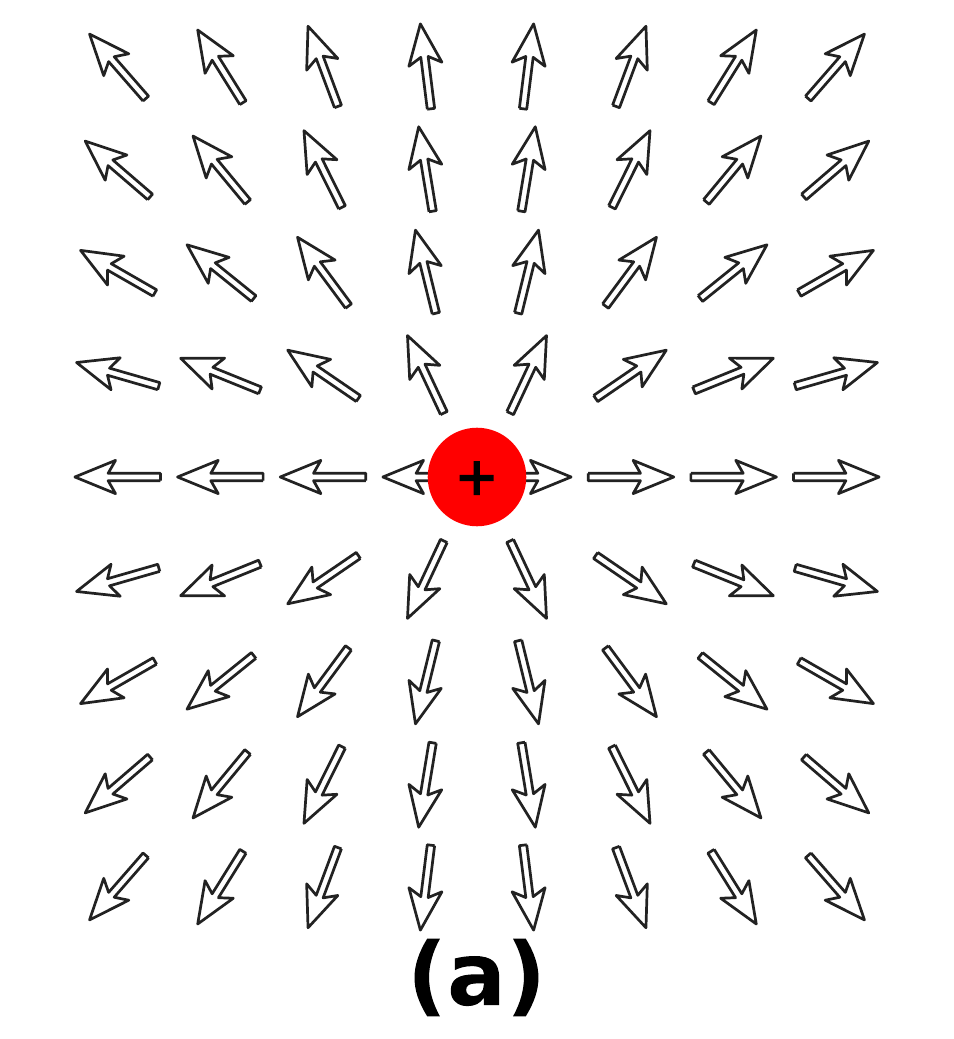} &
        \includegraphics[width=0.32\columnwidth]{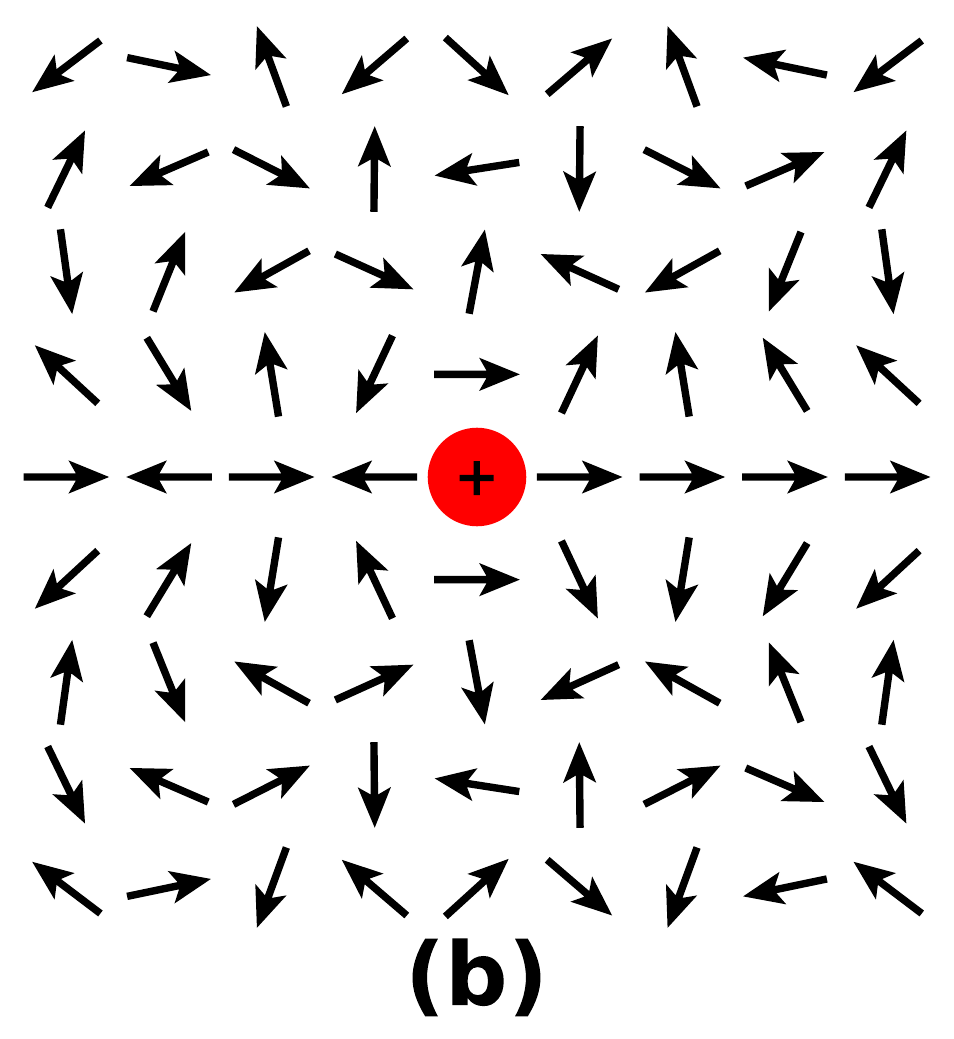} &
        \hspace*{-0.033\columnwidth}\includegraphics[width=0.32\columnwidth]{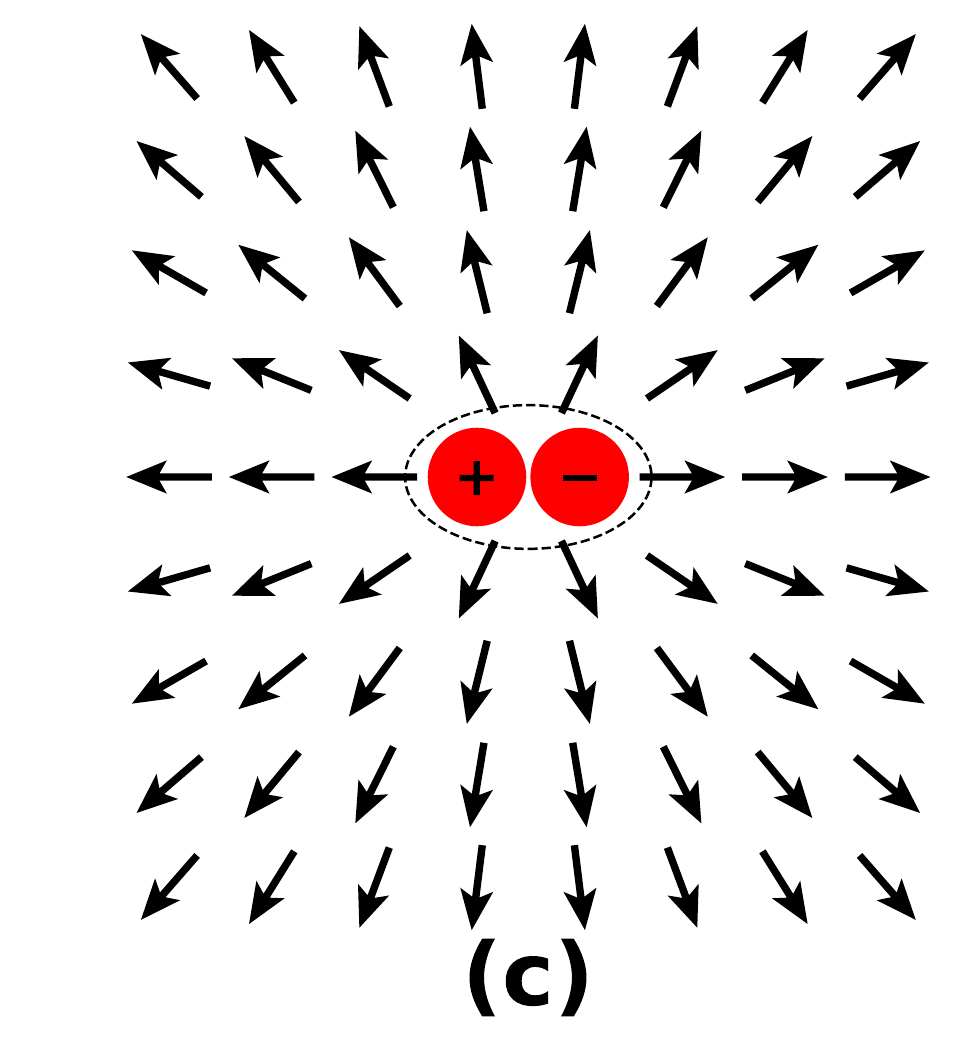} \\[4pt]
        \includegraphics[width=0.32\columnwidth]{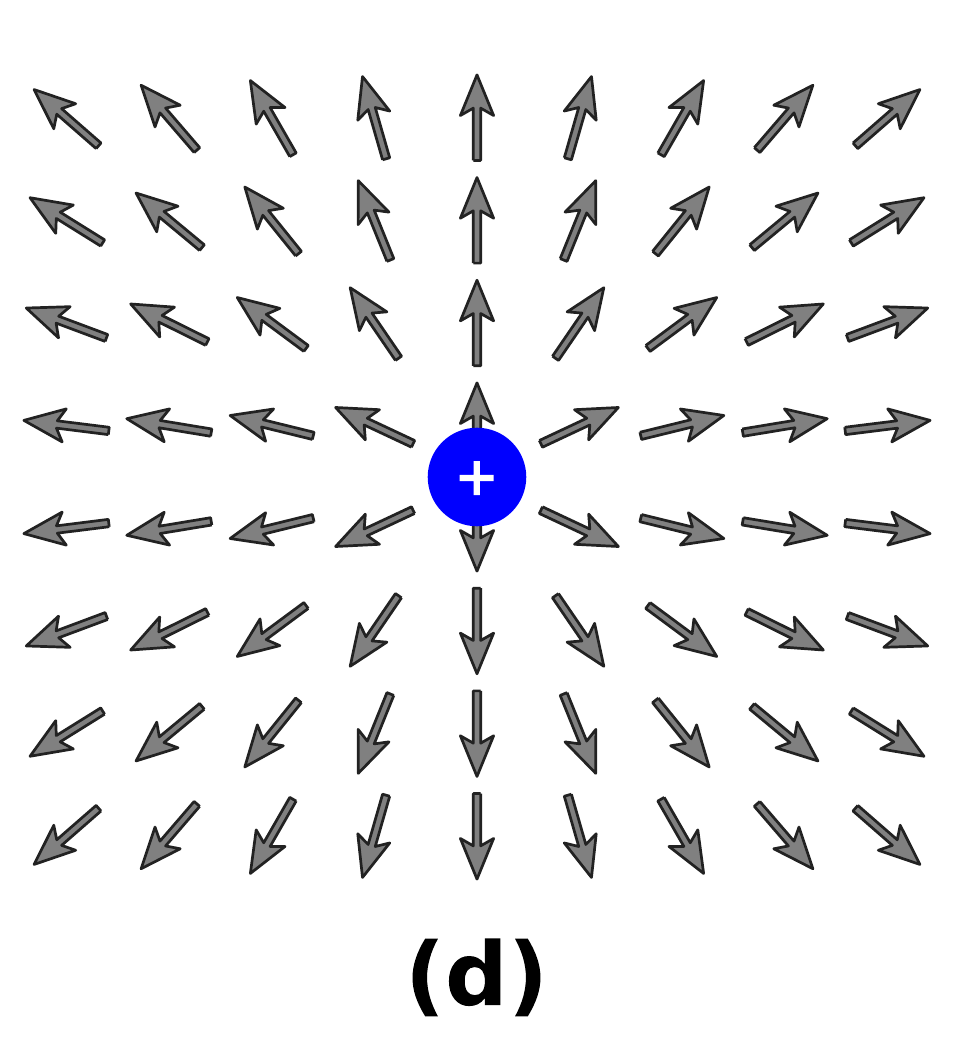} &
        \includegraphics[width=0.32\columnwidth]{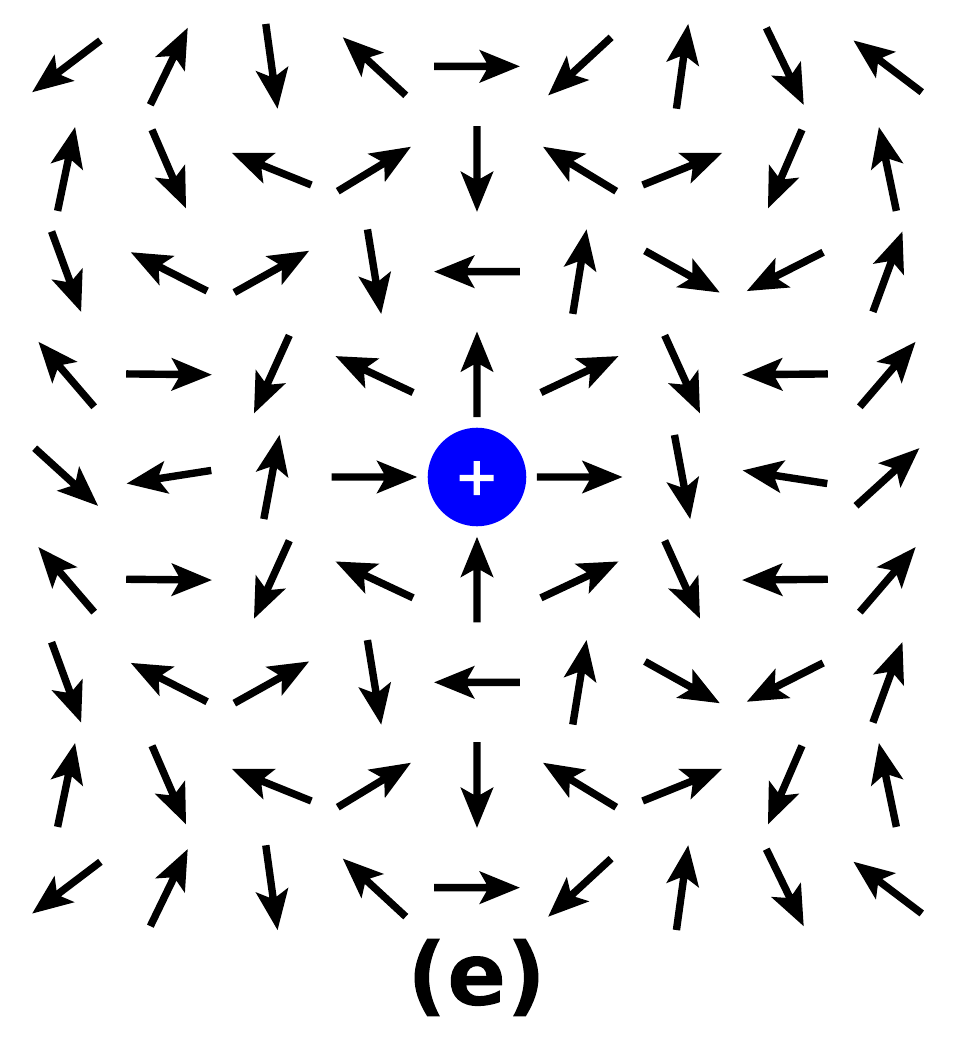} &
        \includegraphics[width=0.3\columnwidth]{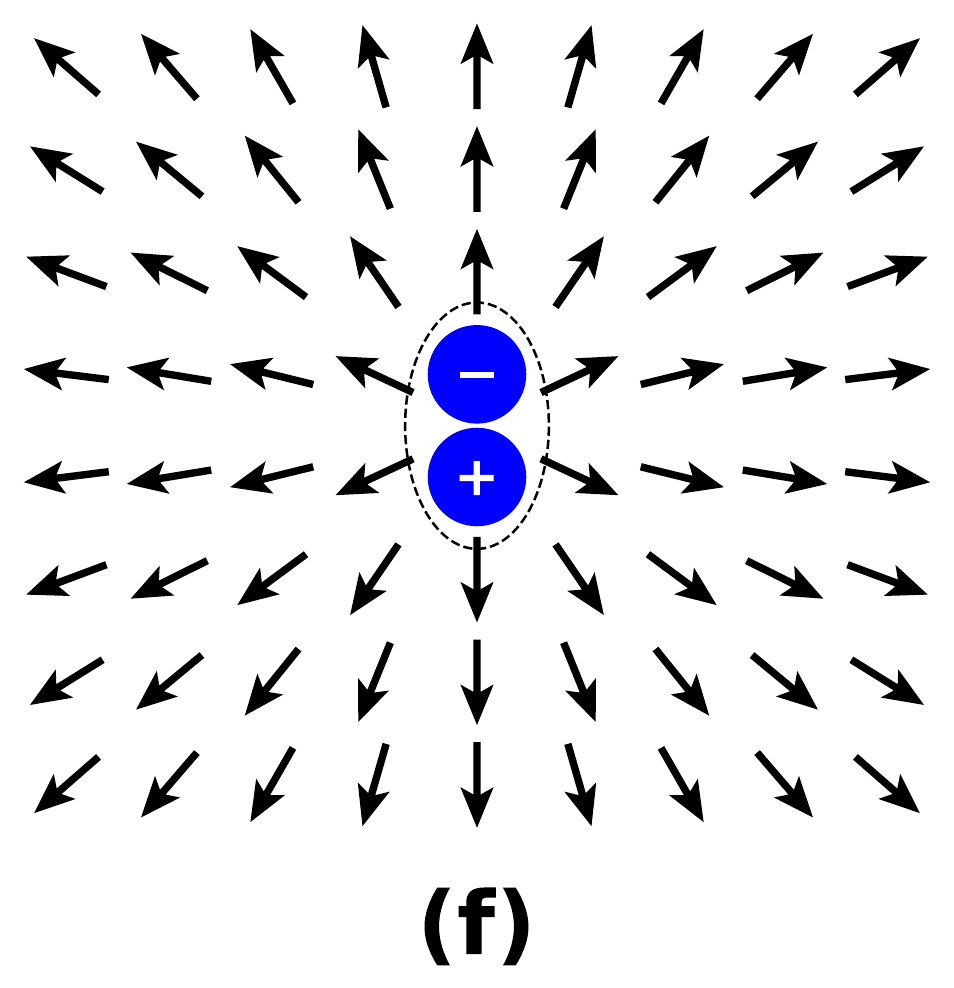}
    \end{tabular}
\caption{Textures of phase-gradient vortices.
    (a)--(c) $\Theta_1^{(1)}$ (red); (d)--(f) $\Theta_1^{(2)}$ (blue).
    Arrows in (a),(d) represent the compact dipole-field angles $\chi_1$ and $\chi_2$; arrows in all other panels represent the original phase $\theta$.
    Thus, (a),(d) are conventional $+1$ vortices of the corresponding dipole fields.
    (b),(e) The same PG vortices in the $\theta$ representation [Eqs.~\eqref{eq:Theta11} and \eqref{eq:Theta12}].
    (c),(f) Neutral nearest-neighbor PG vortex--antivortex pairs ($d=a$), whose far-field textures reproduce a unit phase vortex. For a general pair, the winding is proportional to its dipole moment (see \cite{SM}, Sec.~I).}
    \label{fig:hierarchy}
\end{figure}

The answer is yes, but through a reconstructed
topological mechanism. Combining analytic vortex energetics and Gaussian
continuum theory with Monte Carlo simulations, we establish that in the
2D dipole-conserving XY model, a phase vortex $\Theta_0$ is
not the defect that drives criticality: it has finite self-energy. The relevant
defects are the two phase-gradient (PG) vortices $\Theta_1^{(1)}$ and
$\Theta_1^{(2)}$, which are defined in the dipole fields $\chi_\alpha=a\partial_\alpha\theta$ ($\alpha=1,2$) and have logarithmically divergent self-energies
(Fig.~\ref{fig:hierarchy}); they are therefore capable of driving KT
unbinding. 
Correspondingly, the uniform twist that defines the helicity modulus in the standard
XY model becomes a dipole symmetry here, so $J_s$ vanishes identically
and its universal jump~\cite{nelsonUniversalJumpSuperfluid1977} is absent. KT criticality is instead diagnosed by dipole-field correlation ratios, generalized helicity moduli, and PG-vortex densities.

This PG-vortex mechanism leads to a tunable sequence of thermal
transitions. At the isotropic point $(J_1=J_2=J_3)$, the two PG-vortex species are symmetry related and
unbind together, yielding a single KT transition. Anisotropy $(J_1\neq J_2, J_3>0)$ lifts this
degeneracy, splitting the transition into two and stabilizing an intermediate
phase with quasi-long-range order in only one dipole channel. The two
transitions merge again when $J_3=0$, even for
$J_1\neq J_2$. Such dipole symmetries can emerge in tilted optical lattices~
\cite{lakeDipolarBoseHubbardModel2022,lakeDipoleCondensatesTilted2023,
xuMultipolarCondensatesMultipolar2024a}
and magnetic systems~
\cite{sousFractonsPolarons2020,sousFractonsFrustrationHoledoped2020,
sanyalUnidirectionalSubsystemSymmetry2024}, 
placing our results in a broad physical context. In what follows, we define the model, develop diagnostics tailored to
dipole conservation, and use parallel-tempering Monte Carlo simulations
to determine its finite-temperature phase diagram.

\textit{Model and KT Diagnostics}.---
We study the minimal dipole-conserving XY model on a square lattice:
\begin{align}
    H = & -J_1 \sum_{\mathbf{i}} \cos\!\left(2\theta_{\mathbf{i}} - \theta_{\mathbf{i}-\hat{x}_1} - \theta_{\mathbf{i}+\hat{x}_1}\right) \nonumber                                  \\
        & -J_2 \sum_{\mathbf{i}} \cos\!\left(2\theta_{\mathbf{i}} - \theta_{\mathbf{i}-\hat{x}_2} - \theta_{\mathbf{i}+\hat{x}_2}\right) \nonumber                                  \\
        & -2J_3 \sum_{\mathbf{i}} \cos\!\left(\theta_{\mathbf{i}} + \theta_{\mathbf{i}+\hat{x}_1+\hat{x}_2} - \theta_{\mathbf{i}+\hat{x}_1} - \theta_{\mathbf{i}+\hat{x}_2}\right).
    \label{eq:H_lattice}
\end{align}
Here $\theta_{\mathbf{i}}\sim\theta_{\mathbf{i}}+2\pi$ is a compact phase and
$\hat{x}_1,\hat{x}_2$ are lattice basis vectors of length $a$ (the lattice constant). The three  terms
are ferromagnetic couplings respectively 
along $\hat{x}_1$ and $\hat{x}_2$  and around a plaquette. All terms respect the following
three compact $U(1)$ symmetries: a conventional global charge rotation and two independent
dipole shifts (see Introduction). The continuum limit is given by:
\begin{equation}
    \!\!\! H_{\mathrm{cont}} \!=\! \frac{a^2}{2} \int \!\! d^2\mathbf{x}\,
    \bigl[ J_1 (\partial_1^2\theta)^2 + J_2 (\partial_2^2\theta)^2 + 2J_3 (\partial_1\partial_2\theta)^2 \bigr].
    \label{eq:H_cont}
\end{equation}
Because the propagator of $\theta$ scales as $q^{-4}$, the original phase
correlator decays super-exponentially~\cite{kapustinHohenbergMerminWagnertypeTheoremsDipole2022,
    stahlSpontaneousBreakingMultipole2022}. Quasi-long-range order,
when present, is instead carried by 
$\chi_\alpha=a\partial_\alpha\theta$ ($\alpha=1,2$), which we call the dipole fields.

The compactness of these dipole fields is defined on the lattice:
$\chi_{\alpha,\mathbf{i}}\equiv[\Delta_\alpha\theta_{\mathbf{i}}]_\pi$, where
$\Delta_\alpha\theta_{\mathbf{i}}\equiv\theta_{\mathbf{i}+\hat{x}_\alpha}-\theta_{\mathbf{i}}$ is the forward difference along $\hat{x}_\alpha$, and $[x]_\pi$
denotes reduction to $(-\pi,\pi]$; $\chi_{\alpha,\mathbf{i}}$ reduces to $a\partial_\alpha\theta$ in
the continuum. Thus one can define windings not only of $\theta$, but also of
$\chi_1$ and $\chi_2$, labeled by integer winding numbers $\ell_0,\ell_1,\ell_2\in\mathbb{Z}$, defined by $\oint\nabla\theta\cdot d\mathbf{l}=2\pi\ell_0$, $\oint\nabla\chi_\alpha\cdot d\mathbf{l}=2\pi\ell_\alpha$ ($\alpha=1,2$).
(Fig.~\ref{fig:hierarchy}). In the isotropic case, the minimum-energy continuum
textures, for a vortex centered at the origin, are concisely written using $z=x_1+ix_2$:
\begin{align}
    \Theta_0(\mathbf{x})       & = \ell_0\,\operatorname{Im}\bigl[\ln(z/a)\bigr], \label{eq:Theta0}        \\
    \Theta_1^{(1)}(\mathbf{x}) & = \ell_1\,\operatorname{Im}\bigl[(z/a)\ln(z/a)\bigr], \tag{$4a$}\label{eq:Theta11}  \\
    \Theta_1^{(2)}(\mathbf{x}) & = -\ell_2\,\operatorname{Re}\bigl[(z/a)\ln(z/a)\bigr]. \tag{$4b$}\label{eq:Theta12}
\end{align}
\setcounter{equation}{4}
In Cartesian coordinates, these reduce to~\cite{chenFractonicSuperfluidsII2021a}
$\Theta_0=\ell_0\varphi$,
$\Theta_1^{(1)}=\ell_1\bigl[(x_1/a)\varphi+(x_2/a)\ln(r/a)\bigr]$ and
$\Theta_1^{(2)}=\ell_2\bigl[(x_2/a)\varphi-(x_1/a)\ln(r/a)\bigr]$, respectively,
with $\varphi(\mathbf{x})$ the polar angle and $r=|\mathbf{x}|$.
$\Theta_0$ is the vortex of the original phase field, whereas
$\Theta_1^{(1)}$ and $\Theta_1^{(2)}$ are PG vortices of the
dipole fields $\chi_1$ and $\chi_2$, respectively.

The energetics of the three species are fundamentally different. For the
isotropic model ($J_1=J_2=J_3=J$), evaluating the vortex self-energy
of the continuum vortex configurations (\cite{SM}, Secs.~I and~II) with a short-distance cutoff $a$ on a system of linear size $L$ gives: $E_{\Theta_0} = \pi J\ell_0^2$, $E_{\Theta_1^{(1,2)}} = 2\pi J\ell_{1,2}^{2}\ln\frac{L}{a}$.
Thus, the phase vortex has a \emph{finite} self-energy in the
thermodynamic limit, whereas $\Theta_1^{(1)}$ and $\Theta_1^{(2)}$ retain the
\emph{logarithmically divergent} self-energies required for the
binding-unbinding scenario. Dipole conservation thus reconstructs the defect hierarchy: the
relevant KT defects are not the phase vortices of $\theta$, but the
vortices of its compact gradients.

The three defect sectors are related not by a splitting of phase-vortex charge but through the dipole moment of a PG-vortex pair. Consider two oppositely charged
$\Theta_1^{(1)}$ vortices separated by $d\,\hat{x}_1$. At distances $r\gg d$,
their combined field approaches $(d/a)\varphi(\mathbf{x})$, and hence carries
the far-field conventional winding $\ell_0^{\rm far}=d/a$. The minimal lattice dipole
with $d=a$ therefore reproduces a unit phase vortex, whereas increasing
the separation changes the conventional winding. Consequently, the intrinsic
topological charges of the $\Theta_1^{(\alpha)}$ vortices are the integer
windings $\ell_\alpha$ of the compact dipole fields, not fractions of a
fixed $\ell_0$ (\cite{SM}, Sec.~I for the far-field expansion). Their logarithmic interaction, rather than the
finite-energy phase-vortex sector, controls the KT transitions. The
$\Theta_1^{(\alpha)}$ are subdimensionally
mobile~\cite{pretkoFractonElasticityDuality2018,pretkoCrystaltofractonTensorGauge2019,manojFractonicViewFolding2021,
    gorantla2+1dimensionalCompactLifshitz2023a}, as expected from dipole
conservation, whereas a tightly bound pair is mobile in both lattice
directions. 

In the standard XY model, the helicity modulus $J_s$---the stiffness against a uniform
phase twist---measures the superfluid density and exhibits a universal jump at
the KT transition~\cite{nelsonUniversalJumpSuperfluid1977,hasenbuschTwodimensionalXYModel2005,olssonMinnhagenHelicityModulus1991,prokofevTwoDefinitionsSuperfluid2000}. In the
dipole-conserving model, a uniform twist $\theta(\mathbf{x})\rightarrow\theta(\mathbf{x})+\gamma x_1/a$ is not a physical deformation but a dipole symmetry---it shifts
$\partial_1\theta$ by a constant while leaving every second derivative
unchanged [Eq.~\eqref{eq:H_cont}]. Consequently $J_s$ vanishes identically, and the
standard modulus is blind to the transition.

We therefore employ two complementary diagnostics for locating the KT
transition. The first characterizes the quasi-long-range order of the dipole
fields directly through the correlation functions $C_\alpha(\mathbf r)\equiv
\langle e^{i(\chi_\alpha(\mathbf{r})-\chi_\alpha(\mathbf{0}))}\rangle$.
The correlation ratio $C(L/2)/C(L/4)$ (with $C\equiv C_\alpha$) provides a
scale-invariant crossing at the KT transition; we use the crossing point to
determine the critical temperature $T_c^{(\alpha)}$ of the $\chi_\alpha$ channel.

The second probes the stiffness of each dipole channel independently. To this
end, we introduce a \emph{quadratic} phase twist. Let $F(\gamma_\iota)$ be the
free energy in the
presence of the quadratic twists of amplitudes $\gamma_1,\gamma_2,\gamma_3$:
$\theta\to\theta+\tfrac12\gamma_1(x_1/a)^2$,
$\theta\to\theta+\tfrac12\gamma_2(x_2/a)^2$, and
$\theta\to\theta+\gamma_3(x_1/a)(x_2/a)$. We define three generalized helicity
moduli:
\begin{align}
    J_{s\iota} & = \frac{a^2}{c_\iota L^2}\frac{\partial^2 F}{\partial\gamma_\iota^2}\Bigg|_{\gamma_\iota=0}, \quad \iota=1,2,3,
        \label{eq:J_s}
\end{align}
with $c_1=c_2=1$, $c_3=2$.  These deformations generate uniform shifts of $\partial_1^2\theta$, $\partial_2^2\theta$, and $\partial_1\partial_2\theta$, respectively, and therefore probe the three second-derivative channels appearing in Eq.~\eqref{eq:H_cont} (\cite{SM}, Secs.~III and~IV for the continuum formulation and RG analysis; Sec.~V for the lattice implementation); the transition temperatures obtained from the correlation ratios are independently checked by the moduli, which drop when the corresponding channel loses stiffness.

Within the Gaussian continuum theory at low temperatures, the dipole-field correlations decay as a power law,
$C_\alpha(\mathbf r)\sim r^{-\eta_\alpha}$, with exponents given by
(\cite{SM}, Sec.~III) at inverse temperature $\beta=1/T$:
\begin{align}
   \!\! \eta_\alpha & = \left({2\sqrt{2}\,\pi\beta\sqrt{J_{s\alpha}(J_{s3}+\sqrt{J_{s1}J_{s2}})}}\right)^{-1}\,,  \, \alpha=1,2.
    \label{eq:eta}
\end{align}
The KT transition in the $\chi_\alpha$ channel occurs when the corresponding exponent reaches the universal value $\eta_\alpha=\sfrac{1}{4}$ (\cite{SM}, Sec.~IV).
Beyond serving as a diagnostic, Eq.~\eqref{eq:eta} constrains the possible phase structure.
When the two dipole channels have different moduli, they may lose quasi-long-range order at different temperatures, allowing two distinct transitions and an intermediate phase with quasi-long-range order in only one channel.
However, when $J_3=0$, the Hamiltonian becomes invariant under $\theta\to\theta+\gamma_3(x_1/a)(x_2/a)$, enforcing $J_{s3}=0$, and the two exponents reduce to $\eta_1\propto ({J_{s1}^{3/4}J_{s2}^{1/4}})^{-1}$ and
    $\eta_2\propto({J_{s1}^{1/4}J_{s2}^{3/4}})^{-1}$, respectively. In this case, if either $J_{s1}$ or $J_{s2}$ vanishes, quasi-long-range order is lost in both channels,
excluding such an intermediate phase. Hence only a single transition is expected along the $J_3=0$ line, consistent with the numerical results below.

Finally, we measure vortex densities directly from plaquette winding
numbers. The vorticity for the
phase vortex is~\cite{tobochnikMonteCarloStudy1979,maccariInterplaySpinWaves2020}
$q_{\mathbf{i}}^{(0)} =
    \frac{1}{2\pi}\sum_{{}_{\mathbf{i}}\square}[\Delta\theta]_{\pi}$, where $\sum_{{}_{\mathbf{i}}\square}$ runs
counterclockwise over the four bonds of the elementary plaquette with
lower-left site $\mathbf{i}$, $[\cdot]_{\pi}$ denotes the compact reduction as above, and $\Delta\theta$ denotes the difference of $\theta$ along the corresponding bond. The
vorticities $q_{\mathbf{i}}^{(1)}$ and $q_{\mathbf{i}}^{(2)}$ are defined
analogously for  $\chi_{1,\mathbf{i}}$ and
$\chi_{2,\mathbf{i}}$:
\begin{align}
    q_{\mathbf{i}}^{(\alpha)} = \frac{1}{2\pi}\sum_{{}_{\mathbf{i}}\square}[\Delta\chi_\alpha]_{\pi}\,, \quad \alpha=1,2.\label{eq:vorticity}
\end{align}
The corresponding densities
$\rho(\Theta_0)=L^{-2}\sum_{\mathbf{i}}|q_{\mathbf{i}}^{(0)}|$ and
$\rho(\Theta_1^{(\alpha)})=L^{-2}\sum_{\mathbf{i}}|q_{\mathbf{i}}^{(\alpha)}|$
($\alpha=1,2$). Crucially, the plaquette-level measurement localizes the
topological charge correctly: an isolated PG vortex yields
$q_{\mathbf{i}}^{(0)}=0$ on every plaquette, even though a large contour
encircling it would register a contour-dependent conventional winding (\cite{SM}, Sec.~I). The vortex densities
therefore cleanly identify which species proliferates at each
transition.

Together, these diagnostics probe the PG vortices directly; we now
test numerically the prediction that they, not the phase vortex, control
the KT criticality.

\begin{figure}[t]
    \centering
    \includegraphics[width=0.95\columnwidth]{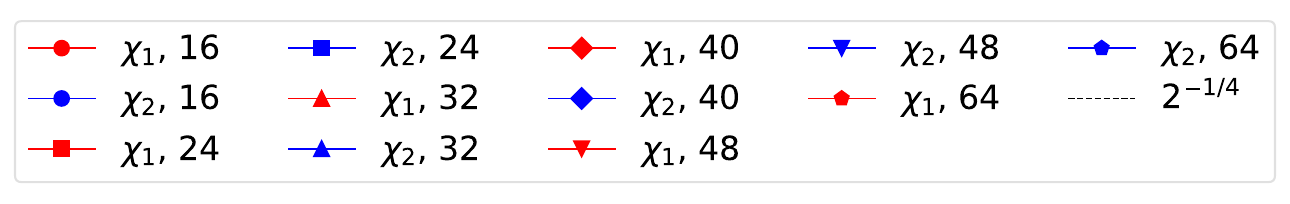}\\[0pt]
    \begin{tabular}{@{}c@{\hspace{2pt}}c@{}}
        \includegraphics[width=0.47\columnwidth]{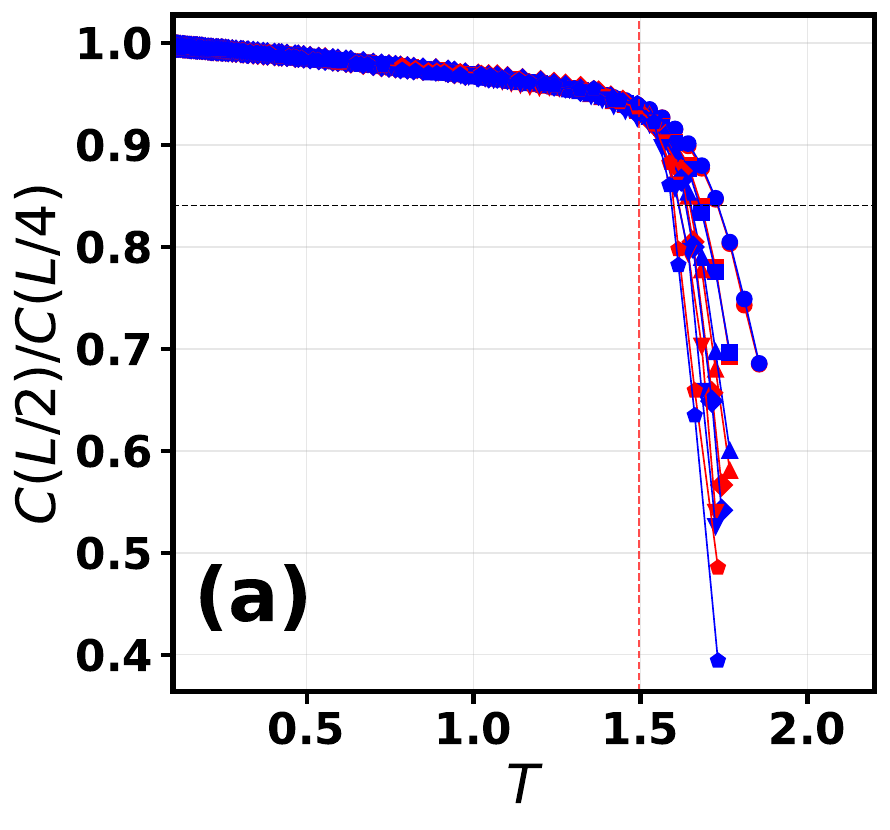} &
        \includegraphics[width=0.47\columnwidth]{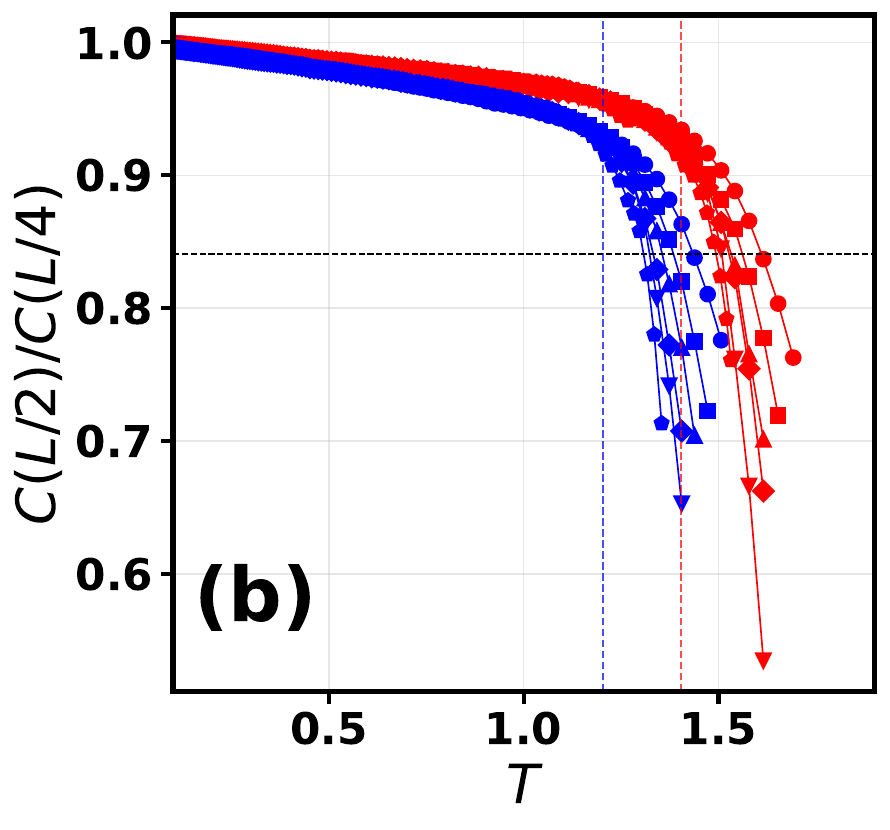}
    \end{tabular}
    \caption{Dipole-field correlation ratio $C(L/2)/C(L/4)$, where $C\equiv C_\alpha(\mathbf r)=\langle e^{i(\chi_\alpha(\mathbf r)-\chi_\alpha(\mathbf 0))}\rangle$.
    The horizontal dash-dotted line marks the KT threshold $2^{-1/4}\approx0.8409$.
    (a)~Isotropic point $J_1{=}J_2{=}J_3{=}1$: a single scale-invariant crossing.
    (b)~Anisotropic case $J_1{=}1$, $J_2{=}0.6$, $J_3{=}1$: two distinct scale-invariant crossings.
    Data for the $J_3{=}0$ and $J_2{=}0$ regimes are shown in the End Matter (Fig.~\ref{fig:em_data}).}
    \label{fig:corr}
\end{figure}

\begin{figure}[t]
    \centering
    \includegraphics[width=0.95\columnwidth]{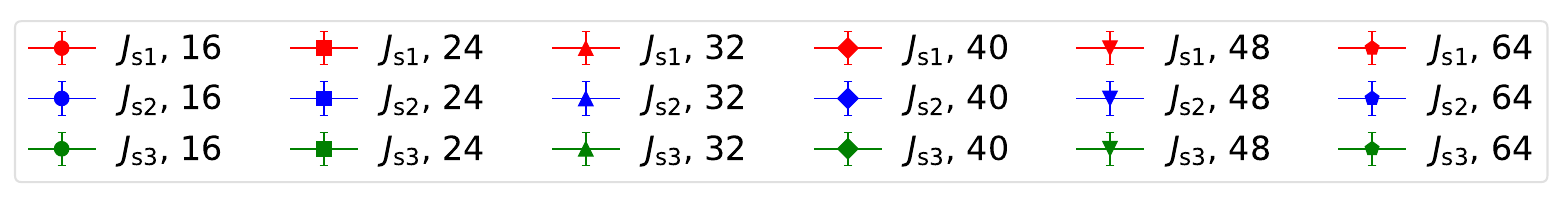}\\[0pt]
    \begin{tabular}{@{}c@{\hspace{2pt}}c@{}}
        \includegraphics[width=0.47\columnwidth]{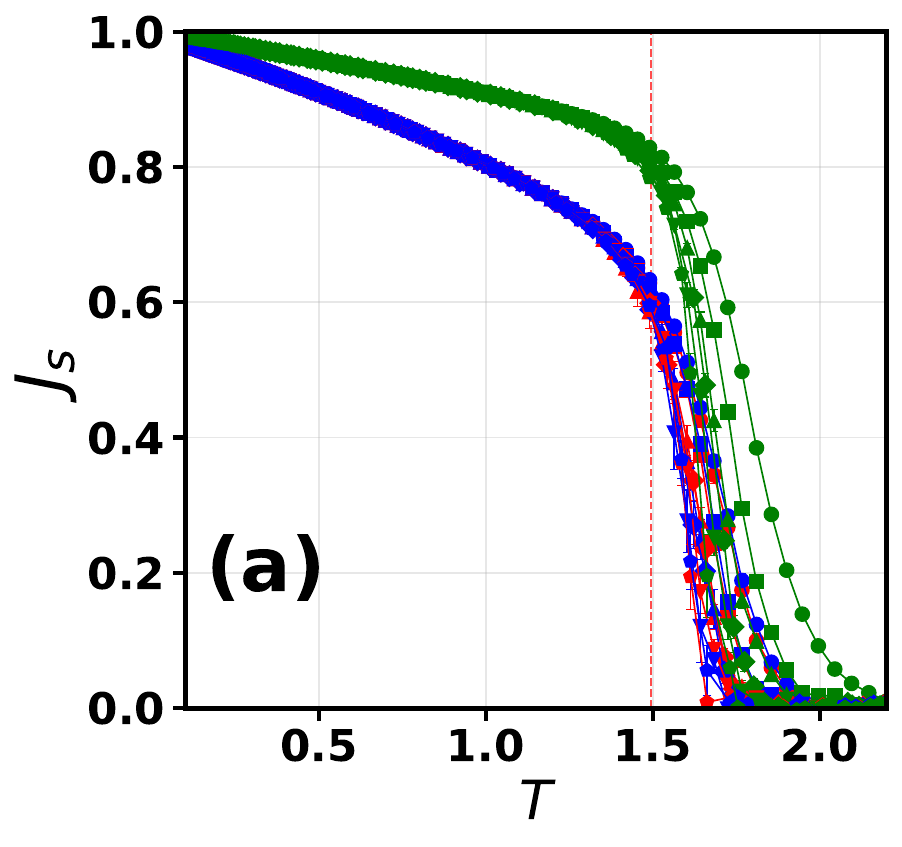} &
        \includegraphics[width=0.47\columnwidth]{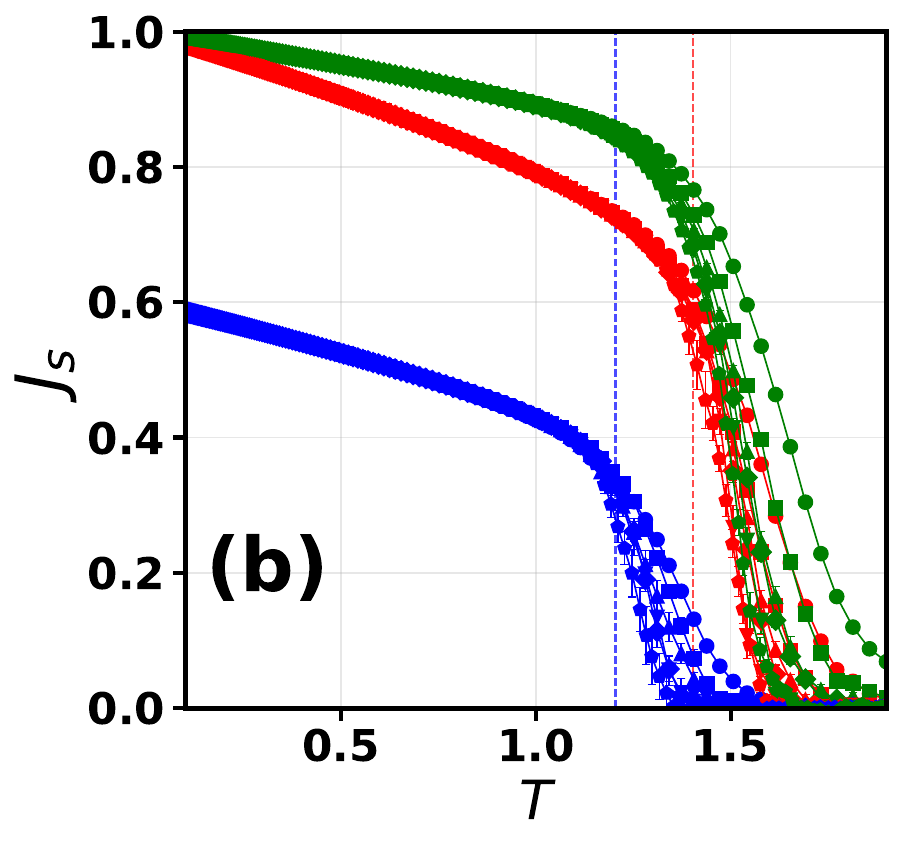}
    \end{tabular}
    \caption{Generalized helicity moduli $J_{s1}$, $J_{s2}$, $J_{s3}$ for the two central regimes.
    Vertical dashed lines mark the thermodynamic-limit transition temperatures from finite-size scaling.
    (a)~Isotropic point $J_1{=}J_2{=}J_3{=}1$: all three moduli drop together.
    (b)~Anisotropic case $J_1{=}1$, $J_2{=}0.6$, $J_3{=}1$: $J_{s2}$ drops first, while $J_{s1}$ and $J_{s3}$ persist to higher temperature.}
    \label{fig:stiffness}
\end{figure}

\begin{figure}[t]
    \centering
    \includegraphics[width=0.95\columnwidth]{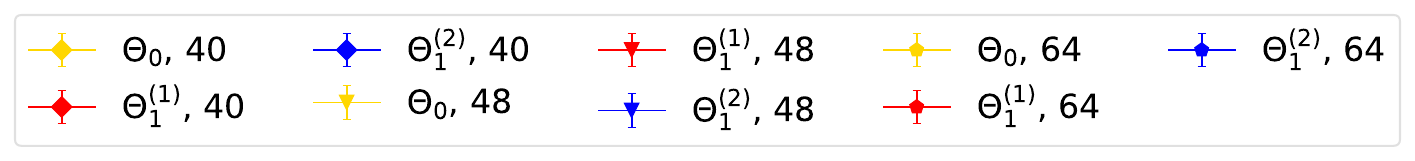}\\[0pt]
    \begin{tabular}{@{}c@{\hspace{2pt}}c@{}}
        \includegraphics[width=0.47\columnwidth]{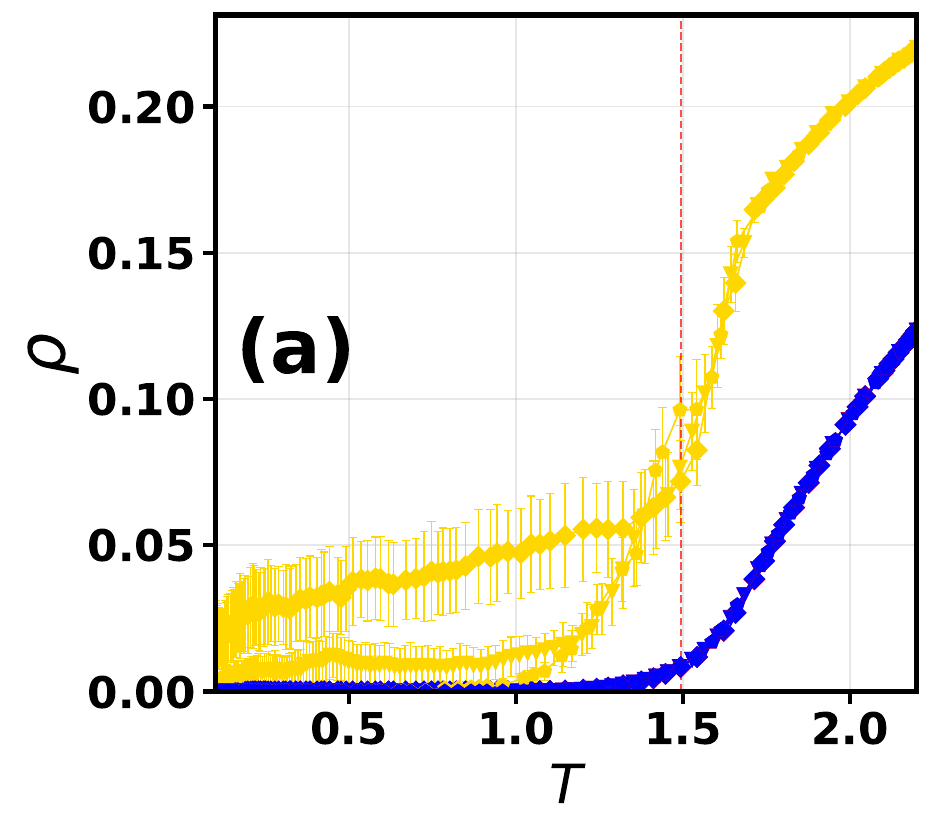} &
        \includegraphics[width=0.47\columnwidth]{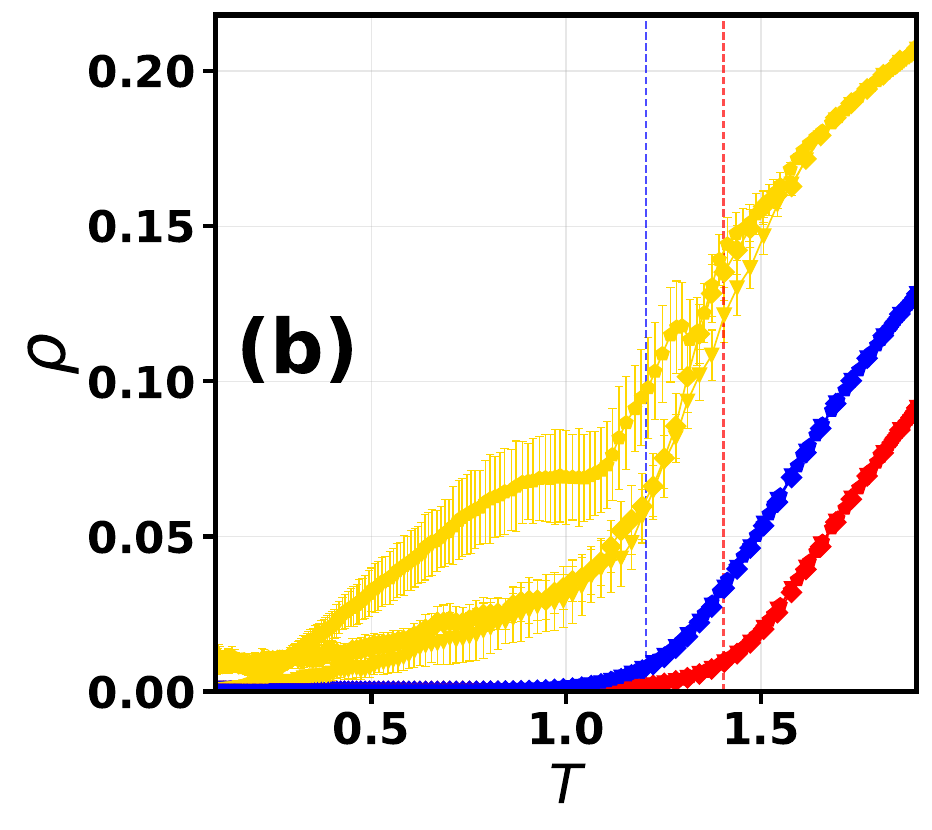}
    \end{tabular}
    \caption{Vortex densities $\rho(\Theta_0)$, $\rho(\Theta_1^{(1)})$, and $\rho(\Theta_1^{(2)})$ as functions of temperature for the two central regimes.
    $\Theta_0$ has finite self-energy, so $\rho(\Theta_0)$ is thermally generated at all temperatures;
    the PG-vortex densities rise rapidly near the independently determined transition temperature(s).
    (a)~Isotropic point: $\Theta_1^{(1)}$ and $\Theta_1^{(2)}$ proliferate simultaneously.
    (b)~Anisotropic case $J_1{=}1$, $J_2{=}0.6$, $J_3{=}1$: $\Theta_1^{(2)}$ proliferates before $\Theta_1^{(1)}$.}
    \label{fig:vortex}
\end{figure}

\textit{Monte Carlo results}.--- Because the Hamiltonian contains three- and four-site couplings, standard cluster algorithms (e.g., Wolff) are not directly applicable; we therefore employ single-spin Metropolis updates supplemented by parallel tempering~\cite{earlParallelTempering2005,hasenbuschTwodimensionalXYModel2005,olssonMonteCarloAnalysis1995I,olssonMonteCarloAnalysis1995} on $L\times L$ lattices with $L=16$--$64$. 
Critical temperatures are extracted
from the KT finite-size scaling form
$T_c(L)=T_c(\infty)+A/(\ln L)^2$~\cite{weberMonteCarloDetermination1988}. We combine
three diagnostics: the dipole-field correlation
ratio $C(L/2)/C(L/4)$~\cite{surungan2019berezinskii,okabe2025bkt,kong2025orientational},
whose KT threshold is $2^{-1/4}$; generalized helicity moduli, whose loss signals the
softening of the corresponding channel; and the vortex densities
$\rho(\Theta_0)$, $\rho(\Theta_1^{(1)})$, and $\rho(\Theta_1^{(2)})$, which identify the defects that
proliferate. The two central regimes are shown in
Figs.~\ref{fig:corr}--\ref{fig:vortex}; the $J_3{=}0$ and $J_2{=}0$
limits are presented in the End Matter, with extended data in \cite{SM}, Sec.~VI.

\noindent\textbf{(a) Isotropic ($J_1{=}J_2{=}J_3{=}1$).} At the isotropic
point, the two PG-vortex species are symmetry related and unbind
together. The dipole-field correlation ratio shows a single
scale-invariant crossing [Fig.~\ref{fig:corr}(a)], which determines the KT
critical temperature; KT finite-size scaling gives
$T_c\approx1.495$, with consistent estimates from the two channels
    [$1.4985(83)$ and $1.4912(78)$]. All three generalized helicity moduli drop at
this same temperature [Fig.~\ref{fig:stiffness}(a)], and
$\Theta_1^{(1)}$ and $\Theta_1^{(2)}$ proliferate simultaneously
    [Fig.~\ref{fig:vortex}(a)]. By contrast, $\rho(\Theta_0)$ is already
thermally populated because $\Theta_0$ has finite self-energy. Thus the single
transition is not driven by phase vortices, but by the simultaneous
unbinding of the two PG-vortex species. 

\noindent\textbf{(b) Anisotropic split ($J_1{=}1,\,J_2{=}0.6,\,J_3{=}1$).}
Anisotropy lifts the degeneracy between the two channels. The
correlation ratio shows two scale-invariant crossings
    [Fig.~\ref{fig:corr}(b)], determining the two critical temperatures
$T_c^{(2)}\approx1.20$ and $T_c^{(1)}\approx1.40$; correspondingly, the
helicity moduli display two distinct drops [Fig.~\ref{fig:stiffness}(b)]:
$J_{s2}$, which is destroyed by the proliferation of $\Theta_1^{(2)}$, drops at
$T_c^{(2)}$, while $J_{s1}$ and $J_{s3}$ remain finite until
$T_c^{(1)}$. 
The modulus $J_{s3}$ follows $J_{s1}$ rather than $J_{s2}$ because $\partial_1\partial_2\theta=\partial_2\chi_1=\partial_1\chi_2$ couples to both dipole channels: its stiffness borrows rigidity from whichever channel retains quasi-long-range order, and vanishes only when the last ordered channel (here $\chi_1$) loses rigidity at $T_c^{(1)}$ (\cite{SM}, Secs.~III and~V).
The vortex densities corroborate the sequence:
$\Theta_1^{(2)}$ proliferates before $\Theta_1^{(1)}$
[Fig.~\ref{fig:vortex}(b)]. Between these two temperatures, $\chi_2$ is
disordered while $\chi_1$ retains quasi-long-range order. This intermediate
partial dipole QLRO is the finite-temperature fingerprint of
selective PG-vortex unbinding: only one of the two
PG-vortex species has become unbound.

\noindent\textbf{(c) $J_3{=}0$ and (d) $J_2{=}0$ limits.}
The limiting cases separate the role of anisotropy from the structure of the
channels. For $J_3{=}0$, the two transitions re-merge into one even
when $J_1\neq J_2$, consistent with Eq.~\eqref{eq:eta}: both exponents then
share the factor $\sqrt{J_{s1}J_{s2}}$, so once either channel
vanishes, both dipole exponents diverge. For $J_2{=}0$, the
$\chi_2$ channel is absent and only $\Theta_1^{(1)}$ drives a single KT
transition. Thus the proliferation sequence is controlled not by anisotropy
alone, but by which higher-moment channels remain active. Detailed
data for these two regimes, including correlation ratios, helicity moduli, and
vortex densities, are given in the End Matter, and the full phase diagram is
shown in \cite{SM}, Sec.~V.
These numerical results complete the finite-temperature
phase diagram; we now discuss their implications.

\textit{Discussion}.--- We have explored a classical statistical model called ``dipole-conserving XY model'' that generalizes the standard XY model.  
Dipole conservation reconstructs, rather than eliminates, KT criticality by shifting the relevant topological defects from the phase field $\theta$ to its compact gradients $\chi_\alpha$. Consequently, the natural probes are dipole-field correlations, generalized helicity moduli, and PG-vortex densities, rather than conventional phase coherence and superfluid stiffness. The selective unbinding of the two PG-vortex species further shows that the finite-temperature phase structure is controlled by the number and coupling of active higher-moment channels, not by spatial anisotropy alone. Possible settings for this physics include  tilted optical lattices~\cite{lakeDipolarBoseHubbardModel2022,lakeDipoleCondensatesTilted2023,xuMultipolarCondensatesMultipolar2024a}, hole-doped antiferromagnets~\cite{sousFractonsPolarons2020,sousFractonsFrustrationHoledoped2020}, and honeycomb-lattice magnets~\cite{sanyalUnidirectionalSubsystemSymmetry2024}.

More broadly, compact higher derivatives of the phase support two higher-moment defect species generated by $\operatorname{Re}[(z/a)^n\ln(z/a)]$ and $\operatorname{Im}[(z/a)^n\ln(z/a)]$~\cite{yuanHierarchicalProliferationHigherrank2023a}. The dipole model is the first nontrivial member of this hierarchy. At the quadrupole level, our energetic analysis finds unequal defect self-energies even at the isotropic point, suggesting two KT transitions without explicit anisotropy. Whether this hierarchy survives coupling to a dynamical electromagnetic field, where gauge screening cuts off logarithmic vortex interactions beyond the penetration depth, remains an open question. Finally, PG-vortex dynamics under a temperature gradient may provide a setting for exploring analogues of vortex-driven Nernst phenomena. 

\textit{Acknowledgments}.--- We sincerely thank Yuan-Ming Lu for his enlightening discussions. This work was supported by the National Natural Science Foundation of China (NSFC) under Grant No.~12474149, the Research Center for Magnetoelectric Physics of Guangdong Province under Grant No.~2024B0303390001, and the Guangdong Provincial Key Laboratory of Magnetoelectric Physics and Devices under Grant No.~2022B1212010008. ZY is supported by the Scientific Research Project (No.~WU2025B011) and the Start-up Funding of Westlake
University.


\renewcommand{\thesection}{EM\arabic{section}}
\renewcommand{\theHsection}{EM\arabic{section}}
\renewcommand{\thesubsection}{\Alph{subsection}}
\renewcommand{\theHsubsection}{\Alph{subsection}}
\renewcommand{\theequation}{EM\arabic{equation}}
\renewcommand{\theHequation}{EM\arabic{equation}}
\renewcommand{\thefigure}{EM\arabic{figure}}
\renewcommand{\theHfigure}{EM\arabic{figure}}
\renewcommand{\thetable}{EM\arabic{table}}
\renewcommand{\theHtable}{EM\arabic{table}}

\setcounter{section}{0}
\setcounter{subsection}{0}
\setcounter{equation}{0}
\setcounter{figure}{0}
\setcounter{table}{0}
\bigskip

\begin{center}
    \textbf{ End Matter}\\[0.3em]
\end{center}

\vspace{1em}

\noindent

This End Matter provides the detailed numerical data and discussion for the
two supplementary parameter regimes---the $J_3=0$ merging case and the $J_2=0$
single-species limit---complementing the two central regimes presented in the
main text (Figs.~\ref{fig:corr}--\ref{fig:vortex}).
The corresponding
full Monte Carlo data with all system sizes, specific heat, and finite-size
scaling are collected in \cite{SM}, Sec.~VI.

\begin{figure}[t]
    \centering
    \includegraphics[width=0.95\columnwidth]{fig_legend_corr_ratio.pdf}\\[0pt]
    \begin{tabular}{@{}c@{\hspace{2pt}}c@{}}
        \includegraphics[width=0.48\columnwidth]{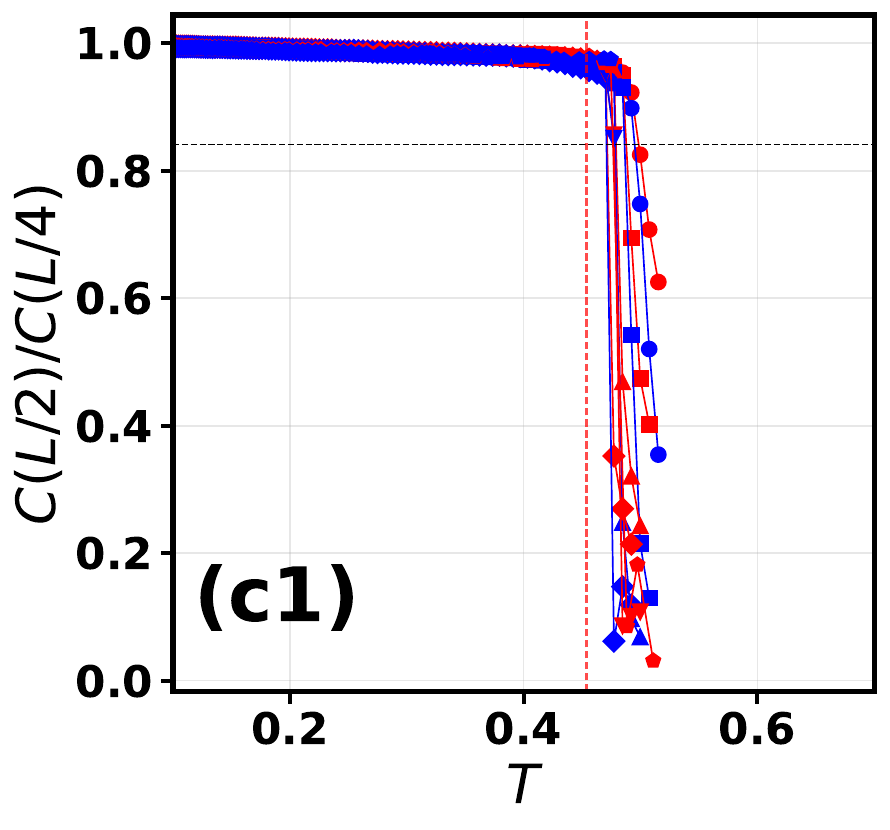} &
        \includegraphics[width=0.5\columnwidth]{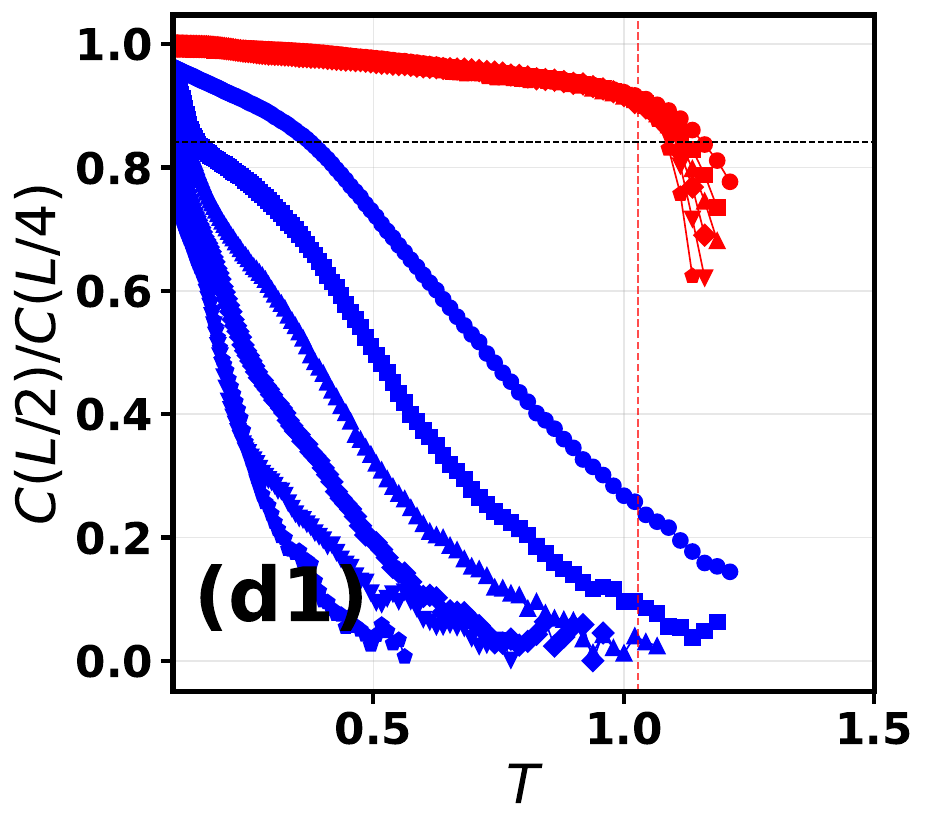}
    \end{tabular}
    \includegraphics[width=0.95\columnwidth]{fig_legend_stiffness.pdf}\\[0pt]
    \begin{tabular}{@{}c@{\hspace{2pt}}c@{}}
        \includegraphics[width=0.48\columnwidth]{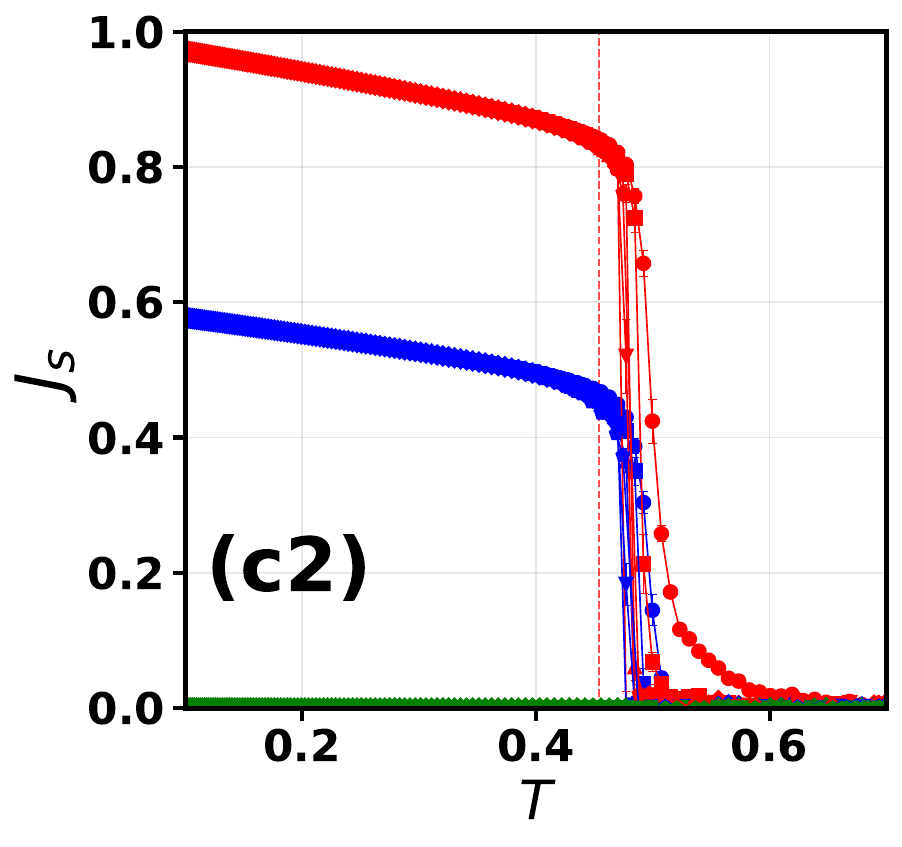} &
        \includegraphics[width=0.5\columnwidth]{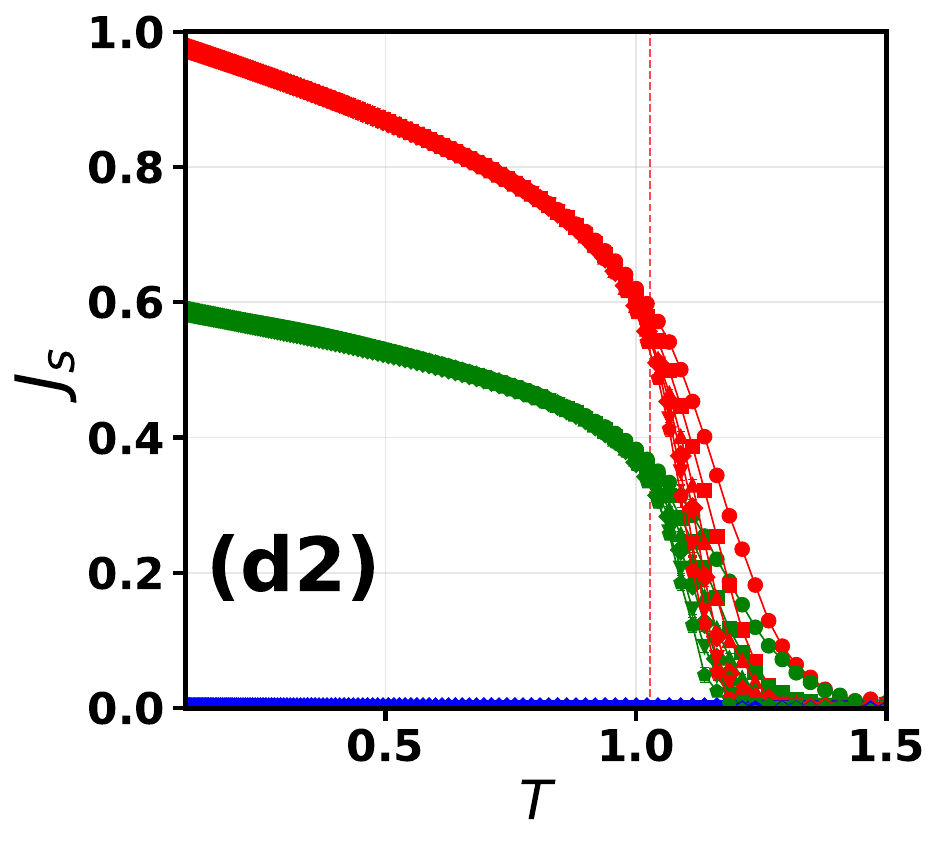}
    \end{tabular}
    \includegraphics[width=0.95\columnwidth]{fig_legend_vortex_density.pdf}\\[0pt]
    \begin{tabular}{@{}c@{\hspace{2pt}}c@{}}
        \includegraphics[width=0.48\columnwidth]{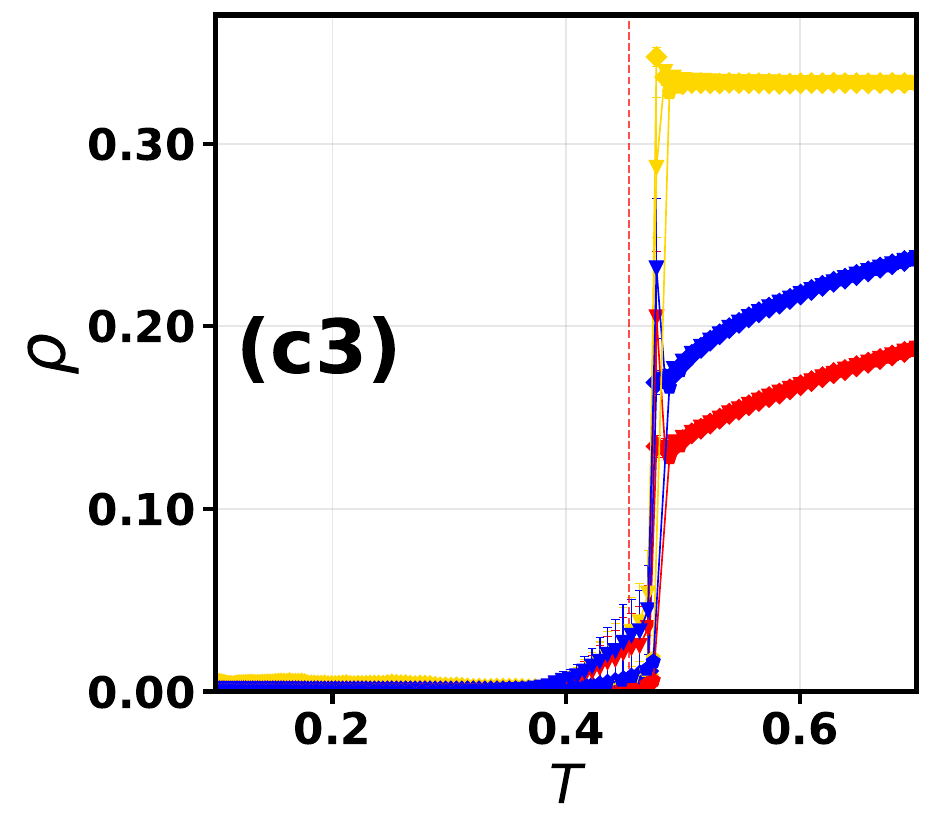} &
        \includegraphics[width=0.5\columnwidth]{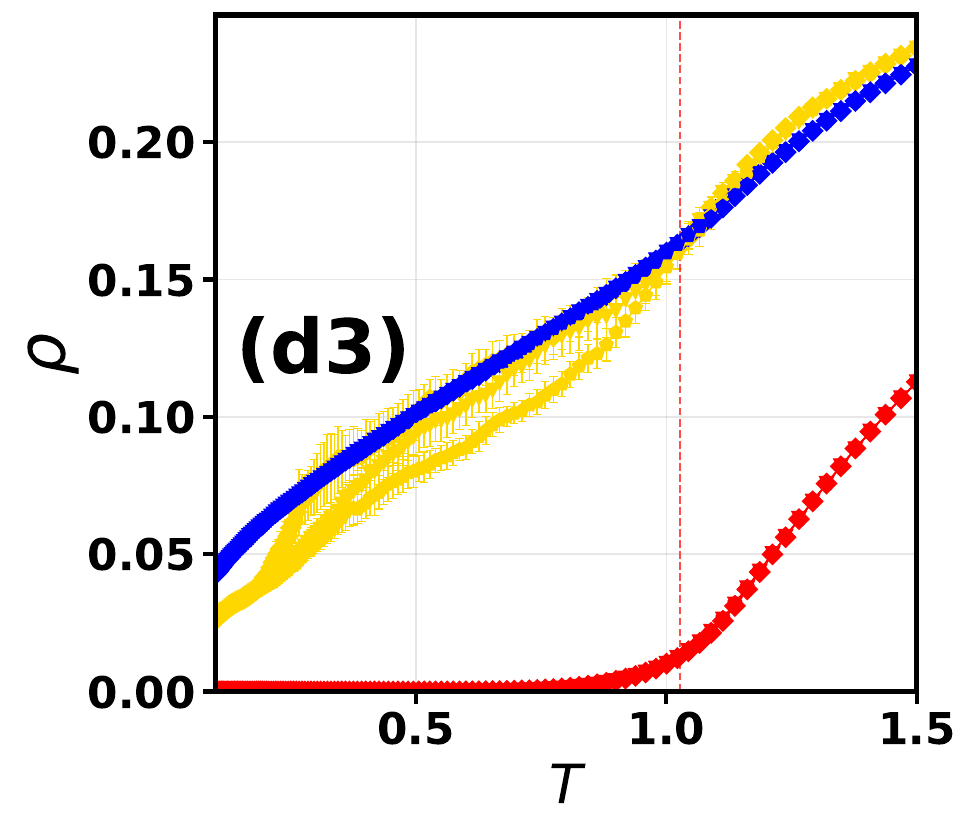}
    \end{tabular}
    \caption{Numerical data for the two limiting regimes: $J_3{=}0$ merging (left column, c1--c3) and $J_2{=}0$ single-species (right column, d1--d3), with $J_1{=}1$.
    Rows from top to bottom: dipole-field correlation ratio (c1,d1), generalized helicity moduli (c2,d2), and vortex densities (c3,d3).
    In the left column both PG-vortex species unbind simultaneously, whereas in the right column only one PG-vortex species controls the transition; all three diagnostics indicate a single KT transition.}
    \label{fig:em_data}
\end{figure}

\noindent\textbf{(c) $J_3{=}0$ merging ($J_1{=}1,\,J_2{=}0.6$).}
We consider $J_3=0$ while keeping $J_1\neq J_2$ (specifically $J_1=1$,
$J_2=0.6$, $J_3=0$). The $J_3$ coupling vanishes, so the continuum
Hamiltonian contains only $(\partial_1^2\theta)^2$ and $(\partial_2^2\theta)^2$.
One might expect that the two PG-vortex species become independent,
potentially preserving two separate transitions as in regime~(b) of the main
text. However, a closer inspection of the dipole correlation exponents [Eq.~\eqref{eq:eta},
$\eta_\alpha = 1/[2\sqrt{2}\,\pi\beta\sqrt{J_{s\alpha}(J_{s3}+\sqrt{J_{s1}J_{s2}})}]$]
reveals a different outcome: the two transitions are expected to merge.
Because the modulus $J_{s3}$ vanishes when $J_3=0$, both $\eta_1$ and $\eta_2$ involve
$\sqrt{J_{s1}J_{s2}}$ in their denominator. Once either $J_{s1}$ or $J_{s2}$
drops to zero---i.e., once either PG-vortex species proliferates---
both exponents diverge simultaneously, destroying dipole quasi-long-range order
in both channels at once. There is therefore no intermediate phase in which one
channel retains power-law correlations while the other is disordered. This is the microscopic
reason why the two unbinding transitions coincide along the $J_3=0$ line,
consistent with the Monte Carlo results.

The correlation ratio Fig.~\ref{fig:em_data}(c1) exhibits a
single scale-invariant crossing, determining the single KT transition;
despite $J_1\neq J_2$, the generalized helicity moduli Fig.~\ref{fig:em_data}(c2) also drop at the same temperature, confirming it, and
the vortex densities Fig.~\ref{fig:em_data}(c3) show
simultaneous proliferation of both PG-vortex species. $\rho(\Theta_0)$ is in principle nonzero because $\Theta_0$ has finite
self-energy, but its magnitude is suppressed below numerical resolution at the
lowest temperatures in this regime.

Put differently, when $J_3$ is removed, the Hamiltonian
becomes invariant under $\theta\to\theta+\gamma_3(x_1/a)(x_2/a)$, which is precisely the
quadratic twist used to define $J_{s3}$ in Eq.~\eqref{eq:J_s}; consequently $J_{s3}=0$ and
the two PG-vortex sectors cannot undergo independent KT unbinding.

\noindent\textbf{(d) $J_2{=}0$ single PG-vortex KT transition ($J_1{=}1,\,J_3{=}0.6$).}
We now turn to the extreme-anisotropy limit $J_1=1$, $J_2=0$, $J_3=0.6$. In
this regime, the $(\partial_2^2\theta)^2$ term is absent from the continuum
Hamiltonian, which fundamentally alters the physics of the $\Theta_1^{(2)}$
vortex. The system acquires a higher-moment symmetry along the $x_2$ direction.
The observables confirm the altered physics: the correlation ratio exhibits a
single scale-invariant crossing, the generalized helicity moduli show a single
drop, and only $\Theta_1^{(1)}$ proliferates at the transition, confirming
a single KT transition.

The correlation ratio Fig.~\ref{fig:em_data}(d1) displays a
single crossing, determining the single KT transition; the generalized
helicity moduli Fig.~\ref{fig:em_data}(d2) show a single
modulus drop, confirming it, and the vortex densities Fig.~\ref{fig:em_data}(d3) confirm that only $\Theta_1^{(1)}$ proliferates, while
$\rho(\Theta_1^{(2)})$ is nonzero at all temperatures and does not control the
transition. This limiting case provides a direct consistency check on the
PG-vortex framework: when a coupling channel is removed, the
higher-moment symmetry of the theory is enhanced, the observables (correlation
ratio, helicity moduli, vortex densities) reflect the altered vortex
energetics, and the second independent KT transition disappears.

Together, these two limiting regimes sharpen the main-text phase diagram.
The $J_3=0$ case shows that it is $J_3$, not
anisotropy alone, that enables independent unbinding of the two PG-vortex
species. The $J_2=0$ case confirms that when the $J_2$ channel is
removed, the corresponding PG-vortex sector
ceases to drive criticality. Both results
reinforce the central conclusion: the finite-temperature behavior of the
dipole-conserving XY model is governed by which higher-moment channels remain
active, consistent with the PG-vortex mechanism established in the main text.


%

\clearpage
\onecolumngrid

\renewcommand{\thesection}{\Roman{section}}
\renewcommand{\theHsection}{SM\Roman{section}}
\renewcommand{\thesubsection}{\Alph{subsection}}
\renewcommand{\theHsubsection}{SM\arabic{section}\Alph{subsection}}
\renewcommand{\theequation}{S\arabic{equation}}
\renewcommand{\theHequation}{SM\arabic{equation}}
\renewcommand{\thefigure}{S\arabic{figure}}
\renewcommand{\theHfigure}{SM\arabic{figure}}
\renewcommand{\thetable}{S\arabic{table}}
\renewcommand{\theHtable}{SM\arabic{table}}

\setcounter{section}{0}
\setcounter{subsection}{0}
\setcounter{equation}{0}
\setcounter{figure}{0}
\setcounter{table}{0}

\begin{center}
    \textbf{\large Supplemental Material}\\[0.3em]

\end{center}

\vspace{1em}
This Supplemental Material provides the technical derivations and extended
numerical results that support the main text and End Matter. Its structure is
organized to parallel the logic of the letter version.

\begin{enumerate}
    \item In Sec.~I, we characterize the intrinsic windings and
          branch-cut structure of the PG vortices, derive how the
          conventional far-field winding of a neutral pair is determined by its dipole
          moment, and extend the construction to quadrupole-conserving models.
    \item In Sec.~II, we present the detailed vortex energetics, showing
          that the phase vortex has finite self-energy while the PG
          vortices have logarithmically divergent self-energies, and we derive the energy
          and far-field winding of a PG-vortex dipole.
    \item In Sec.~III, we derive the Gaussian continuum correlation
          functions used in the main text [Eq.~\eqref{eq:eta}], including the power-law
          dipole-field correlators and the absence of quasi-long-range order in the
          original phase field.
    \item In Sec.~IV, we develop a             leading-order Coulomb-gas picture and
            simplified RG description of the PG-vortex sector. This
            analysis explains the KT criterion $\eta_\alpha=\sfrac{1}{4}$ and how anisotropy can split the isotropic transition into two.
    \item In Sec.~V, we present the full Monte Carlo phase
          diagram and its representative slices. These figures make explicit how the
          single isotropic KT transition splits into two under anisotropy and
          recombines along the $J_3=0$ line.
    \item In Sec.~VI, we collect representative numerical data for all
          four parameter regimes: the isotropic case, the anisotropy-induced
          split-transition case, the $J_3=0$ merging case, and the $J_2=0$ single-species
          limit.
\end{enumerate}

\clearpage

\section{Complex-analysis formulation of PG vortices}
\label{sec:sm_complex}

In this section, we provide a concise mathematical formulation of the three
vortex species using complex analysis. The purpose is to characterize the
intrinsic windings and branch-cut structure of the PG vortices and
to derive how the conventional far-field winding of a neutral pair is
determined by its dipole moment: a single PG vortex carries no
localized, contour-independent phase-vortex charge, while a pair
of them reconstructs a phase vortex in the far field with winding set by
their separation.

\subsection{Complex representation and branch-cut structure}

The three vortex configurations can be written in Cartesian coordinates:
\begin{align}
    \Theta_0(\mathbf{x})       & = \ell_0\,\varphi(\mathbf{x}),                                                         \\
    \Theta_1^{(1)}(\mathbf{x}) & = \ell_1\!\left(\frac{x_1}{a}\,\varphi(\mathbf{x})+\frac{x_2}{a}\ln\frac{r}{a}\right), \\
    \Theta_1^{(2)}(\mathbf{x}) & = \ell_2\!\left(\frac{x_2}{a}\,\varphi(\mathbf{x})-\frac{x_1}{a}\ln\frac{r}{a}\right),
\end{align}
with $\varphi(\mathbf{x})$ the polar angle ($x_1+ix_2=re^{i\varphi}$) and $r=|\mathbf{x}|$.
These three expressions are unified by a single complex
function: with the principal branch $\ln(z/a)=\ln(r/a)+i\varphi$, the main-text forms
\begin{align}
    \Theta_0(\mathbf{x})       & = \ell_0\,\operatorname{Im}\bigl[\ln(z/a)\bigr],      \\
    \Theta_1^{(1)}(\mathbf{x}) & = \ell_1\,\operatorname{Im}\bigl[(z/a)\ln(z/a)\bigr], \\
    \Theta_1^{(2)}(\mathbf{x}) & = -\ell_2\,\operatorname{Re}\bigl[(z/a)\ln(z/a)\bigr]
\end{align}
follow immediately.  The PG vortices involve an extra factor of $z$;
expanding $(z/a)\ln(z/a)$ as
\begin{align}
    \frac{z}{a}\ln\frac{z}{a}
      = (\frac{x_1}{a}+i\frac{x_2}{a})\bigl(\ln\frac{r}{a}+i\varphi\bigr)= \bigl(\frac{x_1}{a}\ln\frac{r}{a} - \frac{x_2}{a}\varphi\bigr)
    \;+\; i\bigl(\frac{x_1}{a}\varphi + \frac{x_2}{a}\ln\frac{r}{a}\bigr)
\end{align}
makes the connection to the real-space expressions above explicit.
These complex forms are single-valued only after a branch of the logarithm
has been chosen, i.e., on the plane with the vortex core and branch cut
removed.

The logarithm $\ln(z/a)$ is multivalued: its imaginary part $\varphi$ jumps by
$2\pi$ when one crosses the branch cut. We adopt the standard principal branch
with the cut on the negative real axis,
\begin{equation}
    -\pi<\varphi\le\pi,
\end{equation}
so that $\varphi$ is the polar angle restricted to $(-\pi,\pi]$.  Denote by $\Delta_C$ the
discontinuity of a function across its branch cut (the value just
above the cut minus that just below it).  For the logarithm itself,
\begin{equation}
    \Delta_C\ln(z/a)=2\pi i.
\end{equation}
Since the three vortex functions are linear in $\ln(z/a)$ or
$(z/a)\ln(z/a)$, their jumps across the cut follow:
\begin{align}
    \Delta_C\Theta_0
     & = \ell_0\,\operatorname{Im}[2\pi i]
    = 2\pi\ell_0,                                               \\
    \Delta_C\Theta_1^{(1)}
     & = \ell_1\,\operatorname{Im}\bigl[(z/a)\cdot2\pi i\bigr]
    = 2\pi\ell_1\,(x_1/a),                                      \\
    \Delta_C\Theta_1^{(2)}
     & = -\ell_2\,\operatorname{Re}\bigl[(z/a)\cdot2\pi i\bigr]
    = 2\pi\ell_2\,(x_2/a).
\end{align}
The phase vortex $\Theta_0$ thus has a constant jump
$2\pi\ell_0$---the familiar $2\pi$ branch cut of an XY vortex---whereas the
PG vortices have position-dependent jumps linear in
$x_1$ or $x_2$.  This is the analytic hallmark of the PG
species: unlike a phase vortex, its branch-cut discontinuity is
not a constant $2\pi$ winding.

The phase topological charge is the winding number of $\theta$, defined by
\begin{equation}
    \frac{1}{2\pi}\oint_{\mathcal C} d\mathbf{l}\cdot\nabla\theta = \ell_0,
    \qquad
    \frac{1}{2\pi}\oint_{\mathcal C} d\mathbf{l}\cdot\nabla\chi_\alpha = \ell_\alpha,
    \quad\alpha=1,2,
\end{equation}
with $\ell_0,\ell_1,\ell_2\in\mathbb{Z}$.
For $\Theta_0$ with $\ell_0=1$, the contour integral yields
$1$ independently of the contour size. For a single $\Theta_1^{(1)}$, however,
a direct evaluation on a circle of radius $R$ gives
\begin{equation}
    \frac{1}{2\pi}\oint_{\mathcal C} d\mathbf{l}\cdot\nabla\Theta_1^{(1)}
    = -\frac{R}{a}\,\ell_1,
\end{equation}
which depends linearly on $R$ and is therefore \emph{not} a
topological invariant.  Even on the lattice, where $R$ takes discrete
values $R=n a$ with $n\in\mathbb{Z}$ and the winding number $n\ell_1$
remains an integer (consistent with the compactness of $\theta$), the
result still varies with the contour size $n$.  A genuine topological
charge must be independent of the integration path, so the phase
winding number fails as a detector for a single PG vortex.
Thus, in the continuum, $\Theta_1^{(1)}$ is not a phase vortex
of the $\theta$ field.  The same conclusion holds for $\Theta_1^{(2)}$.

The continuum analysis above shows that a single PG vortex lacks a
well-defined phase winding number. We now examine how this manifests on
the lattice, where the natural diagnostic is the plaquette winding number.

\subsection{Lattice plaquette winding numbers}

The situation is more subtle and more physically revealing on the lattice,
where the fundamental detector of phase vortices is the compact
plaquette winding number
\begin{equation}
    q_{\mathbf{i}}^{(0)} =
    \frac{1}{2\pi}\sum_{{}_{\mathbf{i}}\square} [\Delta\theta]_{\pi},
\end{equation}
which measures the discrete circulation of $\theta$ around an
elementary plaquette.
For the phase vortex $\Theta_0$ with $\ell_0=1$, the result is
the expected topological signature:
\begin{equation}
    q_{\mathbf{i}}^{(0)} =
    \begin{cases}
        +1, & \text{plaquette containing the core}, \\[2pt]
        0,  & \text{all other plaquettes}.
    \end{cases}
\end{equation}
The winding number is localized at the core.

For $\Theta_1^{(1)}$ with $\ell_1=1$, by contrast, no such localized signal
exists. The outcome depends on the placement of the core relative to lattice
sites. When the core coincides with a lattice site, $q^{(0)} = 0$ on every
plaquette. When the core lies between sites, adjacent $\pm1$ values appear
along the branch cut of the azimuthal angle $\varphi$ (the negative
$x_1$-axis). These are non-topological artifacts of the principal-value
reduction $[\cdot]_{\pi}$: the discontinuity $2\pi(x_1/a)$ falls near $\pm\pi$
when $x_1/a$ is half-integer, and the reduction arbitrarily rounds it to
$\pm\pi$, producing $q^{(0)}=\pm1$. Any infinitesimal fluctuation (thermal,
disorder, or round-off) removes these spurious plaquettes, confirming their
non-topological nature. They are not associated with the core, can be shifted
by moving the core relative to the lattice, and sum to an extensive total.
Crucially, regardless of placement, $q_{\mathbf{i}_{\text{core}}}^{(0)} = 0$,
where $\mathbf{i}_{\text{core}}$ denotes the plaquette containing the
$\Theta_1^{(1)}$ core---the plaquette where the dipole-field winding number
$q^{(1)} = \pm1$ resides.  The same conclusion holds for
$\Theta_1^{(2)}$.

The underlying reason is clear from the definitions. The generalized winding
numbers for the dipole fields are defined analogously by replacing $\theta$
with $\chi_1$ or $\chi_2$ in \eqref{eq:vorticity}:
\begin{equation}
    q_{\mathbf{i}}^{(1)} =
    \frac{1}{2\pi}\sum_{{}_{\mathbf{i}}\square} [\Delta\chi_1]_{\pi},\qquad
    q_{\mathbf{i}}^{(2)} =
    \frac{1}{2\pi}\sum_{{}_{\mathbf{i}}\square} [\Delta\chi_2]_{\pi}.
\end{equation}
While $\nabla\Theta_1^{(a)}$ carries nontrivial circulation in the
$\chi_a$ channel---giving $q_{\text{core}}^{(a)} = \pm1$---the
phase plaquette number $q^{(0)}$, which measures circulation in
$\theta$ alone, registers zero.  In summary,
\begin{align}
    \Theta_0       & : \quad q_{\text{core}}^{(0)} = +1,\quad
    q_{\text{core}}^{(1)} = q_{\text{core}}^{(2)} = 0,        \\[2pt]
    \Theta_1^{(1)} & : \quad q_{\text{core}}^{(0)} = 0,\quad
    q_{\text{core}}^{(1)} = +1,\quad
    q_{\text{core}}^{(2)} = 0,                                \\[2pt]
    \Theta_1^{(2)} & : \quad q_{\text{core}}^{(0)} = 0,\quad
    q_{\text{core}}^{(1)} = 0,\quad
    q_{\text{core}}^{(2)} = +1.
\end{align}

Single PG vortices therefore carry no localized
phase-vortex charge, even though they remain nontrivial vortices of the
generalized theory. Their topological charge is carried by the dipole fields
$\chi_1$ and $\chi_2$, detected by the generalized winding numbers $q^{(1)}$
and $q^{(2)}$, not by $\theta$ itself. This is precisely why $\rho(\Theta_0)$
remains thermally generated (finite self-energy) and does not vanish at the
KT transitions in the Monte Carlo simulations (Fig.~\ref{fig:vortex}): the
phase-vortex density is decoupled from the unbinding of the
PG-vortex species that drives the KT transitions, although its
magnitude may be strongly suppressed in certain parameter regimes (e.g.,
$J_3=0$).

The results above establish that an isolated PG vortex carries no
phase-vortex charge on the lattice. We now turn to the complementary
question: how does a \emph{pair} of PG vortices reconstruct a conventional
phase vortex at long distances?

\subsection{Neutral PG vortex pairs and far-field winding}

Although a single $\Theta_1^{(1)}$ carries no localized,
contour-independent phase-vortex charge, a neutral pair of two opposite
PG vortices reproduces a phase vortex texture in the far
field, with conventional winding set by their separation. Consider two
$\Theta_1^{(1)}$ vortices of opposite unit charge located at $\mathbf{0}$
(charge $+1$) and at $d\hat{x}_1$ (charge $-1$). Using the complex
representation, the combined configuration is
\begin{equation}
    \Theta_{\text{pair}}^{(1)}(\mathbf{x};d)
    = \operatorname{Im}\!\bigl[(z/a)\ln(z/a)\bigr]
    - \operatorname{Im}\!\bigl[((z-d)/a)\ln((z-d)/a)\bigr].
\end{equation}
For $|z|\gg d$, set $w=z/a$ and $\delta=d/a$, and expand
$(w-\delta)\ln(w-\delta)$:
\begin{align}
    (w-\delta)\ln(w-\delta)
      = (w-\delta)\Bigl[\ln w + \ln\!\Bigl(1-\frac{\delta}{w}\Bigr)\Bigr] \approx w\ln w - \delta\ln w - \delta
    + \mathcal O\!\Bigl(\frac{\delta^2}{w}\Bigr).
\end{align}
Substituting this into the pair field gives
\begin{align}
    \Theta_{\text{pair}}^{(1)}
     & \approx \operatorname{Im}\!\Bigl[w\ln w
        - w\ln w + \delta\ln w + \delta\Bigr]
    + \mathcal O\!\Bigl(\frac{d^2}{ar}\Bigr)                         \nonumber\\
     & = \delta\,\operatorname{Im}\!\bigl[\ln(z/a)\bigr]
    + \mathcal O\!\Bigl(\frac{d^2}{ar}\Bigr) = \frac{d}{a}\,\varphi(\mathbf{x})
    + \mathcal O\!\Bigl(\frac{d^2}{ar}\Bigr),
\end{align}
where the real constant $\delta$ drops out of the imaginary part. To leading
order in $d/r$, the pair therefore carries the far-field conventional winding
\begin{equation}
    \ell_0^{\text{far}} = \frac{1}{2\pi}\oint_{\mathcal C}
    \nabla\Theta_{\text{pair}}^{(1)}\cdot d\mathbf{l} = \frac{d}{a},
\end{equation}
with branch-cut discontinuity $\Delta_C\Theta_{\text{pair}}^{(1)}=2\pi d/a$.
The minimal lattice dipole with $d=a$ is the special case that reproduces a
unit phase vortex ($\ell_0^{\text{far}}=1$); increasing the separation
yields proportionally larger conventional winding, reflecting the dipole
moment of the PG-vortex charges. A $\Theta_1^{(2)}$ pair separated
along $\hat{x}_2$ obeys the completely analogous relation.

The branch-cut discontinuities tell the same story. Choose the logarithmic
branch cuts of the two PG vortices to run parallel to the $-x_1$
direction (the negative real axis) from their respective cores. Then, for the
two vortices separately,
\begin{equation}
    \Delta_C\Theta_1^{(1)}\big|_{\mathbf{0}} = 2\pi\,(x_1/a),\qquad
    \Delta_C\Theta_1^{(1)}\big|_{d\hat{x}_1} = -2\pi\,(x_1/a-\delta),
\end{equation}
so their sum is
\begin{equation}
    \Delta_C\Theta_{\text{pair}}^{(1)}
    = 2\pi\,(x_1/a) - 2\pi\,(x_1/a-\delta) = 2\pi\delta = 2\pi\frac{d}{a}.
\end{equation}
The linear position dependence cancels, leaving the constant jump
$2\pi d/a$; for $d=a$ this reduces to the standard jump $2\pi$
characteristic of $\Theta_0$.  Physically, this
cancellation expresses the fact that a neutral pair of PG
vortices is indistinguishable from a phase vortex at long
distances, with the far-field winding set by the pair separation, i.e., by
the dipole moment of its PG-vortex charges.

\begin{figure}[htbp]
    \centering
    \begin{tabular}{@{}c@{\hspace{3pt}}c@{\hspace{3pt}}c@{\hspace{3pt}}c@{}}
        \includegraphics[width=0.23\textwidth]{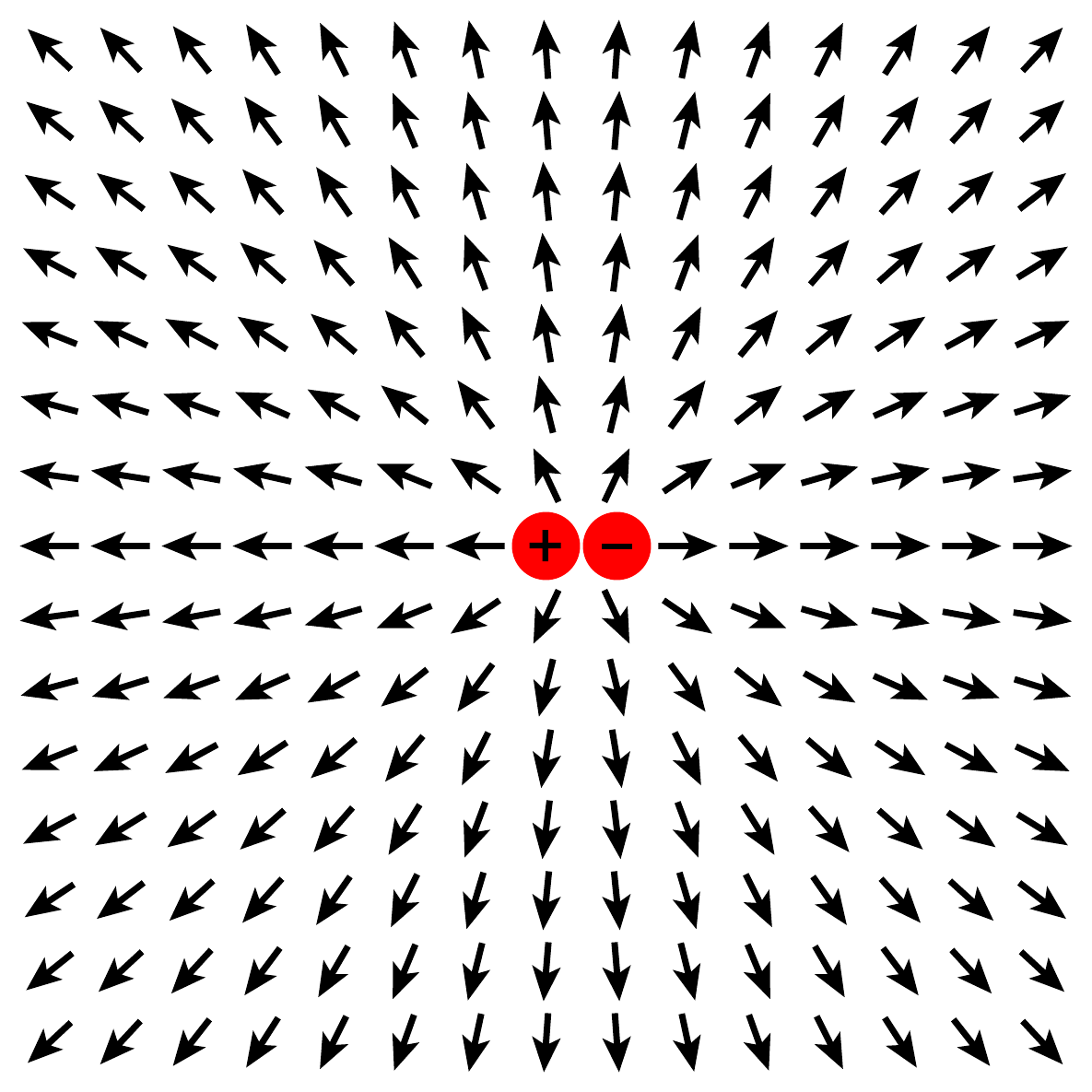} &
        \includegraphics[width=0.23\textwidth]{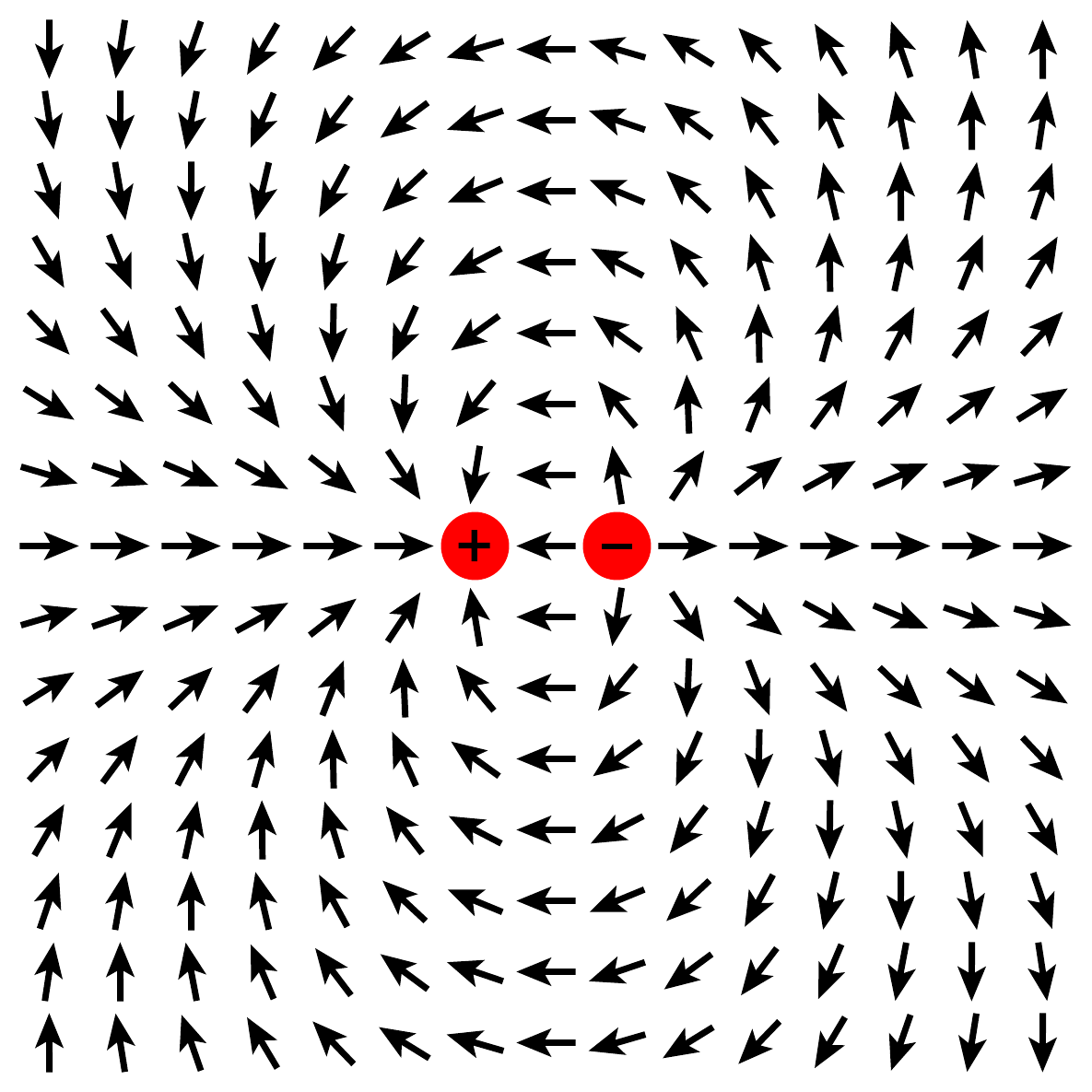} &
        \includegraphics[width=0.23\textwidth]{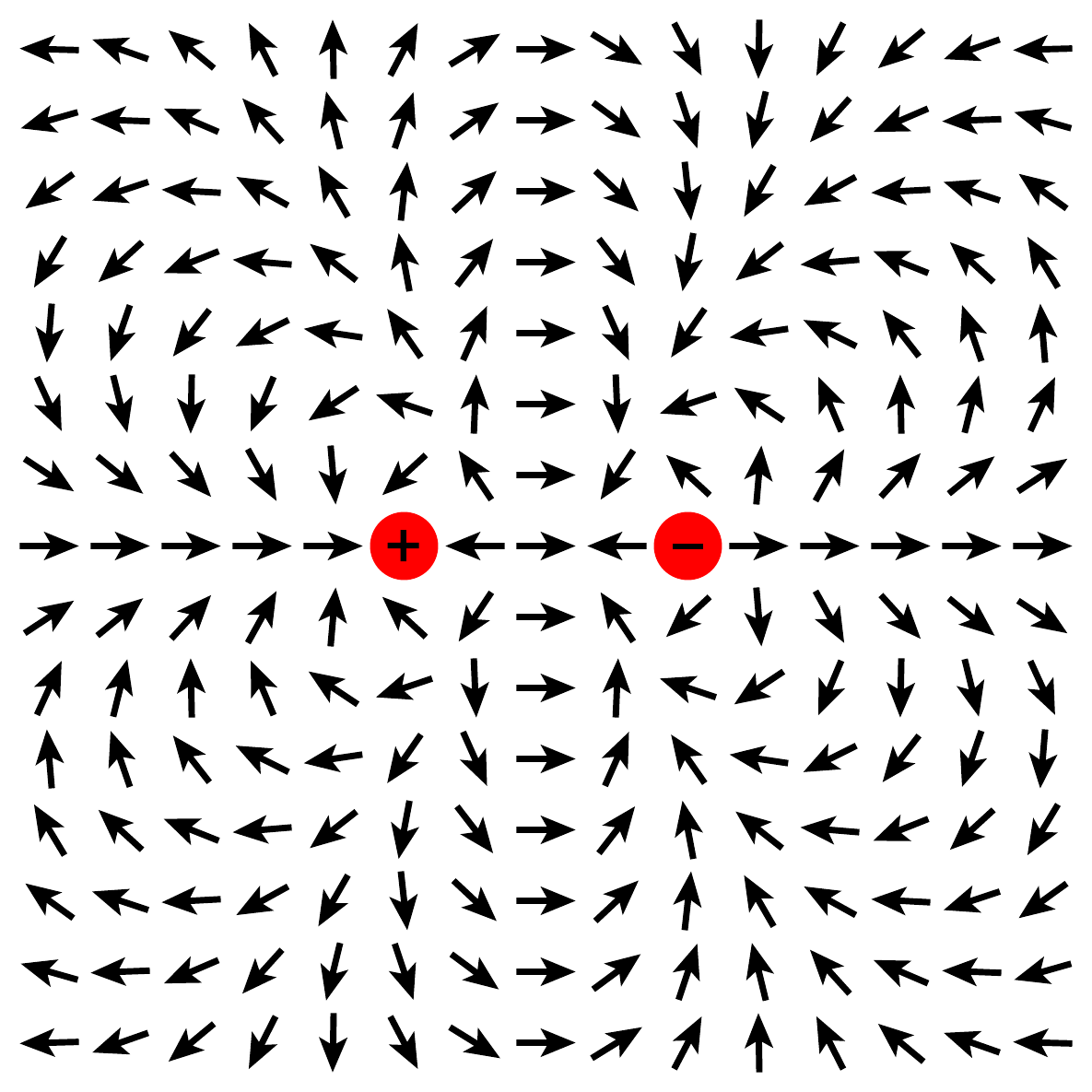} &
        \includegraphics[width=0.23\textwidth]{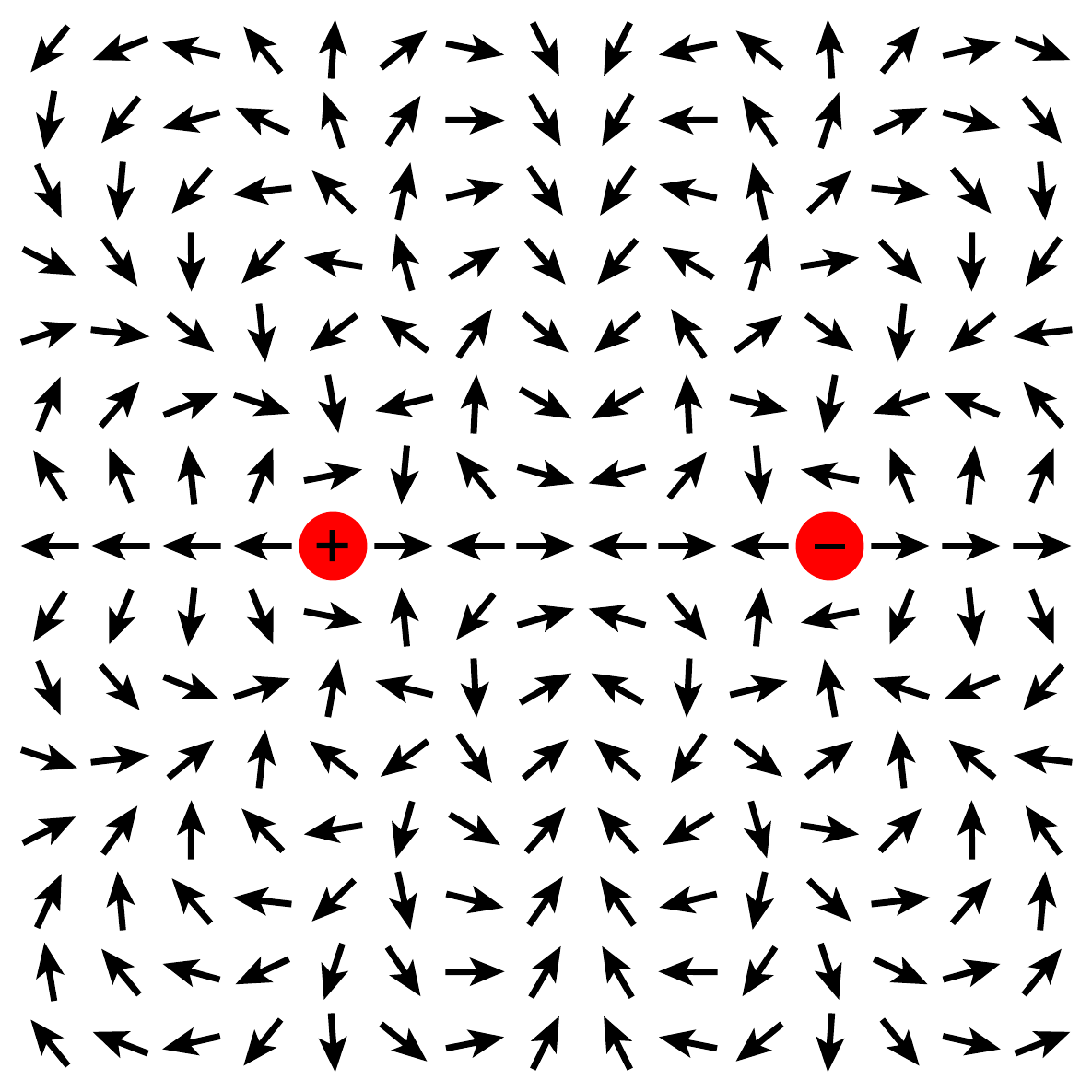} \\[4pt]
        \includegraphics[width=0.23\textwidth]{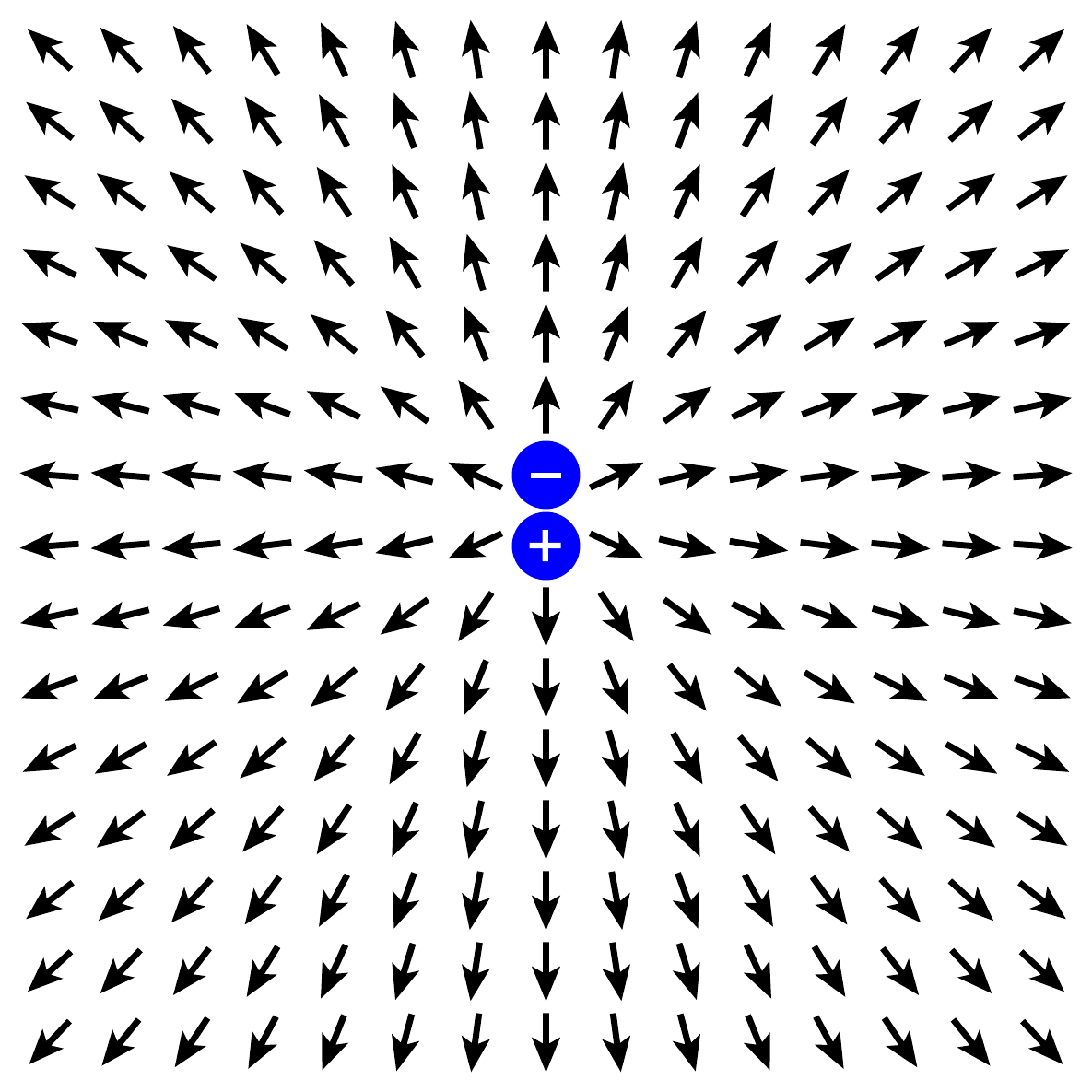} &
        \includegraphics[width=0.23\textwidth]{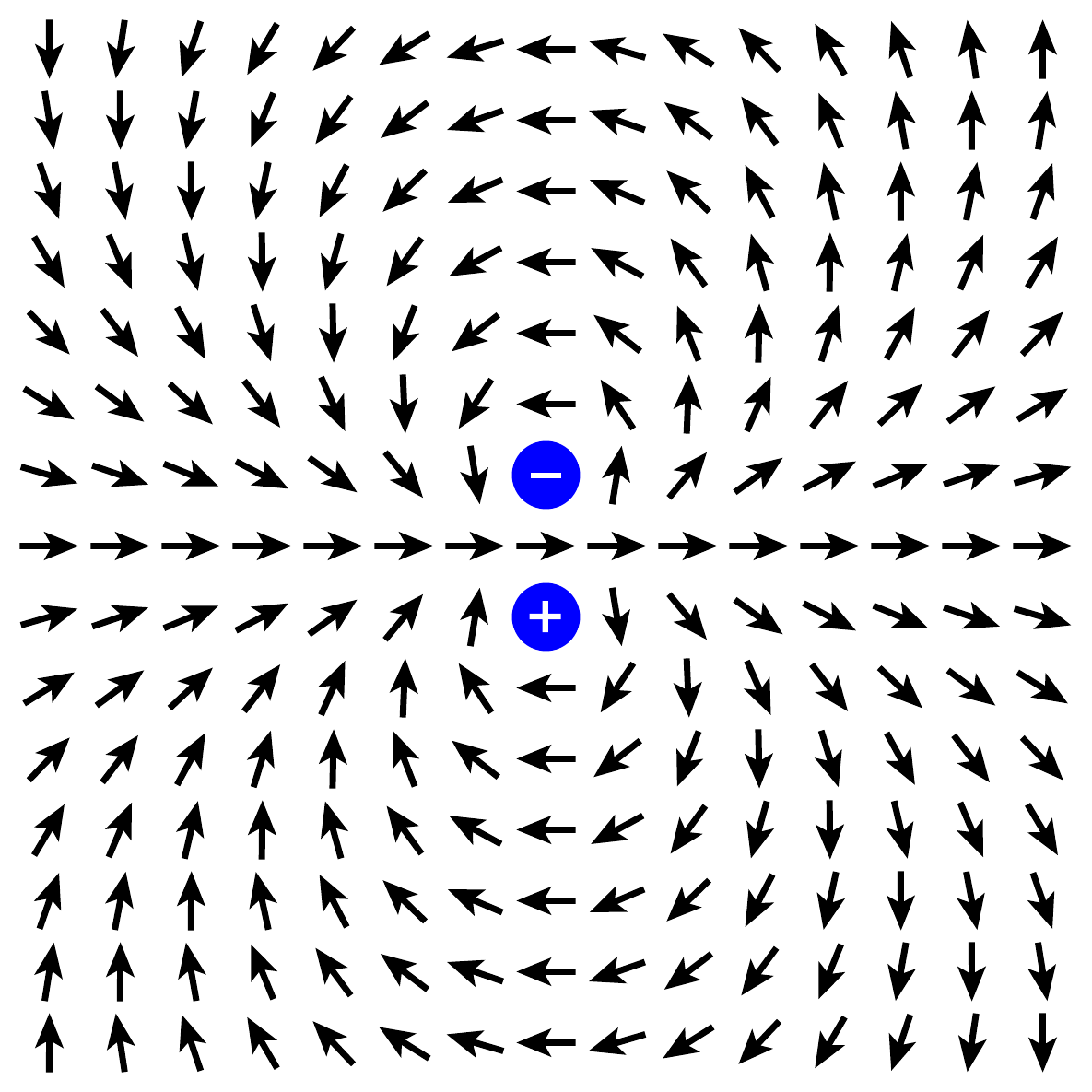} &
        \includegraphics[width=0.23\textwidth]{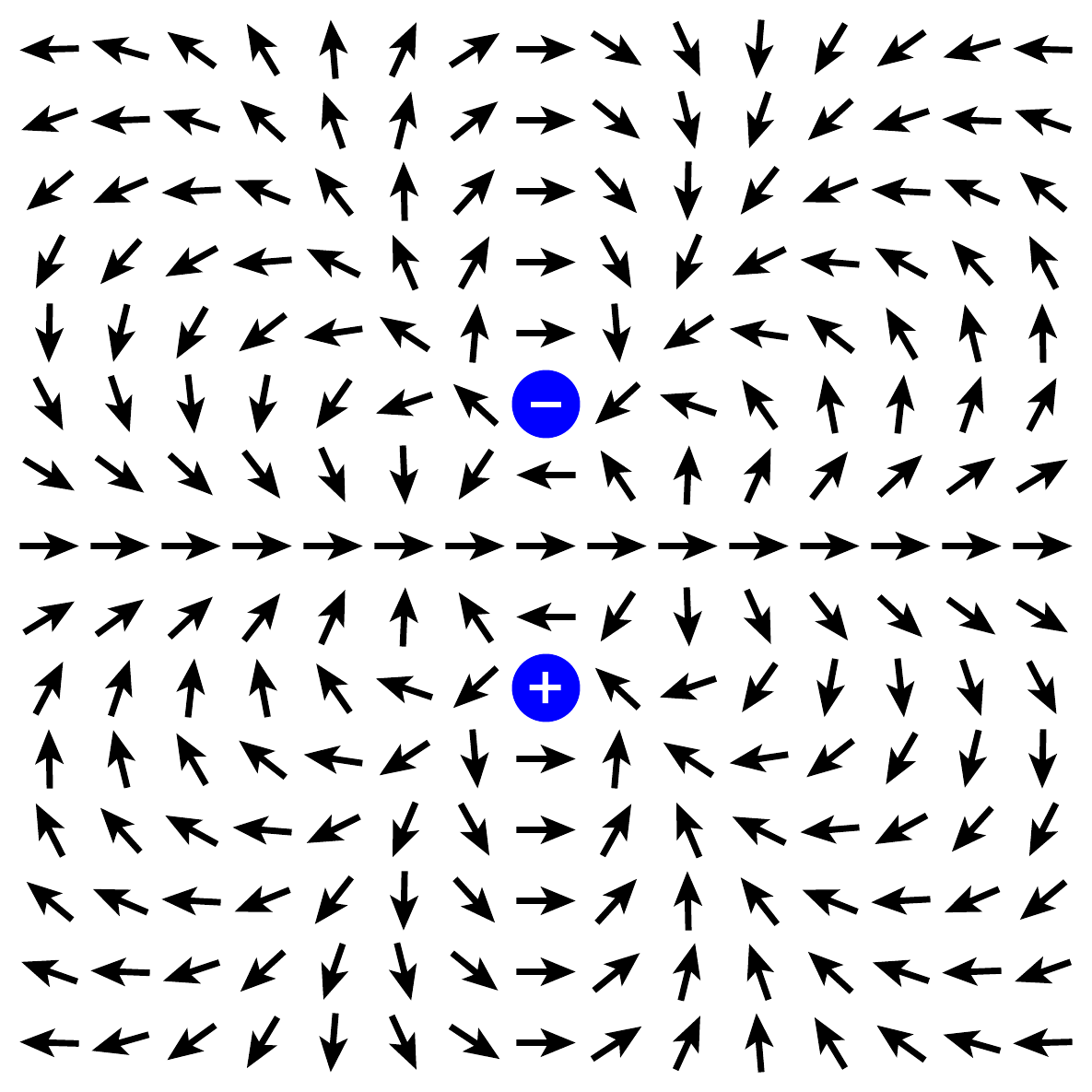} &
        \includegraphics[width=0.23\textwidth]{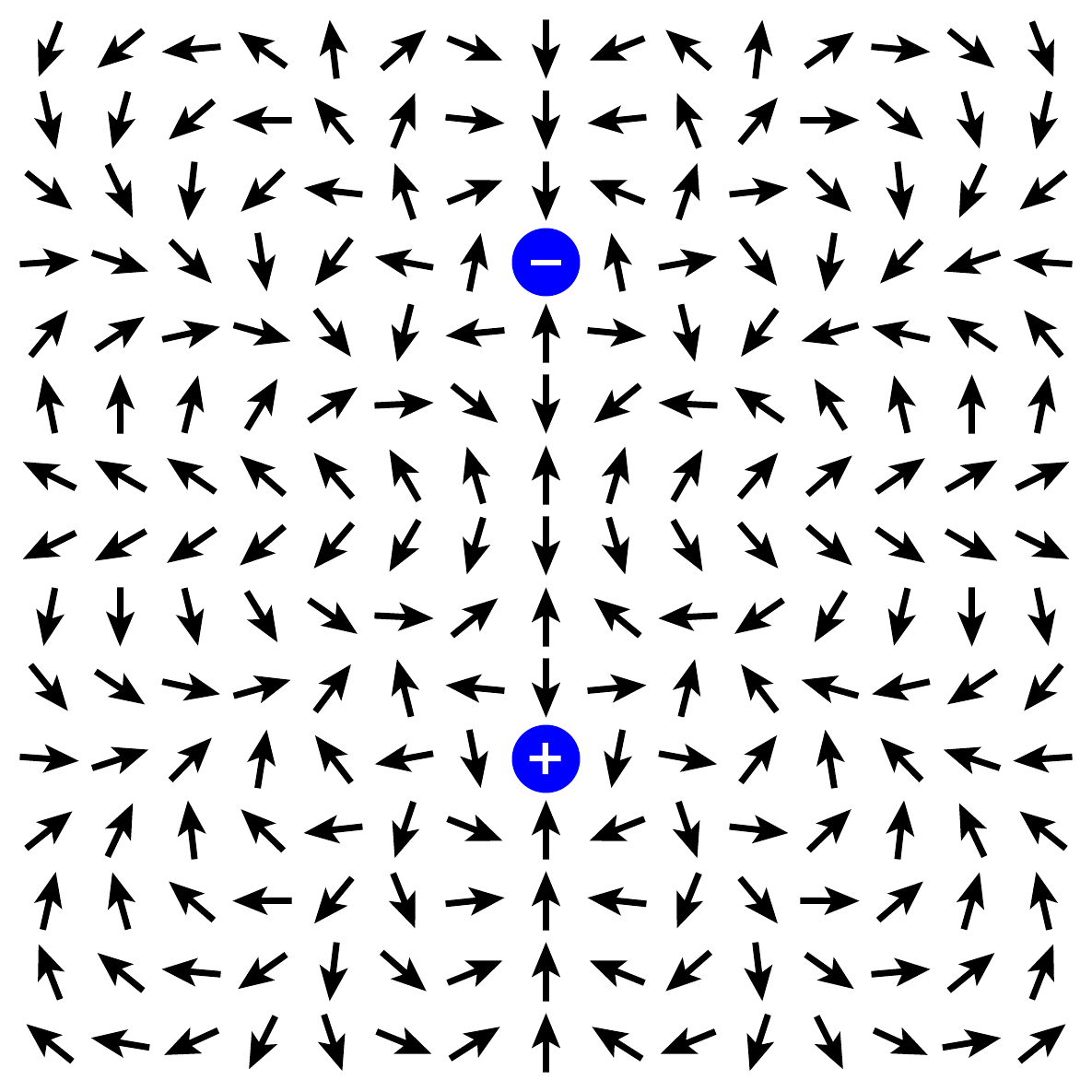}
    \end{tabular}
    \caption{Vector-field visualization of the PG vortex pairs.
    Top row: $\Theta_1^{(1)}$ (red), separated along $\hat{x}_1$.
    Bottom row: $\Theta_1^{(2)}$ (blue), separated along $\hat{x}_2$.
    From left to right, the pair separation increases as $d/a=1,2,4,7$, giving far-field conventional windings $\ell_0^{\rm far}=1,2,4,7$, demonstrating that the far-field winding grows linearly with the dipole moment of the PG-vortex charges.}
    \label{fig:sm_pg_pairs}
\end{figure}

\begin{figure}[htbp]
    \centering
    \begin{tabular}{@{}c@{\hspace{3pt}}c@{\hspace{3pt}}c@{\hspace{3pt}}c@{}}
        \includegraphics[width=0.23\textwidth]{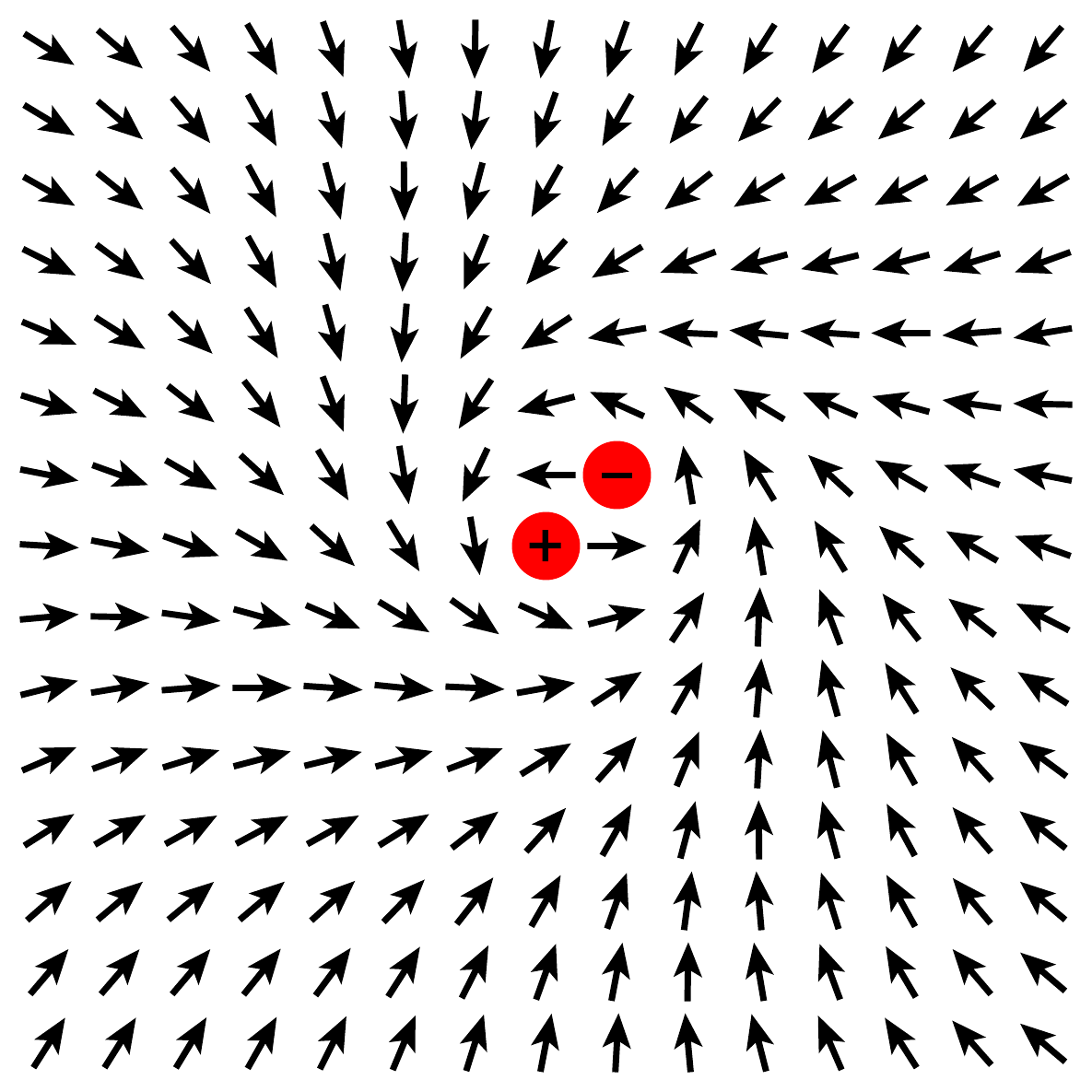} &
        \includegraphics[width=0.23\textwidth]{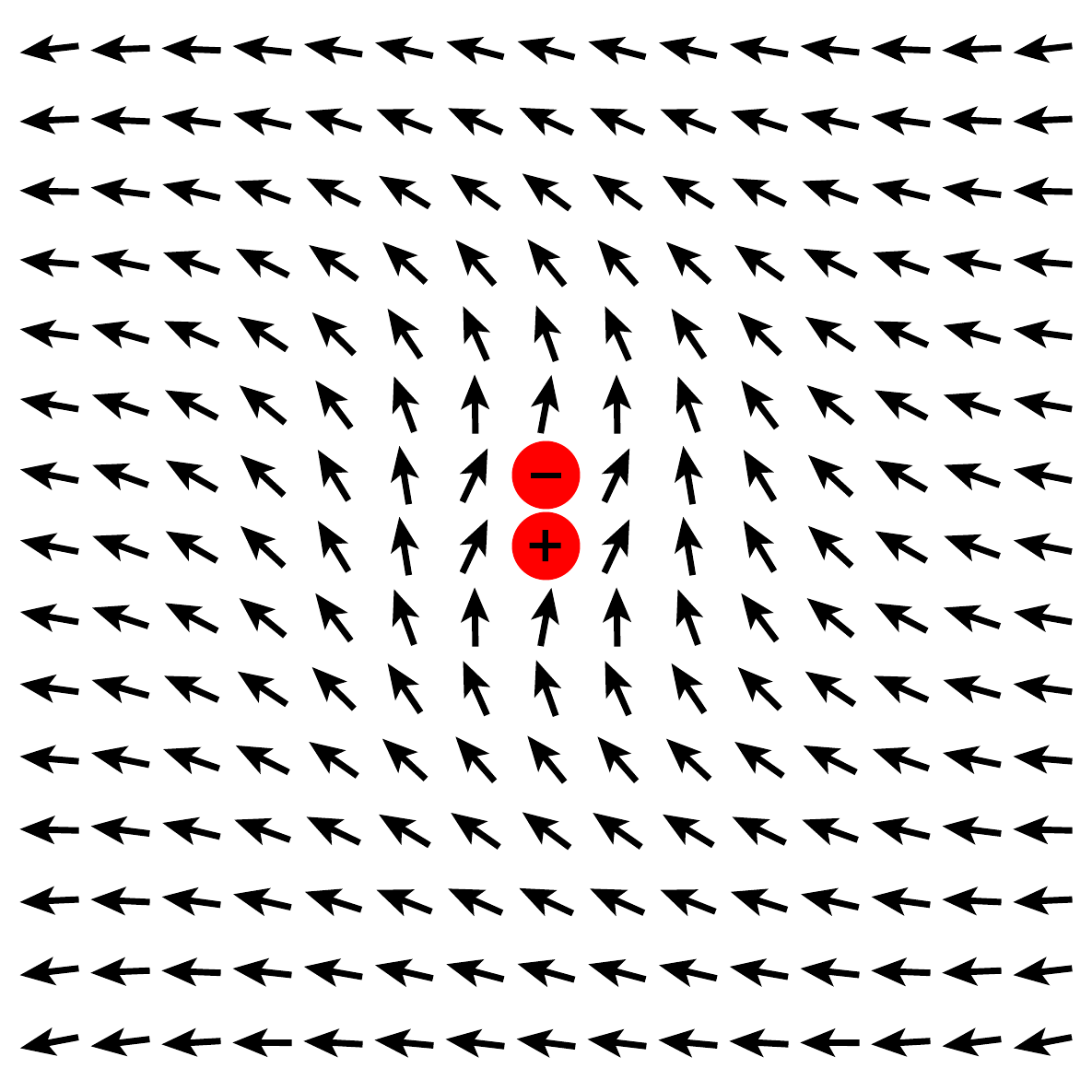} &
        \includegraphics[width=0.23\textwidth]{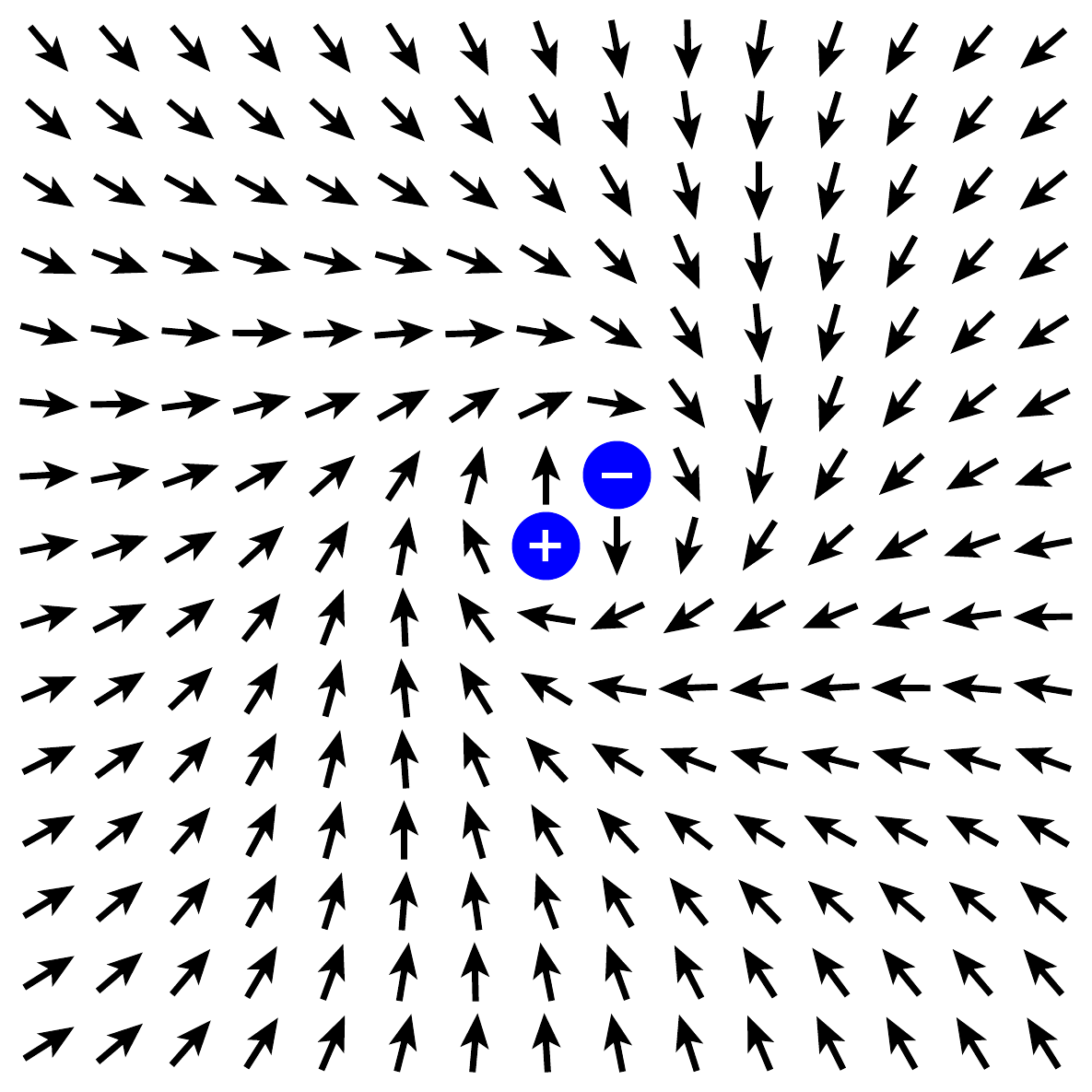} &
        \includegraphics[width=0.23\textwidth]{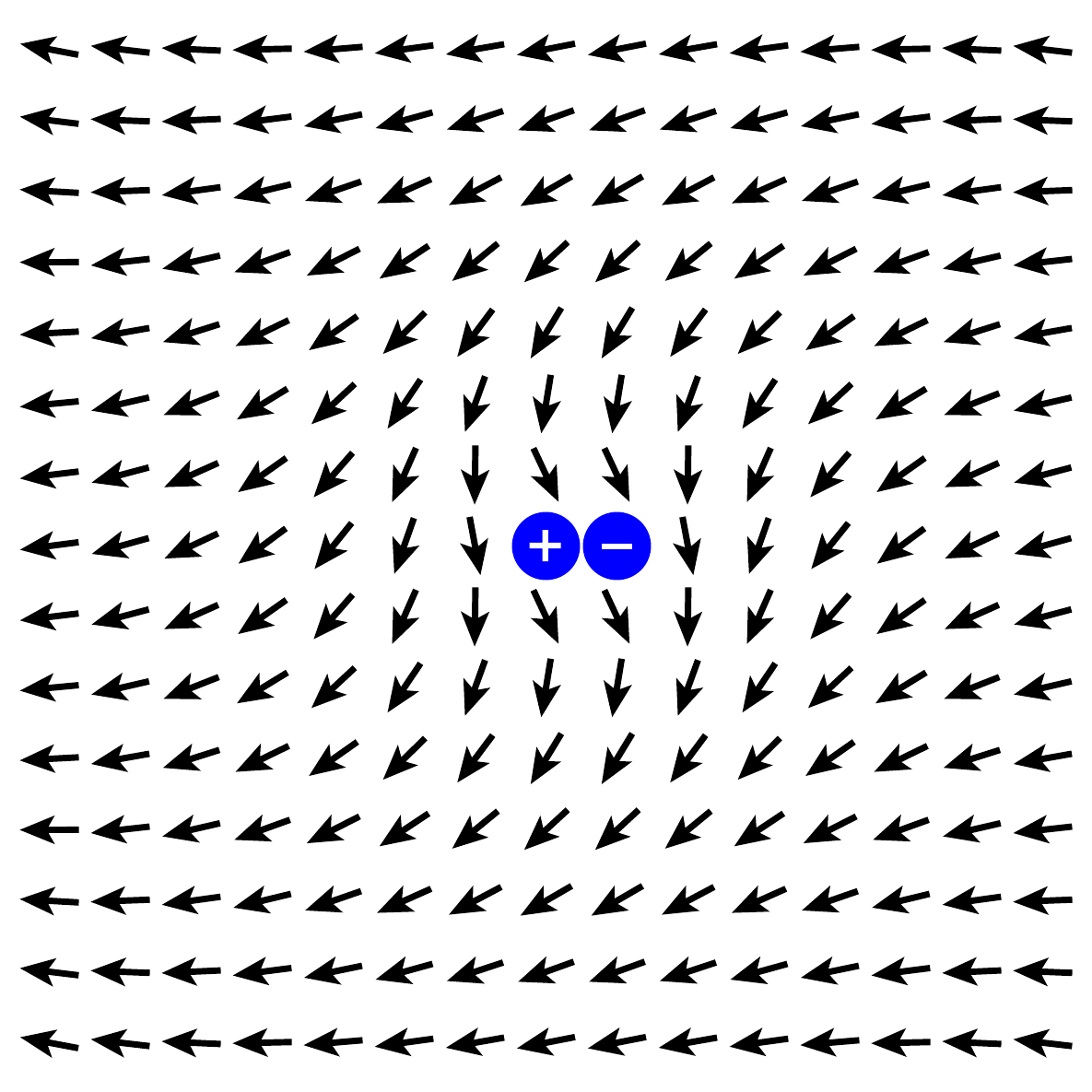}
    \end{tabular}
    \caption{PG vortex pairs with non-standard separations.
    Left: $\Theta_1^{(1)}$ (red) with separation along both $\hat{x}_1$ and $\hat{x}_2$ (diagonal, $d_1=d_2=a$), and along $\hat{x}_2$ only ($d_1=0$, $d_2=a$).
    Right: $\Theta_1^{(2)}$ (blue) with diagonal separation and along $\hat{x}_1$ only.
    Only the separation component along the same direction as the
PG-vortex species contributes to the far-field conventional winding
$\ell_0^{\rm far}$: $d_1/a$ for $\Theta_1^{(1)}$ and $d_2/a$ for
$\Theta_1^{(2)}$.
    A diagonal pair carries $\ell_0^{\rm far}=d_1/a$ ($d_2/a$) for $\Theta_1^{(1)}$ ($\Theta_1^{(2)}$); the orthogonal component does not enter the far-field phase winding.}
    \label{fig:sm_pg_pairs_nonstandard}
\end{figure}

The far-field winding $\ell_0^{\rm far}=d/a$ also explains the subdimensional
mobility of single PG vortices~\cite{pretkoFractonElasticityDuality2018,pretkoCrystaltofractonTensorGauge2019,manojFractonicViewFolding2021,gorantla2+1dimensionalCompactLifshitz2023a}. A
single $\Theta_1^{(1)}$ vortex can move freely along $\hat{x}_2$ but not along
$\hat{x}_1$: displacing it along $\hat{x}_1$ would shift the $x_1$-component of
the total dipole moment, which is forbidden by dipole conservation.
Equivalently, such a displacement would alter the conventional winding measured
by a large enclosing contour, contradicting the fact that the intrinsic
topological charge of $\Theta_1^{(1)}$ resides in $\chi_1$ alone. The allowed
motion along $\hat{x}_2$ preserves both the dipole moment and the far-field
winding. For $\Theta_1^{(2)}$ the roles of the two directions are exchanged.
Only a neutral pair can move freely in both lattice directions without changing its internal dipole moment or far-field winding. This
contrasts with the standard XY model, where single vortices are fully mobile;
here, individual PG vortices inherit their restricted mobility directly from
dipole conservation, and only the bound pair recovers full 2D
mobility.

Having established the far-field behavior of the dipole-level PG vortex
pairs, we now extend the construction to higher multipole orders.

\subsection{Extension to quadrupole-conserving models}

The PG-vortex construction extends to higher-moment-conserving models. The
structure of this hierarchy was established by the classification of
Ref.~\cite{yuanHierarchicalProliferationHigherrank2023a}; what is new here is
its finite-temperature consequence for the vortex-proliferation transitions.
The governing equation for an order-$n$ vortex is
$a^n\partial_1^i\partial_2^{n-i}\theta = \ell\varphi(\mathbf{x})$; the $(n+1)$
compactness conditions at each level yield only \emph{two} independent vortex
species, regardless of $n$. These two species correspond precisely to the real
and imaginary parts of the complex function $(z/a)^n\ln(z/a)$, respectively, each carrying a
distinct coefficient set by the compactness constraints. As a concrete example,
the isotropic quadrupole-conserving model $H=\int
    d^2\mathbf{x}\sum_{i,j,k}(\partial_i\partial_j\partial_k\theta)^2$ has vortex
species
\begin{align}
    \Theta_2^{(1)}(\mathbf{x}) & = \ell_{3}\,
    \operatorname{Im}\bigl[(z/a)^2\ln(z/a)\bigr],        \\
    \Theta_2^{(2)}(\mathbf{x}) & = -\frac{\ell_{4}}{2}\,
    \operatorname{Re}\bigl[(z/a)^2\ln(z/a)\bigr].
\end{align}
Using $z=re^{i\varphi}$, these become
$\Theta_2^{(1)}=\ell_3\bigl(\frac{x_1^2-x_2^2}{a^2}\varphi(\mathbf{x})+\frac{2x_1x_2}{a^2}\ln\frac{r}{a}\bigr)$ and
$\Theta_2^{(2)}=\ell_4\bigl(\frac{x_1x_2}{a^2}\varphi(\mathbf{x})+\frac{x_2^2-x_1^2}{2a^2}\ln\frac{r}{a}\bigr)$, respectively,
whose winding numbers are defined by the quadrupole-field loop integrals
\begin{align}
    &\oint_{\mathcal C} d\mathbf{l}\cdot \nabla\left(a^2\partial_1^2\theta\right)
    = -\oint_{\mathcal C} d\mathbf{l}\cdot \nabla\left(a^2\partial_2^2\theta\right)
    = 4\pi \ell_{3}, \\
    &\oint_{\mathcal C} d\mathbf{l}\cdot \nabla\left(a^2\partial_1\partial_2\theta\right)
    = 2\pi \ell_{4},
\end{align}
Specifically, $\Theta_2^{(1)}$ carries winding $2\ell_3$ in $a^2\partial_1^2\theta$ and $-2\ell_3$ in $a^2\partial_2^2\theta$ (i.e., opposite signs in the two pure second-derivative channels), while $\Theta_2^{(2)}$ carries winding $\ell_4$ in $a^2\partial_1\partial_2\theta$.
The different coefficients in front of $\operatorname{Im}[(z/a)^2\ln(z/a)]$ and $\operatorname{Re}[(z/a)^2\ln(z/a)]$ ensure that in the real-space expansion the coefficient in front of $\varphi$ is an integer at every lattice site; as a result, the two species carry self-energies that differ by a numerical coefficient.
Therefore, even in the isotropic limit, the real and
imaginary parts of $(z/a)^n\ln(z/a)$ enter with different weights, and we expect that \emph{isotropic} quadrupole-conserving models
directly exhibit two distinct KT transitions without requiring anisotropy, in sharp contrast to
the dipole case ($n=1$) where isotropy enforces a single transition.

\begin{figure}[t]
    \centering
    \begin{tabular}{@{}c@{\hspace{2pt}}c@{\hspace{2pt}}c@{}}
        \includegraphics[width=0.31\columnwidth]{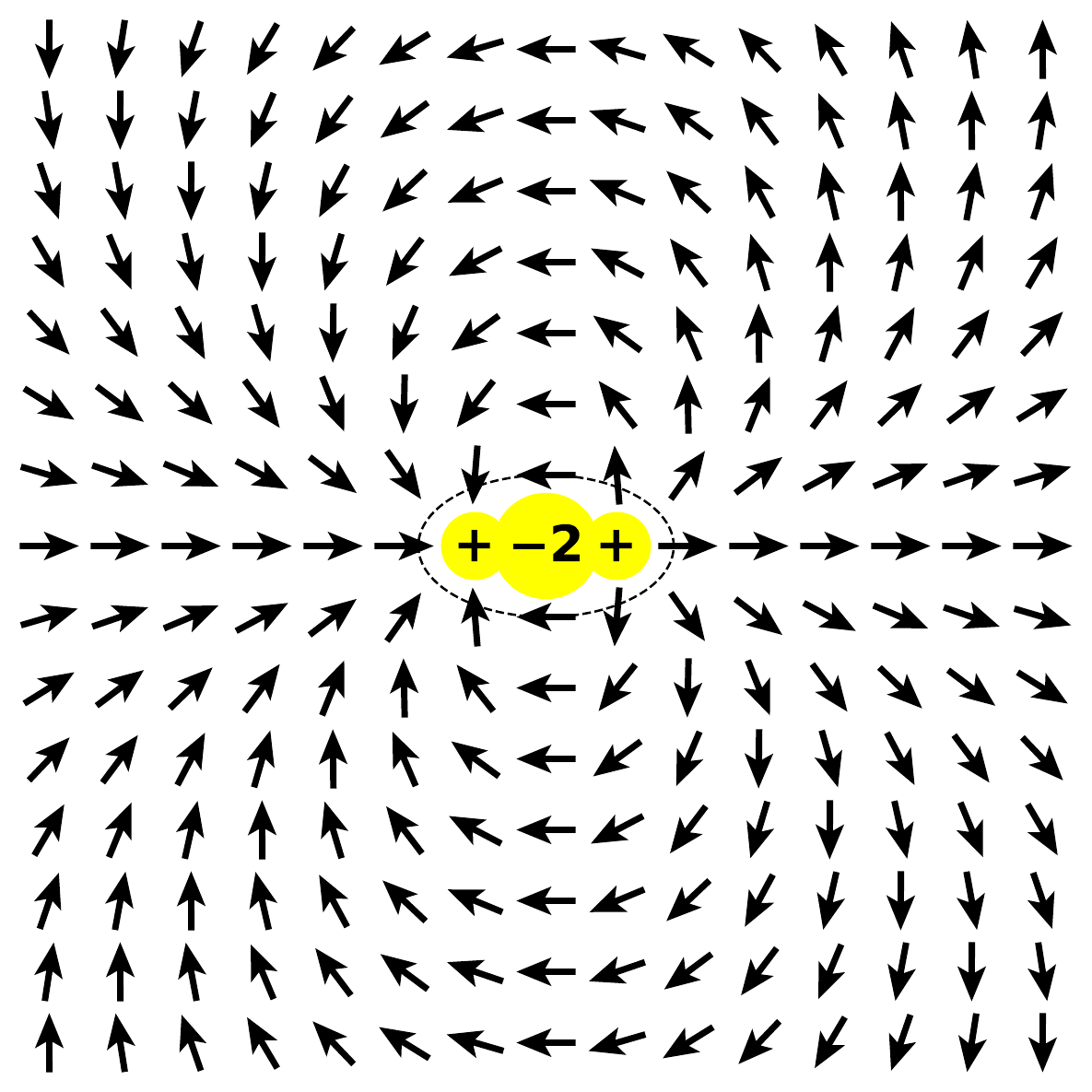} &
        \includegraphics[width=0.31\columnwidth]{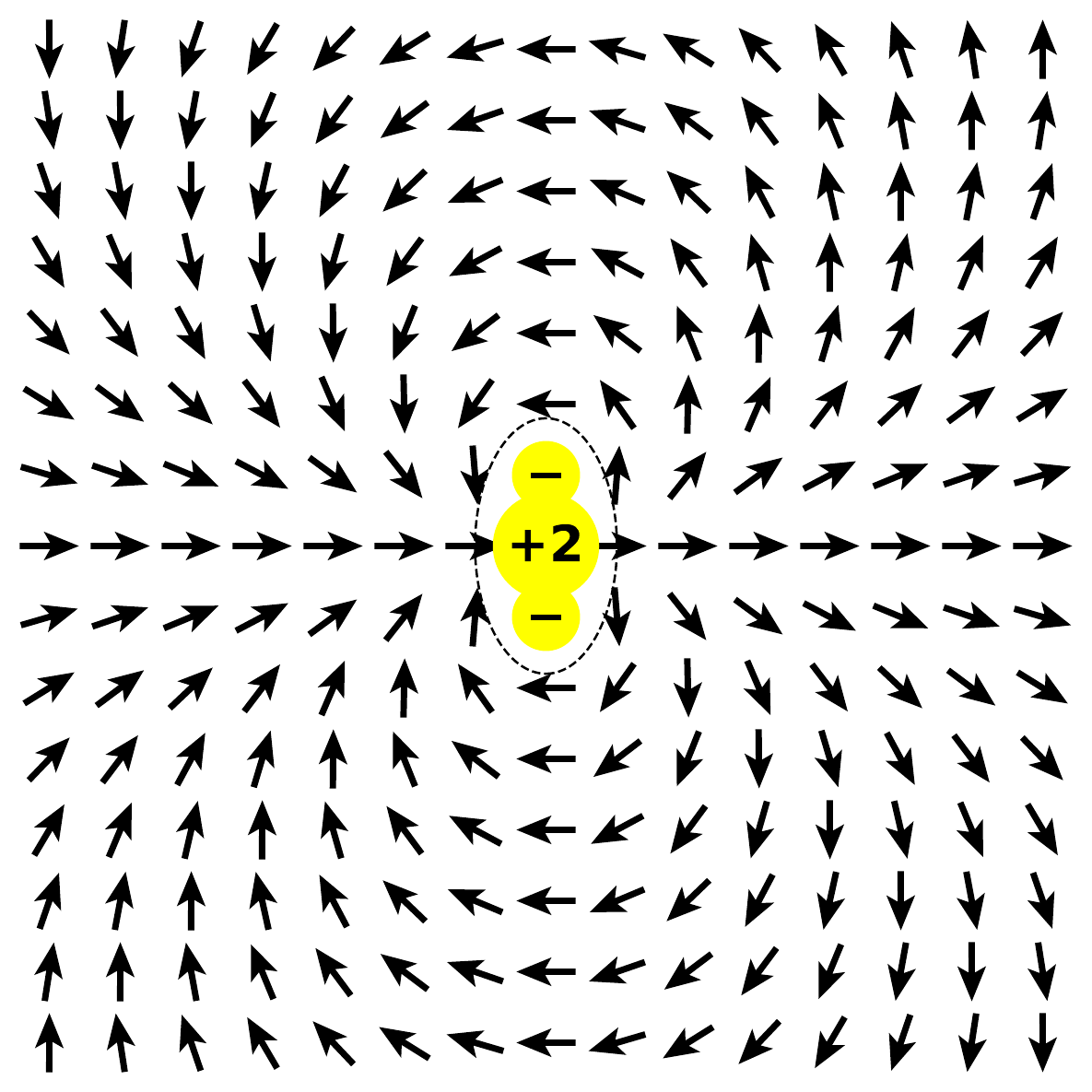} &
        \includegraphics[width=0.31\columnwidth]{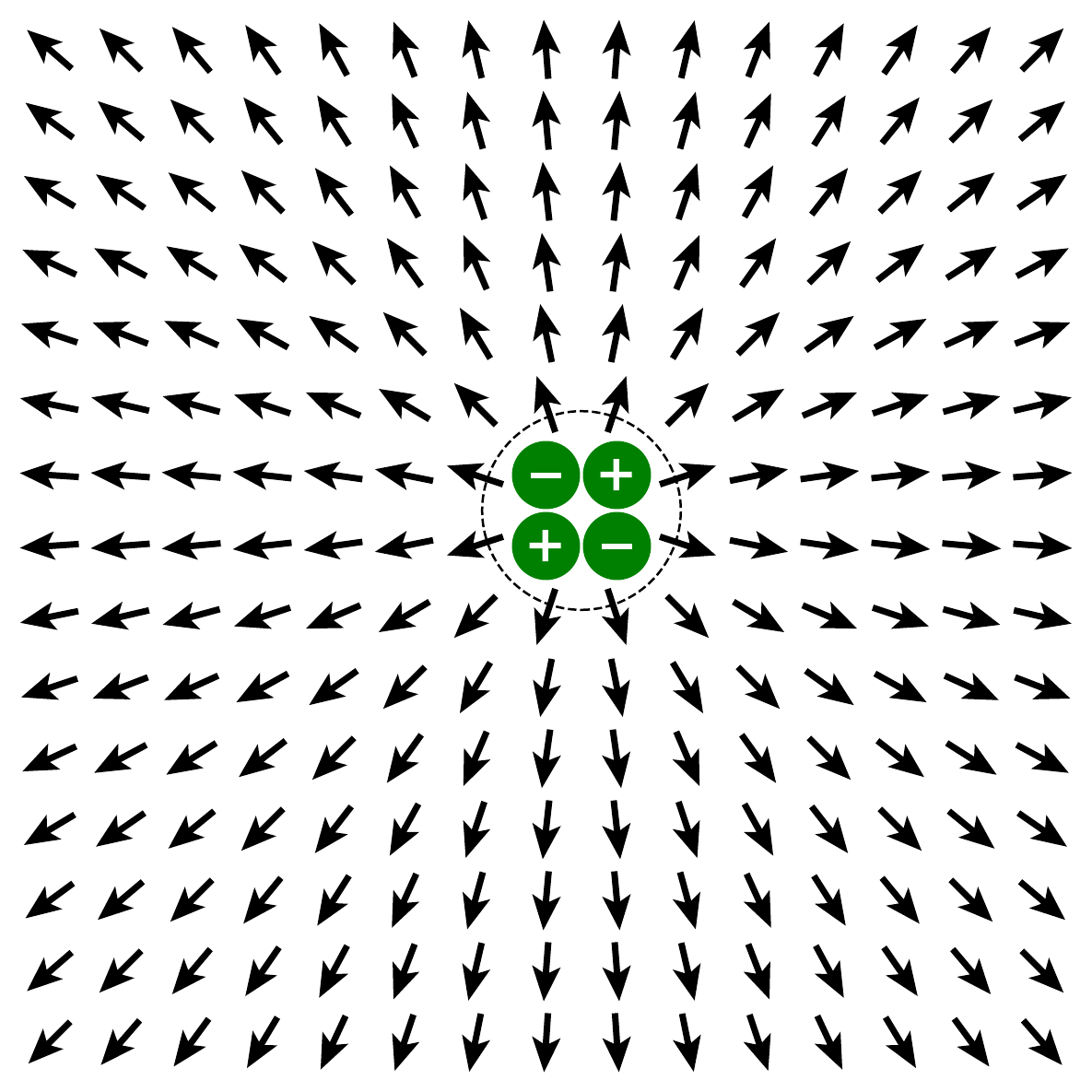}
    \end{tabular}
    \caption{Quadrupole vortex textures showing the
        next level of the higher-moment defect
        hierarchy. Left:  a $(\ell{=}{+1},{-2},{+1})$
        triplet along $\hat{x}_1$ formed by $\Theta_2^{(1)}$. It can be regarded as a phase vortex with a winding number of 2. Center: same species reconstructed along $\hat{x}_2$.
        Right: a $2\times2$ cluster formed by $\Theta_2^{(2)}$.
        It can be regarded as a phase vortex with a winding number of 1.
        In isotropic quadrupole-conserving models $\Theta_2^{(1)}$ and $\Theta_2^{(2)}$
        have \emph{distinct} self-energies. Two KT transitions without anisotropy are expected.}
    \label{fig:hierarchy_full}
\end{figure}

\section{Vortex energetics and the PG-vortex dipole relation}
\label{sec:sm_vortex}

In this section, we present a more explicit analysis of the three vortex
configurations introduced in the main text. The main goal is to show why the
PG vortices $\Theta_1^{(1)}$ and $\Theta_1^{(2)}$ are the
logarithmically costly vortices that control KT physics, while the phase
vortex $\Theta_0$ has only finite energy. We also derive the energy and
far-field winding of a PG-vortex dipole. The three continuum vortex
configurations are
\begin{align}
    \Theta_0(\mathbf{x})       & =\ell_0\varphi(\mathbf{x}),
    \\
    \Theta_1^{(1)}(\mathbf{x}) & = \ell_1 \left( \frac{x_1}{a}\varphi(\mathbf{x}) + \frac{x_2}{a}\ln\frac{r}{a} \right),
    \\
    \Theta_1^{(2)}(\mathbf{x}) & = \ell_2 \left( \frac{x_2}{a}\varphi(\mathbf{x}) - \frac{x_1}{a}\ln\frac{r}{a} \right),
\end{align}
with $\varphi(\mathbf{x})$ the polar angle ($x_1+ix_2=re^{i\varphi}$) and $r=\sqrt{x_1^2+x_2^2}$. These satisfy
\begin{align}
    a\partial_1\Theta_1^{(1)} & = \ell_1 \varphi(\mathbf{x}),
    \\
    a\partial_2\Theta_1^{(2)} & = \ell_2 \varphi(\mathbf{x}),
\end{align}
respectively, up to the usual branch-cut structure.  (The complex-analysis
formulation of Sec.~I unifies these expressions
and makes their branch-cut structure explicit.)  Thus the
PG vortices are simply the familiar KT vortices, but in the compact
dipole fields rather than in $\theta$ itself.

\subsection{Vortex self-energies}

We first evaluate the energy of the phase vortex. For the isotropic Hamiltonian
\begin{align}
    H_{\mathrm{cont}} = \frac{a^2J}{2} \int d^2\mathbf{x}\; \Bigl[ (\partial_1^2\theta)^2 + (\partial_2^2\theta)^2 + 2(\partial_1\partial_2\theta)^2 \Bigr],
\end{align}
the energy of a configuration is
\begin{align}
    E[\theta]
    =
    \frac{a^2J}{2}\int d^2\mathbf{x}\,
    \Bigl[
        (\partial_1^2\theta)^2 + (\partial_2^2\theta)^2 + 2(\partial_1\partial_2\theta)^2
        \Bigr].
\end{align}
For $\Theta_0=\ell\varphi$, the second derivatives decay as $1/r^2$, so the energy density scales as $1/r^4$. Integrating over 2D space,
\begin{align}
    E_{\Theta_0}
    \sim
    a^2J \int_a^L dr\, r\, \frac{1}{r^4}
    \sim
    a^2J\left[\frac{1}{a^2}-\frac{1}{L^2}\right].
\end{align}
The result is finite as $L\to\infty$. Evaluating the coefficient more carefully yields
\begin{align}
    E_{\Theta_0}=\pi J\ell_0^2.
\end{align}
Hence the phase vortex has finite self-energy in the thermodynamic limit.

This already shows why the phase vortex cannot by itself drive a
standard KT transition in the dipole-conserving model. Since its entropy still
grows like $\ln L$ from positional degeneracy, the free energy becomes negative
at any nonzero temperature for large enough $L$, and such vortices are not
sharply tied to a finite-temperature unbinding transition.

For $\Theta_1^{(1)}$, the key point is that $a\partial_1\Theta_1^{(1)}$ has
phase-vortex winding. As a result, the second-derivative energy density
behaves effectively like the gradient energy density of a standard XY
vortex, namely as $1/r^2$. Therefore
\begin{align}
    E_{\Theta_1^{(1)}}
    \sim
    J\int_a^L dr\, r\, \frac{1}{r^2}
    \sim
    J\ln\frac{L}{a}.
\end{align}
A direct calculation gives
\begin{align}
    E_{\Theta_1^{(1)}} = 2\pi J \ell_1^2 \ln\frac{L}{a}.
\end{align}
Similarly,
\begin{align}
    E_{\Theta_1^{(2)}} = 2\pi J \ell_2^2 \ln\frac{L}{a}.
\end{align}
These are precisely the logarithmic self-energies needed for KT binding and unbinding.

Thus the energetic hierarchy of the three vortices is
\begin{align}
    E_{\Theta_0}\sim O(1),
    \qquad
    E_{\Theta_1^{(1)}}\sim \ln L,
    \qquad
    E_{\Theta_1^{(2)}}\sim \ln L.
\end{align}
This is the basic reason why the PG vortices, not the phase vortex, govern the thermal transitions.

A simple free-energy estimate already captures the distinction. The entropy of
placing a single vortex scales as
\begin{align}
    S \sim 2\ln(L/a),
\end{align}
up to nonuniversal constants.

For the phase vortex,
\begin{align}
    F_{\Theta_0}
    =
    E_{\Theta_0}-TS
    \sim
    \pi J\ell_0^2 - 2T\ln(L/a),
\end{align}
which becomes negative at large $L$ for any $T>0$.

For the PG vortices,
\begin{align}
    F_{\Theta_1^{(1)}} & \sim \left(2\pi J\ell_1^2-2T\right)\ln(L/a),
    \\
    F_{\Theta_1^{(2)}} & \sim \left(2\pi J\ell_2^2-2T\right)\ln(L/a).
\end{align}
This suggests a transition when the logarithmic energy is balanced by entropy, yielding the rough estimate
\begin{align}
    T_c^{(1)} = T_c^{(2)} \approx \pi J
\end{align}
for unit winding. As usual, this estimate is only schematic; the transition structure is further clarified by the Gaussian analysis and Monte Carlo results.

\subsection{Energy and far-field winding of a PG-vortex dipole}

We now derive the energy of a PG-vortex dipole and its far-field
conventional winding. Consider two $\Theta_1^{(1)}$ vortices with opposite
winding, located at $\mathbf{0}$ and $a\hat{x}_1$. The
combined field is
\begin{align}
    \Theta_{\rm pair}^{(1)}(\mathbf{x})
    =
    \Theta_1^{(1)}(\mathbf{x};\mathbf{0},+1)
    +
    \Theta_1^{(1)}(\mathbf{x};a\hat{x}_1,-1).
\end{align}
At distances large compared with the separation $a$, the leading far-field contribution of the pair is obtained by expanding in the displacement. Because the two PG vortices have opposite charge, the logarithmic pieces cancel at leading order, and the remaining angular structure reproduces the profile of a phase vortex:
\begin{align}
    \Theta_{\rm pair}^{(1)}(\mathbf{x})
    \sim
    \varphi(\mathbf{x})
    \equiv
    \Theta_0(\mathbf{x})
    \qquad (r\gg a),
\end{align}
up to short-distance and branch-cut dependent details.  (A
rigorous far-field expansion using the complex representation is
given in Sec.~I.)  An analogous construction
using a pair of $\Theta_1^{(2)}$ vortices separated along $\hat{x}_2$
gives the same result.

This is why it is natural to regard $\Theta_0$ as the minimal tightly bound
dipole of PG vortices. In the standard XY model this
substructure is
absent because there are no compact dipole fields. In the dipole-conserving
theory, however, it becomes the correct way to organize the vortex hierarchy.
The thermal transitions are then controlled by whether these more elementary
objects remain bound or proliferate.

\begin{equation}
    E_{\text{pair}}(d) = 2E_{\text{core}} + 4\pi J\ln\frac{d}{a} + \mathcal O(1).
\end{equation}
For $a\lesssim r\lesssim d$, the configuration resolves two individual
PG vortices and accumulates the logarithmic separation energy;
only at $r\gg d$ does it reduce to a phase-vortex far field with winding
$d/a$. The far-field correspondence therefore does not imply energetic
equivalence to an isolated phase vortex at all length scales.

\section{Gaussian continuum theory and correlation functions}
\label{sec:sm_gaussian}

In this section, we derive the long-distance behavior of several correlation
functions in the continuum theory of the anisotropic dipole-conserving XY
model. The purpose is twofold. First, these calculations justify the choice of
observables used in the main text [Eq.~\eqref{eq:eta}]. Second, they clarify why
the low-temperature phase should be viewed as possessing QLRO in the dipole
fields rather than in the original phase field.

We begin from the anisotropic continuum Hamiltonian of the main text [Eq.~\eqref{eq:H_cont}], with the bare couplings $J_1,J_2,J_3$ replaced by the generalized helicity moduli $J_{s1},J_{s2},J_{s3}$ [Eq.~\eqref{eq:J_s}]:
\begin{align}
    H=\frac{a^2}{2}\int d^2\mathbf{x}\,
    \Bigl[
    J_{s1}(\partial_1^2\theta)^2
    +J_{s2}(\partial_2^2\theta)^2
    +2J_{s3}(\partial_1\partial_2\theta)^2
    \Bigr].
    \label{eq:H_gauss_sm}
\end{align}
In momentum space this becomes
\begin{align}
    H= \frac{a^2}{2}\int \frac{d^2\mathbf{q}}{(2\pi)^2}\,
    |\theta_{\mathbf{q}}|^2 K(\mathbf{q}),
\end{align}
with kernel
\begin{align}
    K(\mathbf{q}) = J_{s1} q_1^4 + J_{s2} q_2^4 + 2J_{s3} q_1^2 q_2^2.
\end{align}
At finite temperature $T=1/\beta$, the Gaussian partition function is
\begin{align}
    Z = \int \mathcal{D}\theta\, e^{-\beta H[\theta]},
\end{align}
and the two-point function is
\begin{align}
    \langle \theta_{\mathbf{q}} \theta_{\mathbf{q}'} \rangle
    =
    \frac{(2\pi)^2 \delta(\mathbf{q}+\mathbf{q}')}{\beta a^2 K(\mathbf{q})}.
\end{align}

Introducing polar coordinates $\mathbf{q}=(q\cos\phi,q\sin\phi)$ and
$\mathbf{r}=(r\cos\varphi,r\sin\varphi)$, we write
\begin{align}
    K(\mathbf{q})=q^4 f(\phi),
\end{align}
with
\begin{align}
    f(\phi)=J_{s1}\cos^4\phi+J_{s2}\sin^4\phi+2J_{s3}\cos^2\phi\sin^2\phi.
\end{align}
All correlation functions considered below take the form $\langle e^{iX}\rangle$, where $X$ is linear in the Gaussian field $\theta$. Thus,
\begin{align}
    \langle e^{iX}\rangle = \exp\!\left[-\frac{1}{2}\langle X^2\rangle\right],
\end{align}
and the problem reduces to evaluating the corresponding variance.

\subsection{Dipole-field correlators}

We define the dipole fields
\begin{align}
    \chi_1(\mathbf{x}) = a\,\partial_1\theta(\mathbf{x}),
    \qquad
    \chi_2(\mathbf{x}) = a\,\partial_2\theta(\mathbf{x}).
\end{align}
For $\chi_1$, let
\begin{align}
    X_1(\mathbf r)=\chi_1(\mathbf r)-\chi_1(\mathbf 0)
    =
    a\,[\partial_1\theta(\mathbf r)-\partial_1\theta(\mathbf 0)].
\end{align}
Then
\begin{align}
    \left\langle e^{i(\chi_1(\mathbf r)-\chi_1(\mathbf 0))}\right\rangle
    =
    \exp\!\left[
        -\frac{a^2}{2}
        \left\langle
        \bigl[\partial_1\theta(\mathbf r)-\partial_1\theta(\mathbf 0)\bigr]^2
        \right\rangle
        \right].
\end{align}
Using the Fourier representation of $\partial_1\theta$, we obtain
\begin{align}
    \left\langle
    \bigl[\partial_1\theta(\mathbf r)-\partial_1\theta(\mathbf 0)\bigr]^2
    \right\rangle
    =
    \frac{2}{\beta a^2}
    \int \frac{d^2\mathbf q}{(2\pi)^2}\,
    \frac{q_1^2}{K(\mathbf q)}
    \bigl[1-\cos(\mathbf q\cdot \mathbf r)\bigr],
\end{align}
and hence
\begin{align}
    \left\langle e^{i(\chi_1(\mathbf r)-\chi_1(\mathbf 0))}\right\rangle
    =
    \exp\!\left[
        -\frac{1}{\beta}
        \int \frac{d^2\mathbf q}{(2\pi)^2}\,
        \frac{q_1^2}{K(\mathbf q)}
        \bigl[1-\cos(\mathbf q\cdot \mathbf r)\bigr]
        \right].
    \label{eq:chi1corr-sm}
\end{align}
Similarly,
\begin{align}
    \left\langle e^{i(\chi_2(\mathbf r)-\chi_2(\mathbf 0))}\right\rangle
    =
    \exp\!\left[
        -\frac{1}{\beta}
        \int \frac{d^2\mathbf q}{(2\pi)^2}\,
        \frac{q_2^2}{K(\mathbf q)}
        \bigl[1-\cos(\mathbf q\cdot \mathbf r)\bigr]
        \right].
    \label{eq:chi2corr-sm}
\end{align}
Using
\begin{align}
    \frac{q_1^2}{K(\mathbf q)} = \frac{\cos^2\phi}{q^2 f(\phi)},
    \qquad
    \frac{q_2^2}{K(\mathbf q)} = \frac{\sin^2\phi}{q^2 f(\phi)},
\end{align}
the exponent in Eq.~\eqref{eq:chi1corr-sm} becomes
\begin{align}
    \int \frac{d^2\mathbf q}{(2\pi)^2}\,
    \frac{q_1^2}{K(\mathbf q)}
    \bigl[1-\cos(\mathbf q\cdot \mathbf r)\bigr]
    =
    \frac{1}{(2\pi)^2}
    \int_0^{2\pi} d\phi\,
    \frac{\cos^2\phi}{f(\phi)}
    \int_0^\Lambda \frac{dq}{q}\,
    \Bigl[1-\cos\bigl(qr\cos(\phi-\varphi)\bigr)\Bigr],
\end{align}
with $\Lambda\sim 1/a$ an ultraviolet cutoff. For large $r$, the radial integral behaves as $\ln r + O(1)$ for generic angle, so the leading asymptotics are
\begin{align}
    \int \frac{d^2\mathbf q}{(2\pi)^2}\,
    \frac{q_1^2}{K(\mathbf q)}
    \bigl[1-\cos(\mathbf q\cdot \mathbf r)\bigr]
    \sim
    \frac{\ln r}{(2\pi)^2}
    \int_0^{2\pi} d\phi\,
    \frac{\cos^2\phi}{f(\phi)},
\end{align}
and similarly
\begin{align}
    \int \frac{d^2\mathbf q}{(2\pi)^2}\,
    \frac{q_2^2}{K(\mathbf q)}
    \bigl[1-\cos(\mathbf q\cdot \mathbf r)\bigr]
    \sim
    \frac{\ln r}{(2\pi)^2}
    \int_0^{2\pi} d\phi\,
    \frac{\sin^2\phi}{f(\phi)}.
\end{align}

The angular integrals can be evaluated explicitly:
\begin{align}
    \int_0^{2\pi} d\phi\, \frac{\cos^2\phi}{f(\phi)}
     & =
    \frac{\sqrt{2}\pi}
    {\sqrt{J_{s1}\bigl(J_{s3}+\sqrt{J_{s1}J_{s2}}\bigr)}},
    \\
    \int_0^{2\pi} d\phi\, \frac{\sin^2\phi}{f(\phi)}
     & =
    \frac{\sqrt{2}\pi}
    {\sqrt{J_{s2}\bigl(J_{s3}+\sqrt{J_{s1}J_{s2}}\bigr)}}.
\end{align}
Accordingly,
\begin{align}
    \left\langle e^{i(\chi_1(\mathbf r)-\chi_1(\mathbf 0))}\right\rangle
     & \sim
    r^{-\eta_1},
    \\
    \left\langle e^{i(\chi_2(\mathbf r)-\chi_2(\mathbf 0))}\right\rangle
     & \sim
    r^{-\eta_2},
\end{align}
with
\begin{align}
    \eta_1
     & =
    \frac{1}
    {2\sqrt{2}\pi\beta
        \sqrt{J_{s1}\bigl(J_{s3}+\sqrt{J_{s1}J_{s2}}\bigr)}},
    \label{eq:eta1-sm}
    \\
    \eta_2
     & =
    \frac{1}
    {2\sqrt{2}\pi\beta
        \sqrt{J_{s2}\bigl(J_{s3}+\sqrt{J_{s1}J_{s2}}\bigr)}}.
    \label{eq:eta2-sm}
\end{align}
Thus the dipole fields exhibit power-law correlations at long distances.

In the isotropic limit $J_{s1}=J_{s2}=J_{s3}=J$, the exponents coincide and
reduce to
\begin{align}
    \eta_1=\eta_2=\eta=\frac{1}{4\pi\beta J},
\end{align}
which is the expression used in the main text [Eq.~\eqref{eq:eta}].

\subsection{The phase correlator does not define quasi-long-range order}

We now examine the correlation function of the original phase field,
\begin{align}
    \left\langle e^{i(\theta(\mathbf r)-\theta(\mathbf 0))}\right\rangle
    =
    \exp\!\left[
        -\frac{1}{2}
        \left\langle
        \bigl[\theta(\mathbf r)-\theta(\mathbf 0)\bigr]^2
        \right\rangle
        \right].
\end{align}
The corresponding variance is
\begin{align}
    \left\langle
    \bigl[\theta(\mathbf r)-\theta(\mathbf 0)\bigr]^2
    \right\rangle
    =
    \frac{2}{\beta a^2}
    \int \frac{d^2\mathbf q}{(2\pi)^2}\,
    \frac{1-\cos(\mathbf q\cdot \mathbf r)}{K(\mathbf q)}.
\end{align}
Since $K(\mathbf q)=q^4 f(\phi)$, we may rewrite this as
\begin{align}
    \left\langle
    \bigl[\theta(\mathbf r)-\theta(\mathbf 0)\bigr]^2
    \right\rangle
    =
    \frac{2}{\beta a^2(2\pi)^2}
    \int_0^{2\pi}\frac{d\phi}{f(\phi)}
    \int_{1/L}^{1/a}\frac{dq}{q^3}
    \Bigl[
        1-\cos\bigl(qr\cos(\phi-\varphi)\bigr)
        \Bigr],
    \label{eq:theta-var-sm}
\end{align}
where we introduced an infrared cutoff $1/L$.

The radial integral now contains $dq/q^3$ rather than $dq/q$, which makes the
infrared behavior much more singular. In the regime $a\ll r\ll L$, the dominant
contribution comes from momenta $q\lesssim 1/r$, where
\begin{align}
    1-\cos\bigl(qr\cos(\phi-\varphi)\bigr)
    \approx
    \frac{1}{2}
    q^2 r^2 \cos^2(\phi-\varphi).
\end{align}
Substituting this into Eq.~\eqref{eq:theta-var-sm}, one finds
\begin{align}
    \left\langle
    \bigl[\theta(\mathbf r)-\theta(\mathbf 0)\bigr]^2
    \right\rangle
    \approx
    \mathcal C(\varphi)\,
    r^2\ln\!\left(\frac{L}{r}\right),
\end{align}
with
\begin{align}
    \mathcal C(\varphi)
    =
    \frac{1}{\beta a^2(2\pi)^2}
    \int_0^{2\pi}d\phi\,
    \frac{\cos^2(\phi-\varphi)}{f(\phi)}
    >0.
\end{align}
Therefore,
\begin{align}
    \left\langle e^{i(\theta(\mathbf r)-\theta(\mathbf 0))}\right\rangle
    \sim
    \exp\!\left[
        -\frac{1}{2}\mathcal C(\varphi)\,
        r^2\ln\!\left(\frac{L}{r}\right)
        \right].
\end{align}
This decay is much stronger than a power law and is explicitly dependent on the infrared cutoff $L$. For any fixed nonzero separation $r$, the correlator vanishes in the thermodynamic limit $L\to\infty$. Thus the original phase field does not support quasi-long-range order. This is why the physically relevant low-temperature order must be diagnosed in the dipole fields instead.

\section{Leading-order Coulomb-gas picture and simplified RG discussion}
\label{sec:sm_coulomb}

In this section, we summarize a simplified Coulomb-gas and renormalization-group description of the PG-vortex sector. We keep only the structural ingredients needed to support the main text \eqref{eq:eta}. The purpose is to explain in a concise form why the KT transitions are controlled by $\Theta_1^{(1)}$ and $\Theta_1^{(2)}$ and why anisotropy can split the isotropic transition into two.

\subsection{Coulomb-gas structure in the isotropic case}

The free-energy analysis already shows that the two PG-vortex species are the only defects with logarithmically divergent self-energy. Following the standard treatment of the XY model, one may absorb the finite-energy phase-vortex sector into short-distance renormalization effects and focus on the singular part of the partition function associated with $\Theta_1^{(1)}$ and $\Theta_1^{(2)}$.

For unit charges $\ell_{i^{(\alpha)}}=\pm 1$, the interaction between two opposite vortices of the same PG-vortex species separated by $\mathbf r$ is logarithmic:
\begin{align}
V_{\alpha\alpha}(\mathbf r)\approx 4\pi J \ln\frac{|\mathbf r|}{a},
\qquad \alpha=1,2.
\end{align}
The interaction between different PG-vortex species is weaker in the sense relevant to KT scaling. Schematically,
\begin{align}
V_{12}(\mathbf r)\sim \frac{r_1r_2}{|\mathbf r|^2},
\end{align}
so it depends only on direction at large distances and does not generate the logarithmic interaction responsible for vortex confinement. 
Although the interspecies interaction is not absent, its lack of logarithmic growth means that it does not modify the leading logarithmic scaling dimension of either vortex fugacity.
To leading order in the KT analysis, one therefore treats the singular partition function as approximately factorized:
\begin{align}
Z_s \approx Z_1 Z_2,
\end{align}
where each $Z_\alpha$ is the partition function of a 2D Coulomb gas.
For each species,
\begin{align}
Z_{\alpha}
=
\sum_{N=0}^{\infty}
\frac{1}{(N!)^2}\left(\frac{y^{(\alpha)}}{a^2}\right)^{2N}
\int \prod_{i^{(\alpha)}=1}^{2N} d^2\mathbf{x}_{i^{(\alpha)}}\,
\exp\!\left[
2\pi K^{(\alpha)}
\sum_{i^{(\alpha)}<j^{(\alpha)}}
\ell_{i^{(\alpha)}}\ell_{j^{(\alpha)}}
\ln\frac{|\mathbf{x}_{i^{(\alpha)}}-\mathbf{x}_{j^{(\alpha)}}|}{a}
\right],
\end{align}
where $\ell_{i^{(\alpha)}}=\pm1$ and the sum is restricted to
charge-neutral configurations satisfying
$\sum_{i^{(\alpha)}=1}^{2N}\ell_{i^{(\alpha)}}=0$, as in the standard KT
Coulomb gas.
The fugacity $y^{(\alpha)}=e^{-\beta E_0^{(\alpha)}}$ (with $E_0^{(\alpha)}$ the core energy of species $\alpha$) and the effective coupling $K^{(\alpha)}=2\beta J$ in the isotropic limit.

This reproduces the standard KT structure, except that it now appears twice, once for each PG-vortex species. At the isotropic point the two gases are symmetry related and therefore critical at the same temperature. This is the long-form version of the statement made in the main text (Figs.~\ref{fig:corr}(a), \ref{fig:stiffness}(a), and \ref{fig:vortex}(a)): the isotropic system exhibits only one observable KT transition, but the defects driving it are two degenerate PG-vortex species rather than one phase-vortex species.

\subsection{Simplified RG equations}

Each Coulomb gas obeys the familiar leading-order RG equations.
Denoting the scale-dependent (renormalized) coupling by $K_R^{(\alpha)}$,
\begin{align}
\frac{d K_R^{(\alpha)-1}}{dl}
&=
4\pi^3 y^{(\alpha)2}
+O(y^{(\alpha)4}),
\\
\frac{d y^{(\alpha)}}{dl}
&=
\bigl(2-\pi K_R^{(\alpha)}\bigr)y^{(\alpha)}
+O(y^{(\alpha)3}).
\end{align}
Its bare (microscopic) value, obtained by matching to the lattice Hamiltonian,
is $K_R^{(\alpha)}(0)=K^{(\alpha)}=2\beta J$ in the isotropic limit
[see the Coulomb gas partition function above].
Although a single PG vortex is subdimensionally mobile in the kinetic sense, the mobility restriction is dynamical rather than configurational: in equilibrium the vortex core can occupy any lattice site, so its positional entropy is $2\ln(L/a)$ and the 2D Coulomb-gas treatment is justified.
In the isotropic limit the three generalized helicity moduli
coincide, $J_{s1}=J_{s2}=J_{s3}\equiv J_{s0}$ [Eq.~\eqref{eq:J_s}];
the renormalized coupling is then
$K_R^{(\alpha)} = 2\beta J_{s0}$.
The KT point is reached when $K_R^{(\alpha)}=\frac{2}{\pi}$, the standard KT
critical coupling, which through $K_R^{(\alpha)} = 2\beta J_{s0}$ gives the
corresponding critical value of the generalized helicity modulus in the
normalization of Eq.~\eqref{eq:J_s}:
\begin{align}
J_{s0}(T_c^-)=\frac{T_c}{\pi}.
\end{align}
Substituting $J_{s0}(T_c^-)=T_c/\pi$ into the isotropic dipole-correlation exponent $\eta = \frac{1}{4\pi\beta J_{s0}}$ yields
\begin{align}
\eta_c=\frac14.
\end{align}
This is formally the same critical value as in the standard XY model, but
the field whose exponent reaches this value is now the dipole field rather than
the original phase field, and the vortices whose proliferation drives the
transition are the PG ones.

\subsection{Anisotropy and two KT transitions}

In the anisotropic model, the two PG-vortex species are no longer equivalent. Their core energies and effective couplings differ, and the two Coulomb gases acquire distinct parameters:
$K^{(1)} \neq K^{(2)}$,
$y^{(1)} \neq y^{(2)}$.
As a result, the two species need not cross their KT separatrices at the same temperature. The most direct continuum signature of this splitting appears in the two dipole exponents,
\begin{align}
\eta_1=\frac{1}{2\sqrt{2}\pi\beta\sqrt{J_{s1}(J_{s3}+\sqrt{J_{s1}J_{s2}})}},\qquad
\eta_2=\frac{1}{2\sqrt{2}\pi\beta\sqrt{J_{s2}(J_{s3}+\sqrt{J_{s1}J_{s2}})}},
\end{align}
respectively. A practical criterion for the two KT transitions is then
$\eta_\alpha(T_c^{(\alpha)})=\frac14$,
or equivalently, the corresponding modulus combinations take their critical values at different temperatures.

This leads to the physical scenario emphasized in the main text (Figs.~\ref{fig:corr}(b), \ref{fig:stiffness}(b), and \ref{fig:vortex}(b)). At low temperature, both PG-vortex species are bound and the system has dipole quasi-long-range order in both channels. At the first transition, one species proliferates, destroying the corresponding dipole order while leaving the other channel comparatively intact. At the second transition, the remaining species also proliferates, and the system enters the fully disordered high-temperature regime. Thus anisotropy turns the single isotropic KT transition into two defect-selective KT transitions.

A particularly important exception is the $J_3=0$ line. There the two exponents are not independent in the sense relevant for sustaining quasi-long-range order, because both channels require the simultaneous presence of $J_{s1}$ and $J_{s2}$. This is the continuum explanation for the merger of the two transitions seen numerically.

\subsection{Relation to the main numerical picture}

The simplified Coulomb-gas/RG framework does not replace the numerical analysis in the main text (Figs.~\ref{fig:corr}--\ref{fig:vortex}); rather, it provides an organizing interpretation for it. Its main role is to justify three points that are central in the body of the PRL manuscript:
\begin{enumerate}
    \item The relevant logarithmically interacting defects are the PG vortices $\Theta_1^{(1)}$ and $\Theta_1^{(2)}$, not the finite-energy phase vortex $\Theta_0$.
    \item In the isotropic model, the two PG-vortex species are symmetry related and therefore drive a single KT transition together.
    \item In the anisotropic model, the two species acquire distinct effective scales and can therefore drive two separate KT transitions, except on the special merger line $J_3=0$.
\end{enumerate}

These are precisely the features borne out by the Monte Carlo observables discussed in the main text (Figs.~\ref{fig:corr}--\ref{fig:vortex}) and displayed in expanded form above.

\section{Full phase diagram and additional Monte Carlo overview}
\label{sec:sm_phase_diagram}

\subsection{Lattice implementation of generalized helicity moduli}

The generalized helicity moduli $J_{s\iota}$ defined in the main text [Eq.~\eqref{eq:J_s}]
can be measured directly from Monte Carlo data via the second derivative of the
free energy with respect to the twist amplitude (we set $a{=}1$ throughout).
Introducing quadratic twists
$\theta\to\theta+\tfrac12\gamma_1x_1^{\,2},
\theta\to\theta+\tfrac12\gamma_2x_2^{\,2},
\theta\to\theta+\gamma_3x_1x_2$
shifts the lattice second-order differences $\partial_1^2\theta$, $\partial_2^2\theta$,
and $\partial_1\partial_2\theta$ uniformly, with the shifts set by the amplitudes $\gamma_1,\gamma_2,\gamma_3$,
leaving first-order differences unchanged.
On a periodic lattice, these polynomial expressions are understood as
shorthand for uniform shifts of the corresponding compact second-order
differences; no non-periodic site field is imposed.
Expanding $F(\gamma_\iota)$ to second
order in $\gamma_\iota$ gives
$\frac{\partial^2F}{\partial\gamma_\iota^2}
=\big\langle\frac{\partial^2H}{\partial\gamma_\iota^2}\big\rangle
-\beta\big[\big\langle\big(\frac{\partial H}{\partial\gamma_\iota}\big)^2\big\rangle
-\big\langle\frac{\partial H}{\partial\gamma_\iota}\big\rangle^2\big]$.
For the dipole-conserving XY model [Eq.~\eqref{eq:H_lattice}], the relevant derivatives are
\begin{align}
    \frac{\partial^2 H}{\partial\gamma_1^2} &=  J_1\sum_{\mathbf{x}}\cos(\partial_1^2\theta)_{\mathbf{x}}, &
    \frac{\partial H}{\partial\gamma_1}  &= -J_1\sum_{\mathbf{x}}\sin(\partial_1^2\theta)_{\mathbf{x}},\\
    \frac{\partial^2 H}{\partial\gamma_2^2} &=  J_2\sum_{\mathbf{x}}\cos(\partial_2^2\theta)_{\mathbf{x}}, &
    \frac{\partial H}{\partial\gamma_2}  &= -J_2\sum_{\mathbf{x}}\sin(\partial_2^2\theta)_{\mathbf{x}},\\
    \frac{\partial^2 H}{\partial\gamma_3^2} &= 2J_3\sum_{\mathbf{x}}\cos(\partial_1\partial_2\theta)_{\mathbf{x}}, &
    \frac{\partial H}{\partial\gamma_3}  &= 2J_3\sum_{\mathbf{x}}\sin(\partial_1\partial_2\theta)_{\mathbf{x}}.
\end{align}
The Monte Carlo estimator for each channel is therefore
\begin{equation}
    J_{s\iota} = \frac{1}{c_\iota L^2}
    \Big[\big\langle\frac{\partial^2 H}{\partial\gamma_\iota^2}\big\rangle
    - \beta\big[\big\langle\big(\frac{\partial H}{\partial\gamma_\iota}\big)^2\big\rangle
    -\big\langle\frac{\partial H}{\partial\gamma_\iota}\big\rangle^2\big]\Big],
\end{equation}
with $c_1=c_2=1$, $c_3=2$; the factor $c_3=2$ compensates the factor of $2$
in the $J_3$ coupling relative to $J_1$ and $J_2$, placing all three moduli
on the same scale.

\subsection{Phase diagram from Monte Carlo}

The main text described the phase structure of the anisotropic
dipole-conserving XY model in terms of the unbinding temperatures of the
two PG-vortex species, $\Theta_1^{(1)}$ and $\Theta_1^{(2)}$. Here
we present the full phase diagram extracted from Monte Carlo
simulations, together with representative parameter cuts. We fix $J_1=1.0$ and,
without loss of generality, focus on $J_1\ge J_2$; this choice merely sets a
convention for labeling the two spatial directions. Under this convention,
when the two KT transitions split, the intermediate phase
$T_c^{(2)}<T<T_c^{(1)}$ retains quasi-long-range order in $\chi_1$ while
$\chi_2$ is already disordered. The opposite case $J_2>J_1$ is obtained by the
exchange $J_1\leftrightarrow J_2$ together with $\chi_1\leftrightarrow\chi_2$,
so that the intermediate phase instead preserves quasi-long-range order in
$\chi_2$ while $\chi_1$ becomes short-ranged first.

The simulations use classical Metropolis Monte Carlo with parallel
tempering~\cite{earlParallelTempering2005} on $L\times L$ lattices
($L{=}16,24,32,40,48,64$) at $N_{\text{temp}}{\ge}128$ geometrically spaced
temperatures between $T_{\min}{=}0.1$ and a parameter-dependent $T_{\max}$
chosen to cover the full transition region (e.g., $T_{\max}{=}2.2$ at the
isotropic point $J_1{=}J_2{=}J_3{=}1$, but $T_{\max}{=}0.7$ for
$J_2{=}0.6,J_3{=}0$). Each replica runs for at least $1.2{\times}10^6$ Monte Carlo
steps, with at least $6.0{\times}10^5$ steps used for thermalization; measurements
are taken every 5 steps over the subsequent $6.0{\times}10^5$ steps, yielding 
$1.2{\times}10^5$ samples per replica. Errors are estimated by bootstrap. The
dipole-field correlation ratios
$C(L/2)/C(L/4)$ for $\chi_1,\chi_2$, generalized helicity moduli
$J_{s\iota}$, and vortex densities
$\rho(\Theta_0),\rho(\Theta_1^{(1)}),\rho(\Theta_1^{(2)})$ are computed as
described in the main text. Critical temperatures are extracted from
$T_c(L){=}T_c(\infty){+}A/(\ln L)^2$~\cite{weberMonteCarloDetermination1988} by
fitting the crossing of $C(L/2)/C(L/4)$ with $2^{-1/4}$. The two dipole
channels are fitted independently; when their extrapolated $T_c(\infty)$ agree
within $1\sigma$ they are merged by inverse-variance weighting.

\begin{figure}[htbp]
    \centering
    \includegraphics[width=0.31\textwidth]{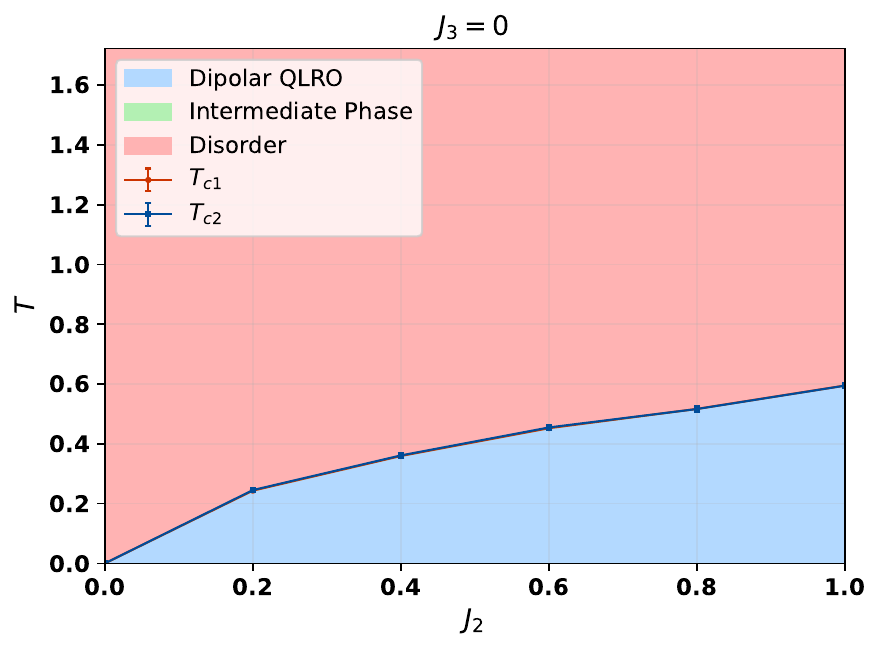}\hfill
    \includegraphics[width=0.31\textwidth]{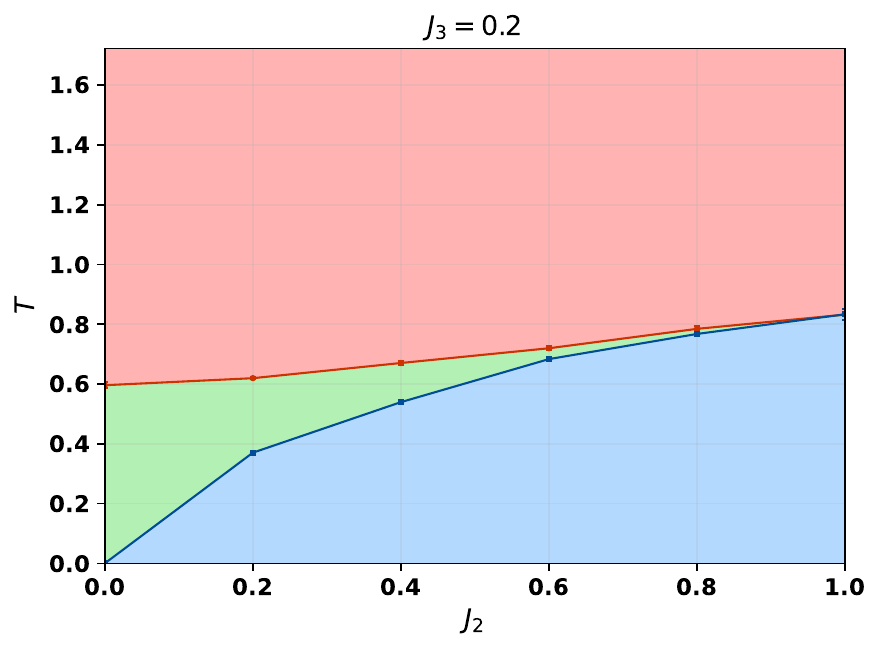}\hfill
    \includegraphics[width=0.31\textwidth]{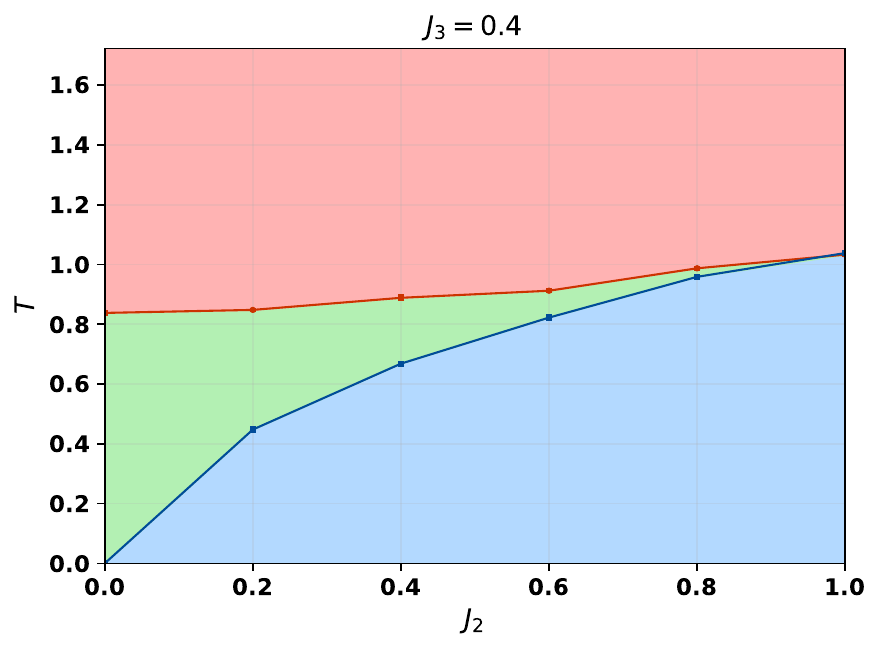}

    \medskip

    \includegraphics[width=0.31\textwidth]{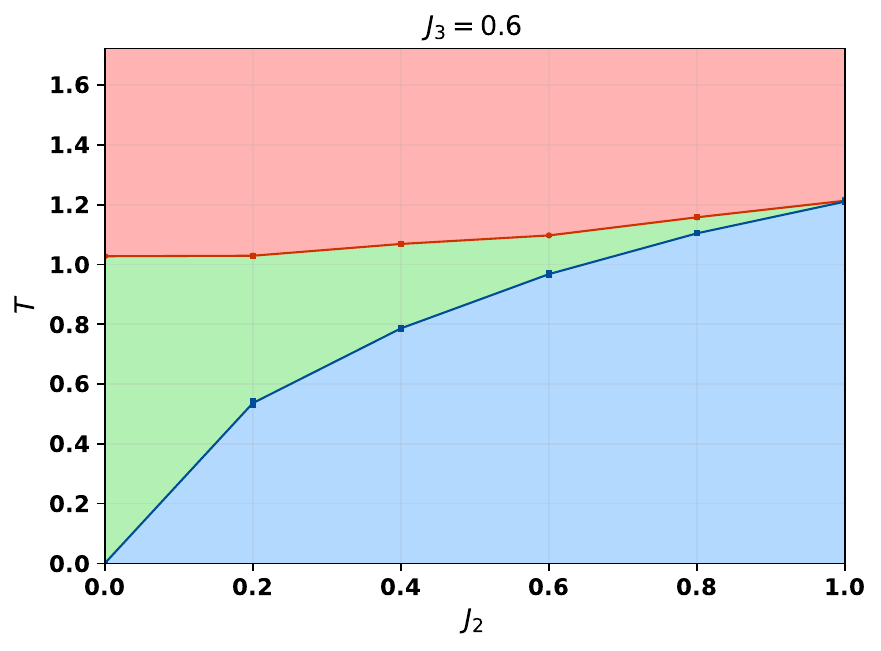}\hfill
    \includegraphics[width=0.31\textwidth]{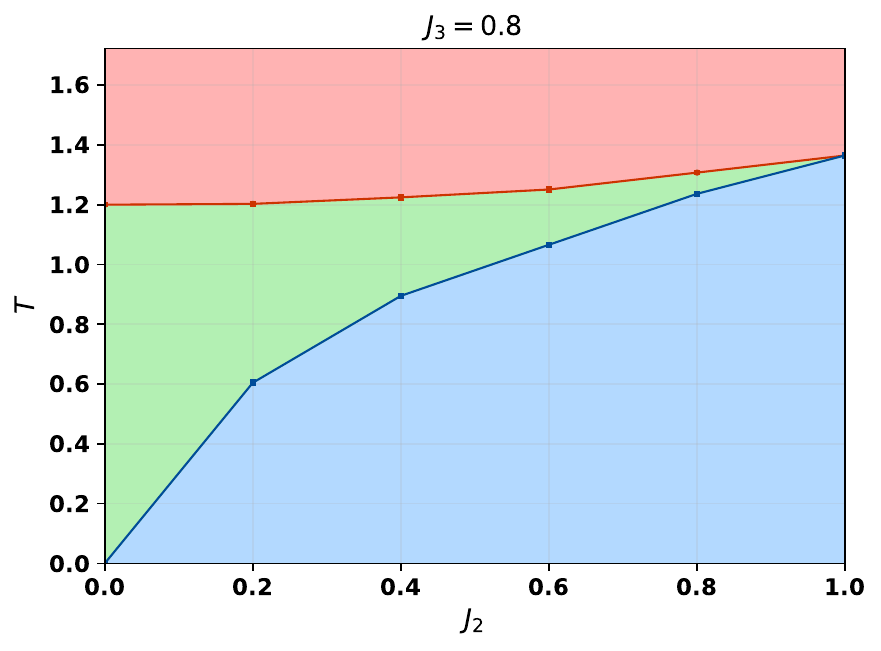}\hfill
    \includegraphics[width=0.31\textwidth]{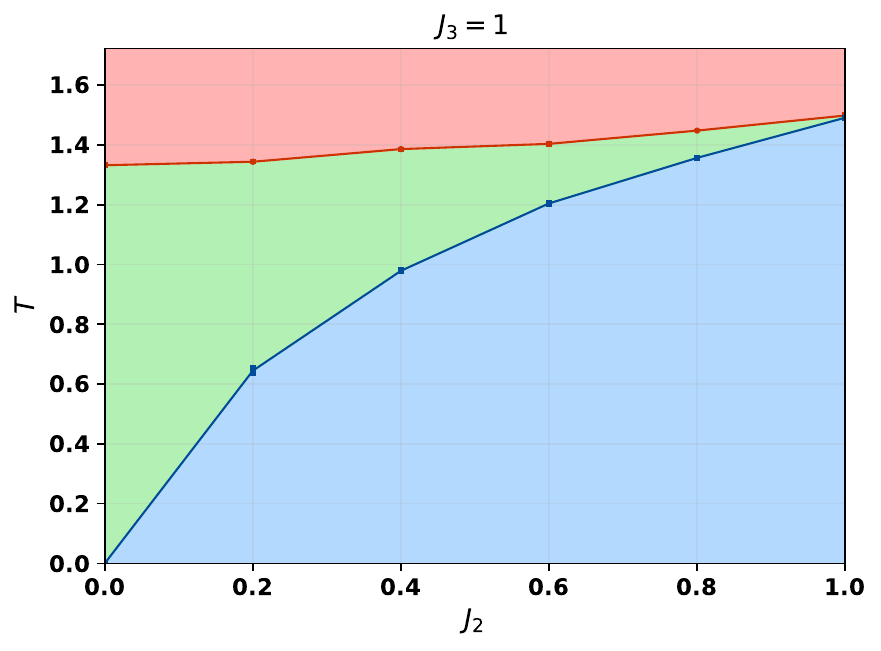}
    \caption{Phase-diagram slices at fixed $J_3$ ($J_3=0,0.2,0.4,0.6,0.8,1$), showing $T_c$ as a function of $J_2$ with $J_1=1$ fixed, obtained from Monte Carlo simulations. For $J_3>0$ and $J_1\neq J_2$, the two transition temperatures split, giving two distinct KT transitions. At $J_3=0$ they merge back into a single transition.}
    \label{fig:phase_slices_J3_sm}
\end{figure}

\begin{figure}[htbp]
    \centering
    \includegraphics[width=0.31\textwidth]{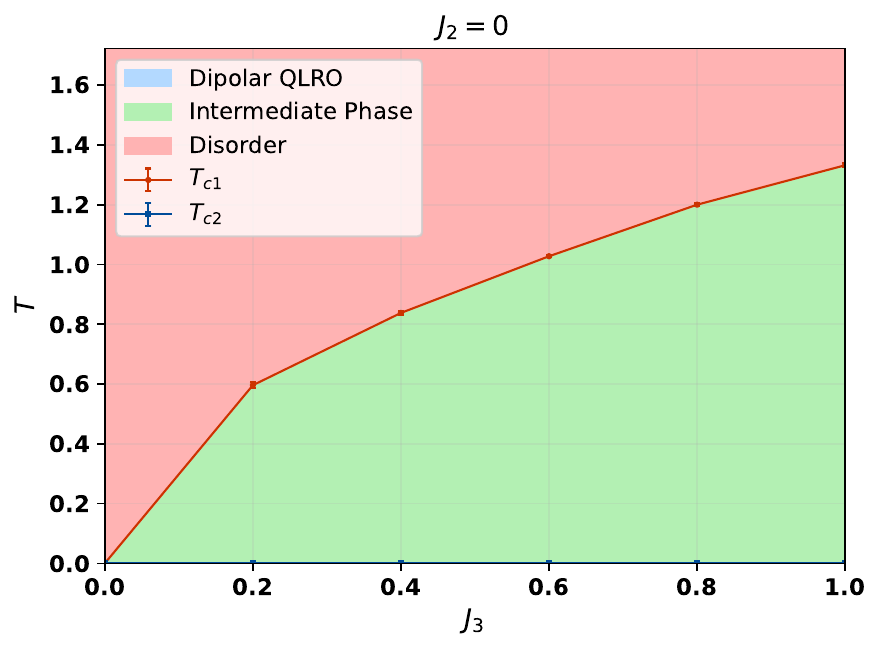}\hfill
    \includegraphics[width=0.31\textwidth]{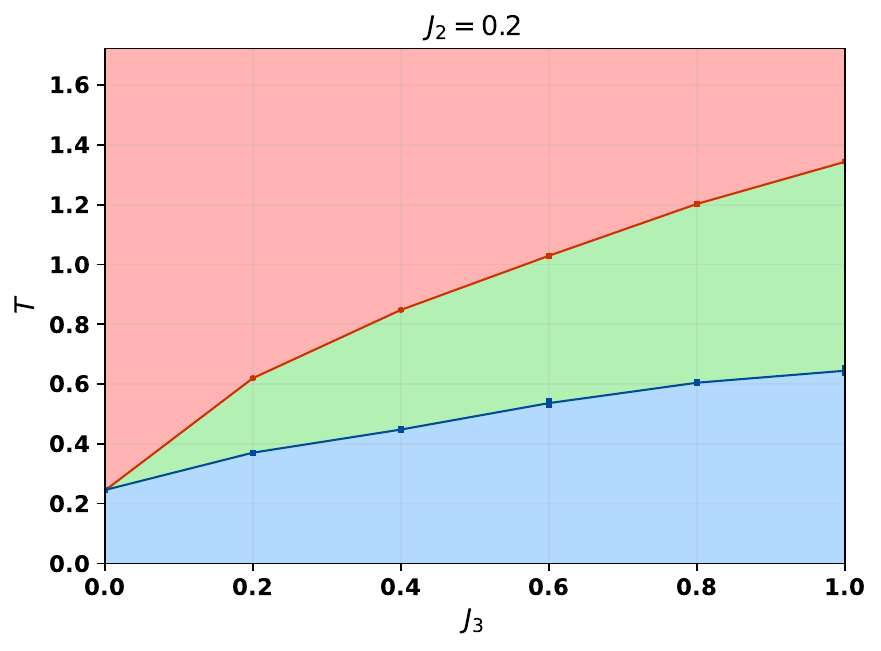}\hfill
    \includegraphics[width=0.31\textwidth]{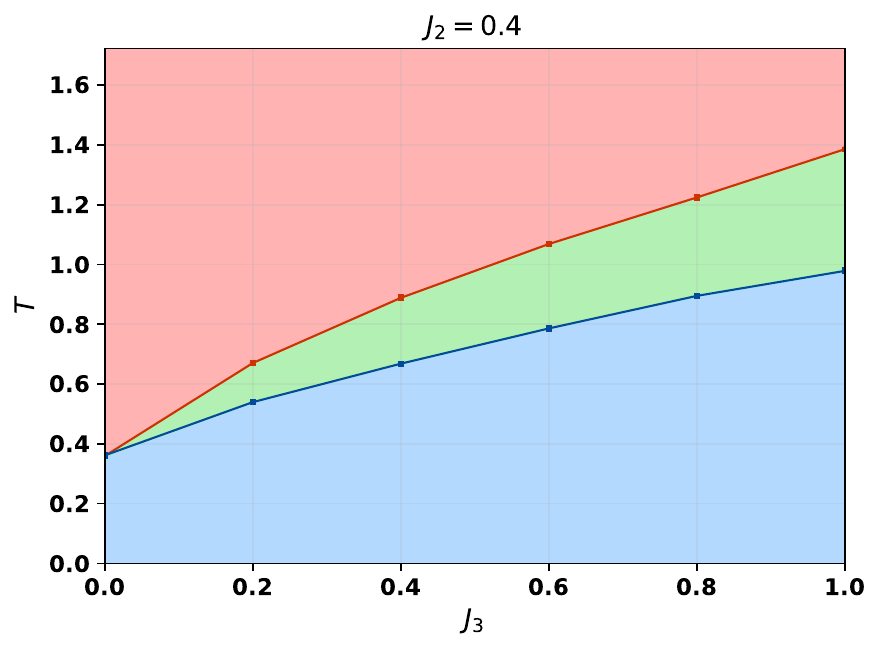}

    \medskip

    \includegraphics[width=0.31\textwidth]{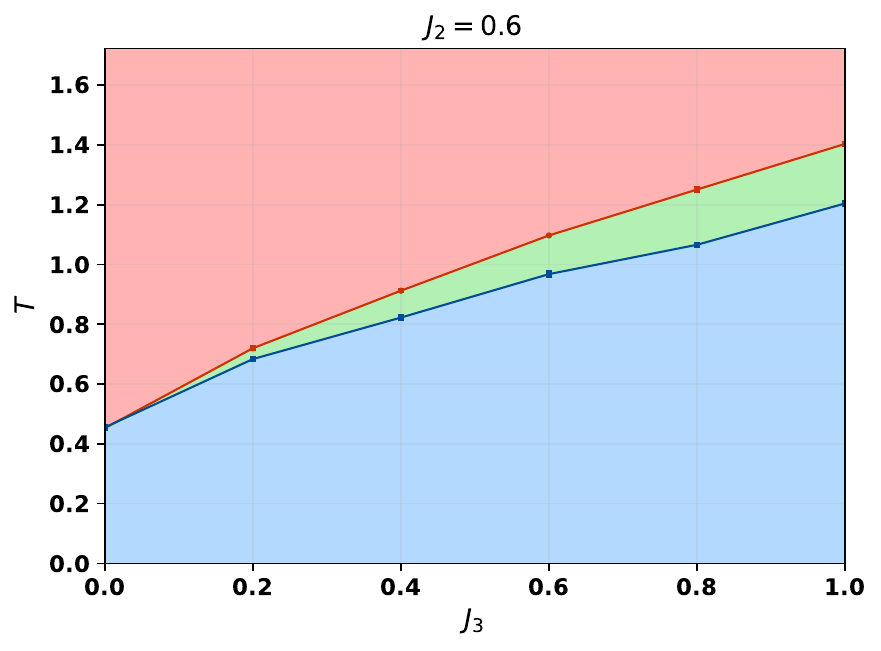}\hfill
    \includegraphics[width=0.31\textwidth]{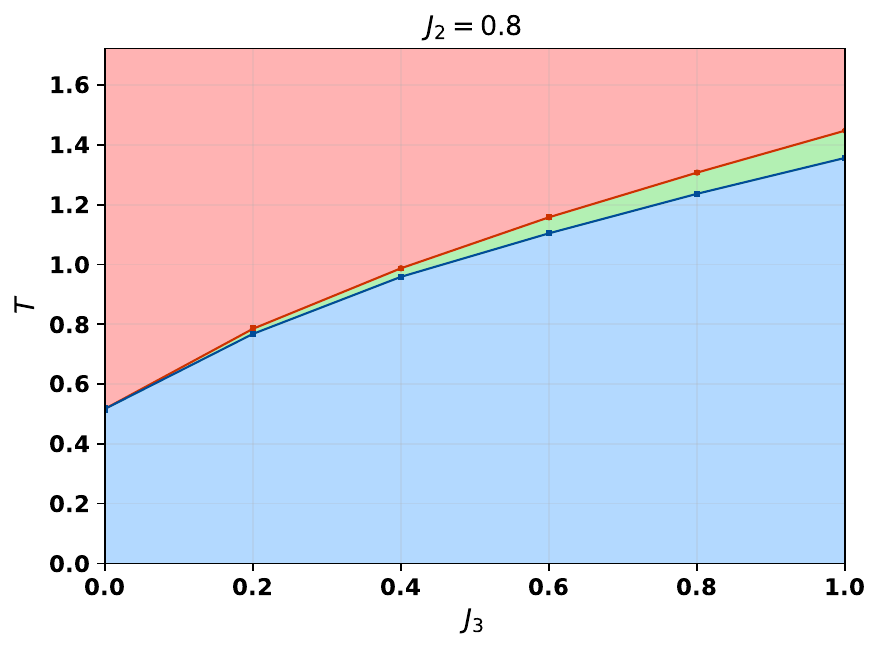}\hfill
    \includegraphics[width=0.31\textwidth]{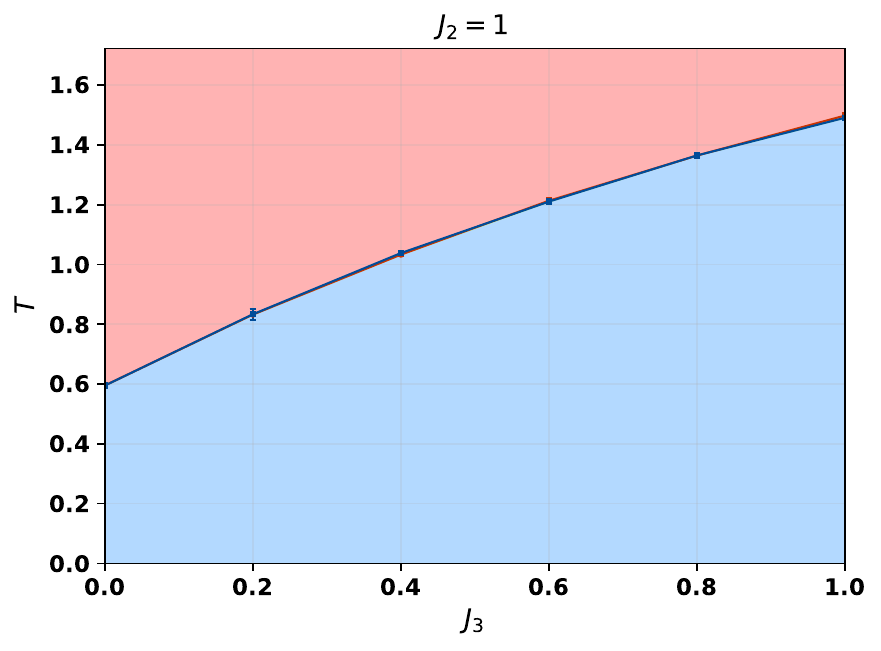}
    \caption{Phase-diagram slices at fixed $J_2$ ($J_2=0,0.2,0.4,0.6,0.8,1$), showing $T_c$ as a function of $J_3$ with $J_1=1$ fixed, obtained from Monte Carlo simulations. At $J_2=0$ only one KT transition survives, consistent with only one PG-vortex species driving the transition.}
    \label{fig:phase_slices_J2_sm}
\end{figure}

An overview of the phase diagram is presented in
Figs.~\ref{fig:phase_slices_J3_sm} and~\ref{fig:phase_slices_J2_sm}.
Fig.~\ref{fig:phase_slices_J3_sm} displays slices at fixed
$J_3$ ($J_3=0,0.2,0.4,0.6,0.8,1$), plotting $T_c$ versus $J_2$. When
$J_3$ is finite and $J_1\neq J_2$, the two transition temperatures split,
producing distinct KT transitions; at $J_3=0$, they merge into a single
transition. Fig.~\ref{fig:phase_slices_J2_sm} shows slices at fixed $J_2$
($J_2=0,0.2,0.4,0.6,0.8,1$), plotting $T_c$ versus $J_3$. The $J_2=0$
case exhibits only one transition, consistent with only one
PG-vortex species driving the transition when the
$(\partial_2^2\theta)^2$ term vanishes.

These data numerically realize the phase diagram outlined in
the main text and illustrate three qualitatively distinct
mechanisms:
\begin{enumerate}
    \item \textit{Degenerate unbinding:} at the isotropic point the two PG-vortex species are symmetry related and unbind together, producing a single KT transition.
    \item \textit{Anisotropy-driven splitting:} for generic $J_3>0$ and $J_1\neq J_2$, the two species become inequivalent and unbind at distinct temperatures, yielding an intermediate phase with only one channel quasi-long-range ordered.
    \item \textit{Merger at $J_3=0$:} despite $J_1\neq J_2$, the two transitions recombine when $J_3$ vanishes, because the two dipole channels then share a common structural condition for quasi-long-range order.
\end{enumerate}
The three phases identified in the anisotropic dipole-conserving XY model
($J_3>0$, $J_1\neq J_2$) are summarized in Table~\ref{tab:phases}.
\textbf{(i) Low-temperature dipole QLRO phase.}
Below $T_c^{(1)}$, both dipole fields $\chi_1$ and $\chi_2$ exhibit
quasi-long-range order: their correlators decay as a power law,
$\langle
    e^{i(\chi_\alpha(\mathbf r)-\chi_\alpha(\mathbf 0))}\rangle\sim r^{-\eta_\alpha}$ with $\eta_\alpha<\sfrac{1}{4}$ fixed by the
renormalized helicity moduli.  The two PG-vortex species
$\Theta_1^{(1)},\Theta_1^{(2)}$ are bound, as their logarithmic self-energies
outweigh the entropic gain, and all three generalized helicity moduli
$J_{s1},J_{s2},J_{s3}$ remain finite.  By contrast, the original phase field
$\theta$ never develops quasi-long-range order (QLRO)---its correlator decays super-exponentially
$\sim\exp[-c\,r^2\ln(L/r)]$ at all temperatures---and the standard
helicity modulus $J_s$ is strictly zero at all temperatures.

\textbf{(ii) Intermediate partial dipole QLRO phase.}
When $T_c^{(2)}<T<T_c^{(1)}$, $\Theta_1^{(2)}$ unbinds and destroys the
$\chi_2$ QLRO, turning $\langle
    e^{i(\chi_2(\mathbf r)-\chi_2(\mathbf 0))}\rangle$ short-ranged and driving $J_{s2}$ to
zero.  The $\chi_1$ channel retains QLRO because
$\Theta_1^{(1)}$ stays bound; accordingly $J_{s1}$ and $J_{s3}$ remain
positive; the modulus $J_{s3}$ survives because $\partial_1\partial_2\theta=\partial_2\chi_1$ inherits stiffness from the surviving $\chi_1$ quasi-long-range order.  This partial dipole QLRO---one channel rigid, the other
soft---is a direct consequence of the anisotropy that lifts the degeneracy
of the two PG-vortex species.

\textbf{(iii) High-temperature disordered phase.}
Above $T_c^{(1)}$, $\Theta_1^{(1)}$ also proliferates, destroying the
remaining $\chi_1$ QLRO and driving all three generalized helicity moduli to
zero.  Both dipole correlators are short-ranged, and the system is fully
disordered.
Throughout all three phases, the phase vortex $\Theta_0$ appears
thermally at any nonzero temperature (its self-energy is finite) and plays no
role in the critical physics; likewise $J_s$ vanishes identically
across both transitions, confirming that the standard KT diagnostic fails
in dipole-conserving systems.
\begin{table}[htbp]
    \centering
    \caption{Evolution of the three correlation functions across the
        dipole-conserving KT transitions ($J_3>0$, $J_1> J_2$).
        The phase field $\theta$ never supports QLRO;
        criticality is carried entirely by the dipole fields $\chi_1,\chi_2$.}
    \label{tab:phases}
    \begin{tabular*}{\columnwidth}{@{\extracolsep{\fill}}lccc}
        \toprule
        \textbf{Correlator}
        & \textbf{Dipolar QLRO}
        & \textbf{Intermediate Phase}
        & \textbf{Disorder} \\
        & $T<T_c^{(2)}$
        & $T_c^{(2)}<T<T_c^{(1)}$
        & $T>T_c^{(1)}$ \\
        \midrule
        $\langle e^{i(\theta(\mathbf r)-\theta(\mathbf 0))}\rangle$
        & $\sim\exp\!\bigl[{-c\,r^{2}\ln(L/r)}\bigr]$
        & $\sim\exp\!\bigl[{-c\,r^{2}\ln(L/r)}\bigr]$
        & $\sim\exp\!\bigl[{-c\,r^{2}\ln(L/r)}\bigr]$ \\
        & \footnotesize (super-exponential, no QLRO)
        & \footnotesize (super-exponential, no QLRO)
        & \footnotesize (super-exponential, no QLRO) \\
        $\langle
            e^{i(\chi_1(\mathbf r)-\chi_1(\mathbf 0))}\rangle$
        & $\sim r^{-\eta_1}$ \;(\textsc{qlro})
        & $\sim r^{-\eta_1}$ \;(\textsc{qlro})
        & short-ranged \\
        $\langle
            e^{i(\chi_2(\mathbf r)-\chi_2(\mathbf 0))}\rangle$
        & $\sim r^{-\eta_2}$ \;(\textsc{qlro})
        & short-ranged
        & short-ranged \\
        \bottomrule
    \end{tabular*}
\end{table}

\section{Representative numerical data for all four regimes}
\label{sec:sm_numerics}

In this section, we collect the representative Monte Carlo data corresponding to
all four parameter regimes discussed in the main text and End Matter:

\begin{minipage}{0.43\textwidth}
(a) the isotropic point (main text);\\
(c) the $J_3=0$ merger regime (End Matter);
\end{minipage}
\hfill
\begin{minipage}{0.53\textwidth}
(b) the anisotropy-induced split-transition regime (main text);\\
(d) the $J_2=0$ single-species regime (End Matter).
\end{minipage}

These figures serve as the detailed numerical backbone with all system sizes,
specific heat, and finite-size-scaling analyses behind the concise discussions
in the main text (Figs.~\ref{fig:corr}--\ref{fig:vortex}) and End Matter
(Fig.~\ref{fig:em_data}).

\subsection{Isotropic case: one transition from two degenerate PG-vortex species}

\begin{figure}[htbp]
    \centering

    \begin{minipage}[b]{0.35\textwidth}
        \centering\includegraphics[width=\textwidth]{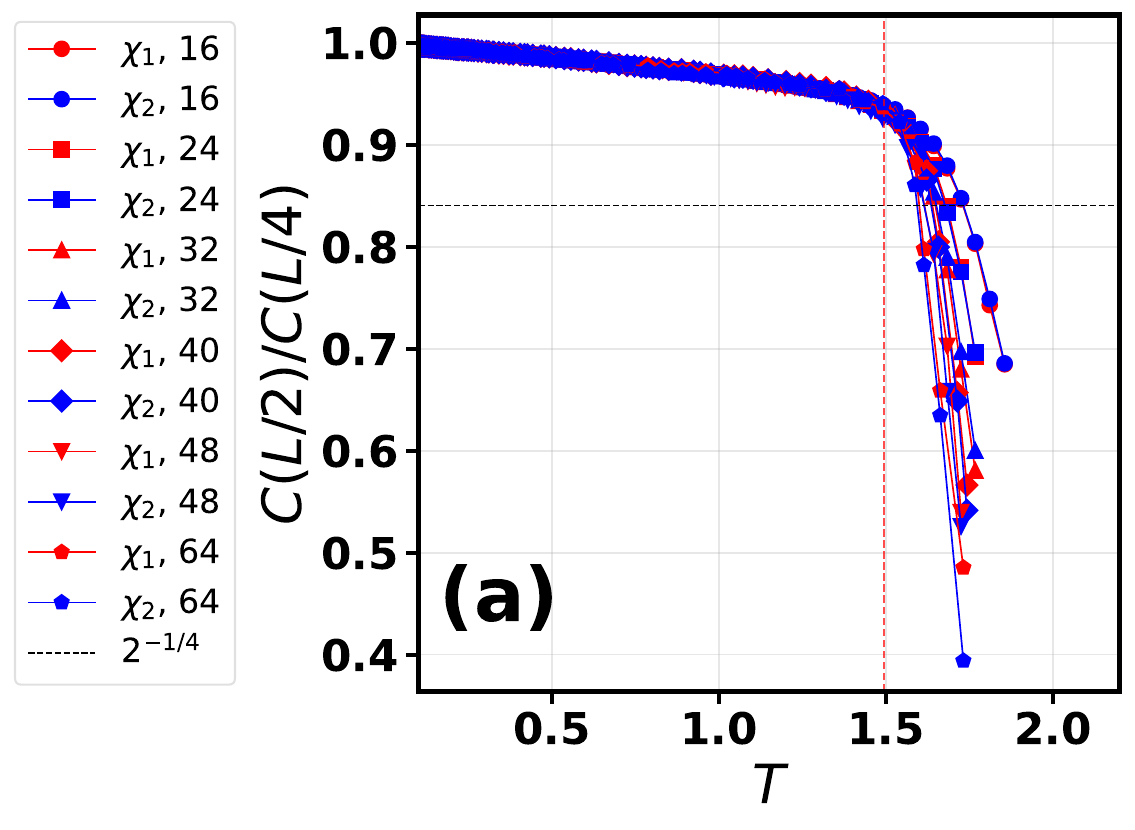}
    \end{minipage}
    \begin{minipage}[b]{0.35\textwidth}
        \centering\includegraphics[width=\textwidth]{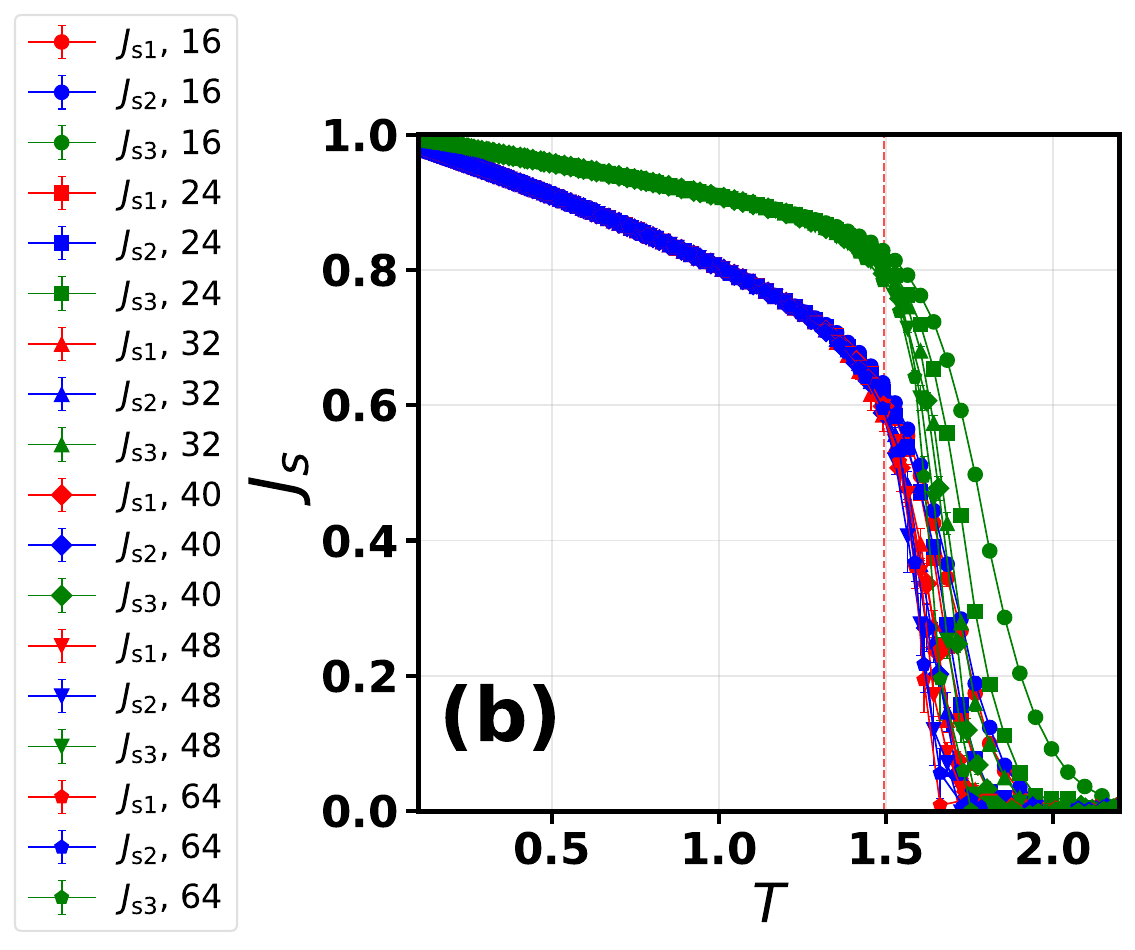}
    \end{minipage}

    \bigskip
    \begin{minipage}[b]{0.219\textwidth}
        \centering\includegraphics[width=\textwidth]{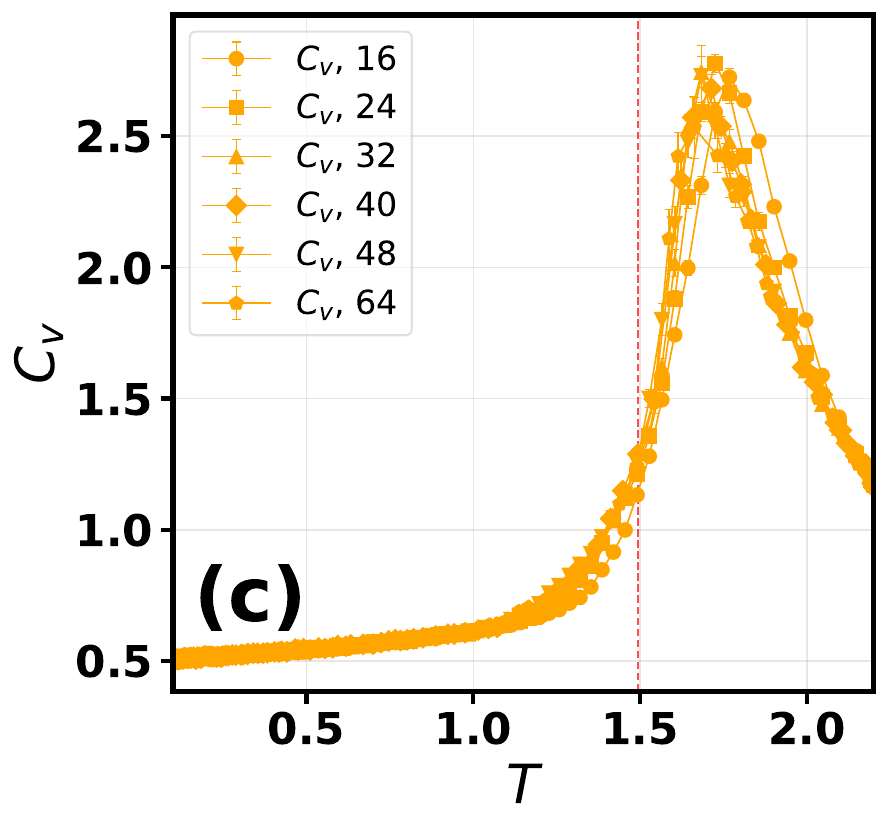}
    \end{minipage}
    \hfill
    \begin{minipage}[b]{0.229\textwidth}
        \centering\includegraphics[width=\textwidth]{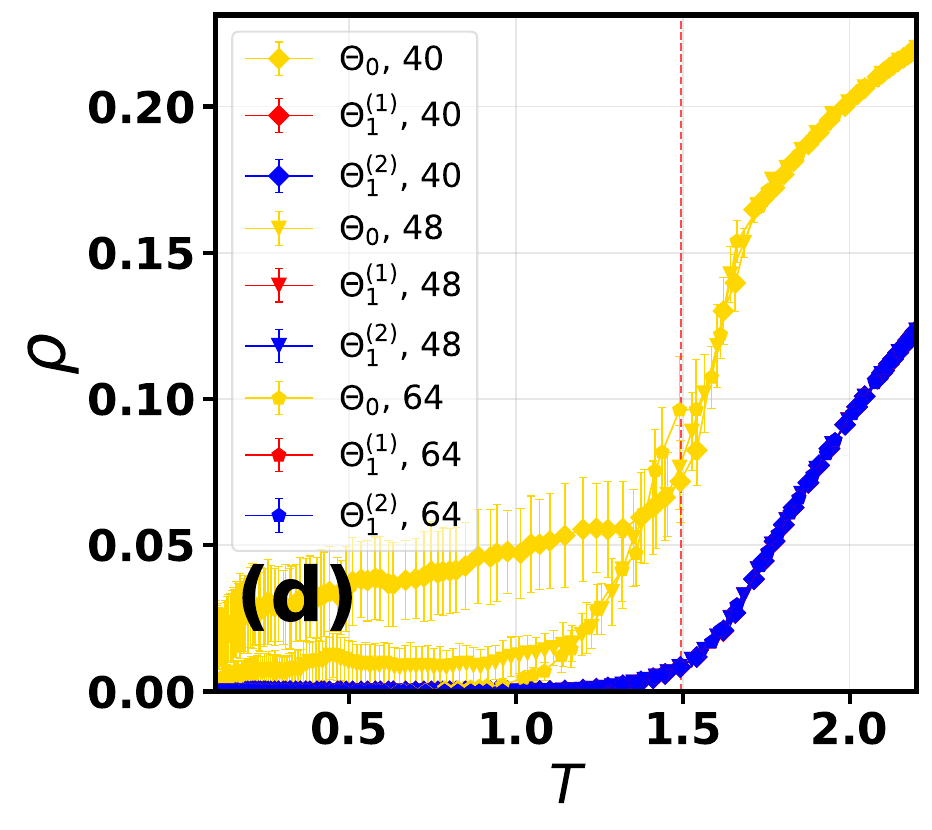}
    \end{minipage}
    \hfill
    \begin{minipage}[b]{0.219\textwidth}
        \centering\includegraphics[width=\textwidth]{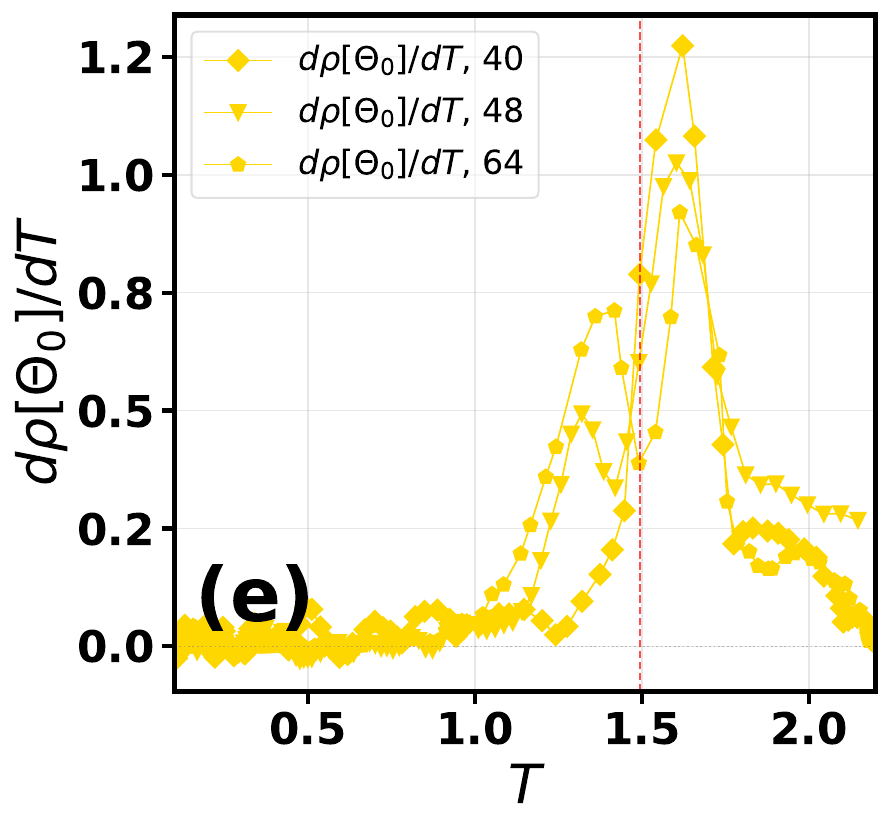}
    \end{minipage}
    \hfill
    \begin{minipage}[b]{0.3\textwidth}
        \centering\includegraphics[width=\textwidth]{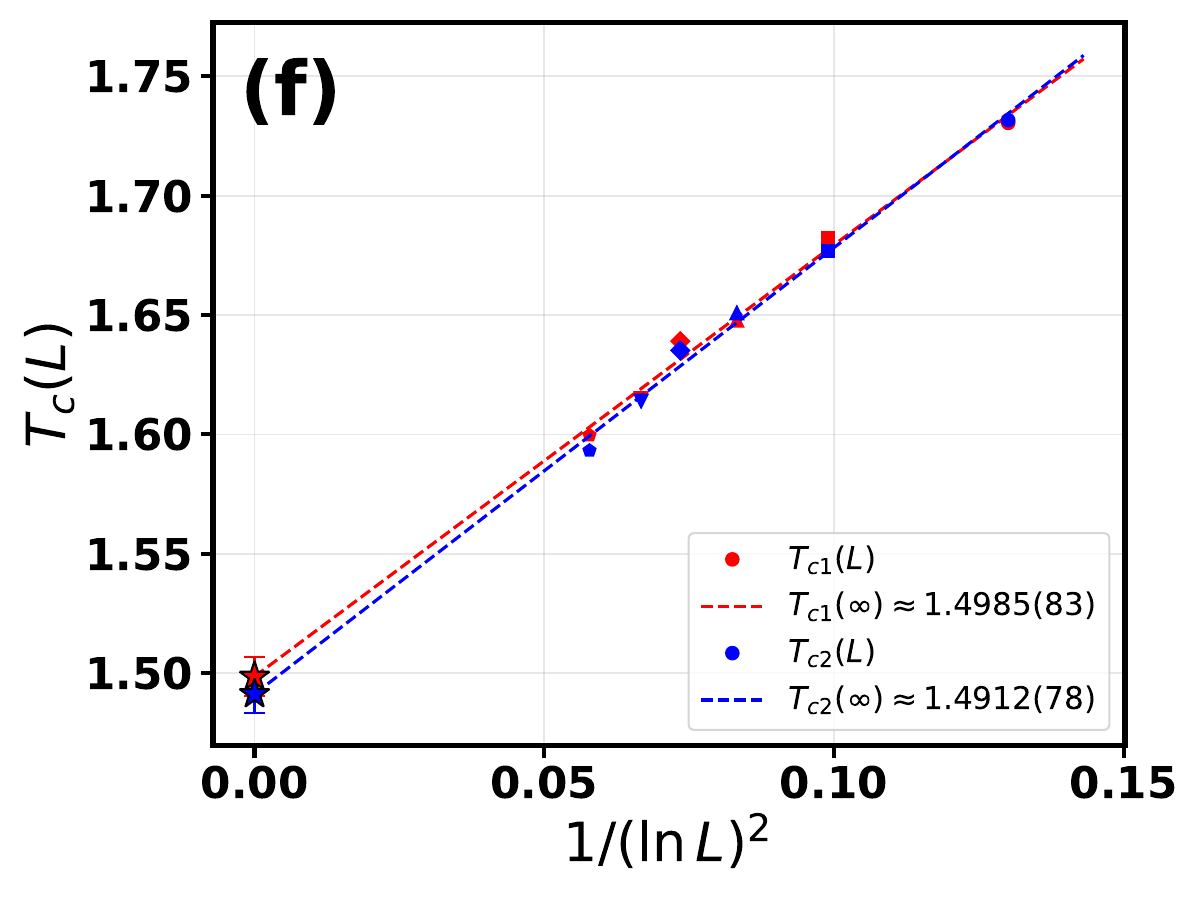}
    \end{minipage}
    \caption{Monte Carlo data for the isotropic dipole-conserving XY model at $J_1=J_2=J_3=1$. (a) Correlation ratio $C(L/2)/C(L/4)$ of the dipole field $\chi$, whose scale-invariant crossing point locates the critical temperature $T_c$. (b) Generalized helicity moduli $J_{s1}$, $J_{s2}$, $J_{s3}$ as functions of temperature for multiple system sizes $L$; all three moduli drop together at a single temperature scale, indicating a common KT transition. (c) Specific heat $C_v$ as a function of temperature, showing a broad peak near the transition. (d) Vortex densities $\rho(\Theta_0)$, $\rho(\Theta_1^{(1)})$, and $\rho(\Theta_1^{(2)})$ of all three species. The two PG-vortex species proliferate simultaneously, confirming their degeneracy under isotropic couplings, while $\rho(\Theta_0)$ remains finite at all temperatures due to the finite self-energy of $\Theta_0$. (e) Temperature derivative $\mathrm{d}\rho(\Theta_0)/\mathrm{d}T$, whose peak signals critical enhancement of phase-vortex generation. (f) Finite-size scaling analysis of the pseudocritical temperatures. Together these panels demonstrate that the single KT transition is driven by the simultaneous unbinding of the two degenerate PG-vortex species.}
    \label{fig:isotropic_results_sm}
\end{figure}

We first consider the isotropic point $J_1=J_2=J_3=1$. The two PG-vortex
species are degenerate, giving a single KT transition signaled by a
scale-invariant crossing of the dipole-field correlation ratio $C(L/2)/C(L/4)$
and a common drop in the generalized helicity moduli.
The phase-vortex density $\rho(\Theta_0)$ does not track this transition
owing to its finite self-energy, confirming that the critical behavior is
governed by the PG-vortex sector. The corresponding Monte Carlo
data are shown in Fig.~\ref{fig:isotropic_results_sm}, which presents the
complete set of observables for this regime, complementing the overview in the
main-text Figs.~\ref{fig:corr}, \ref{fig:stiffness}, and \ref{fig:vortex}(a).

\subsection{Anisotropic case I: $J_1\neq J_2$ and $J_3>0$ --- splitting into two KT transitions}

\begin{figure}[htbp]
    \centering
    \begin{minipage}[b]{0.35\textwidth}
        \centering\includegraphics[width=\textwidth]{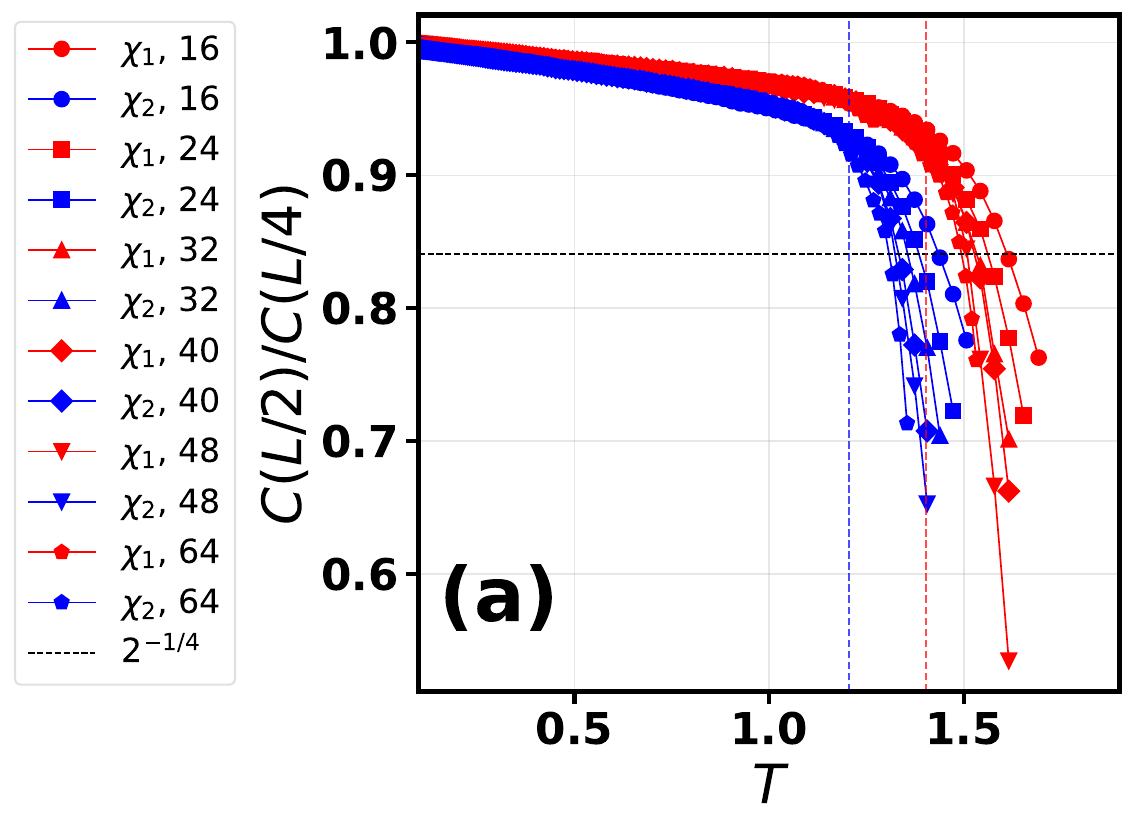}
    \end{minipage}
    \begin{minipage}[b]{0.35\textwidth}
        \centering\includegraphics[width=\textwidth]{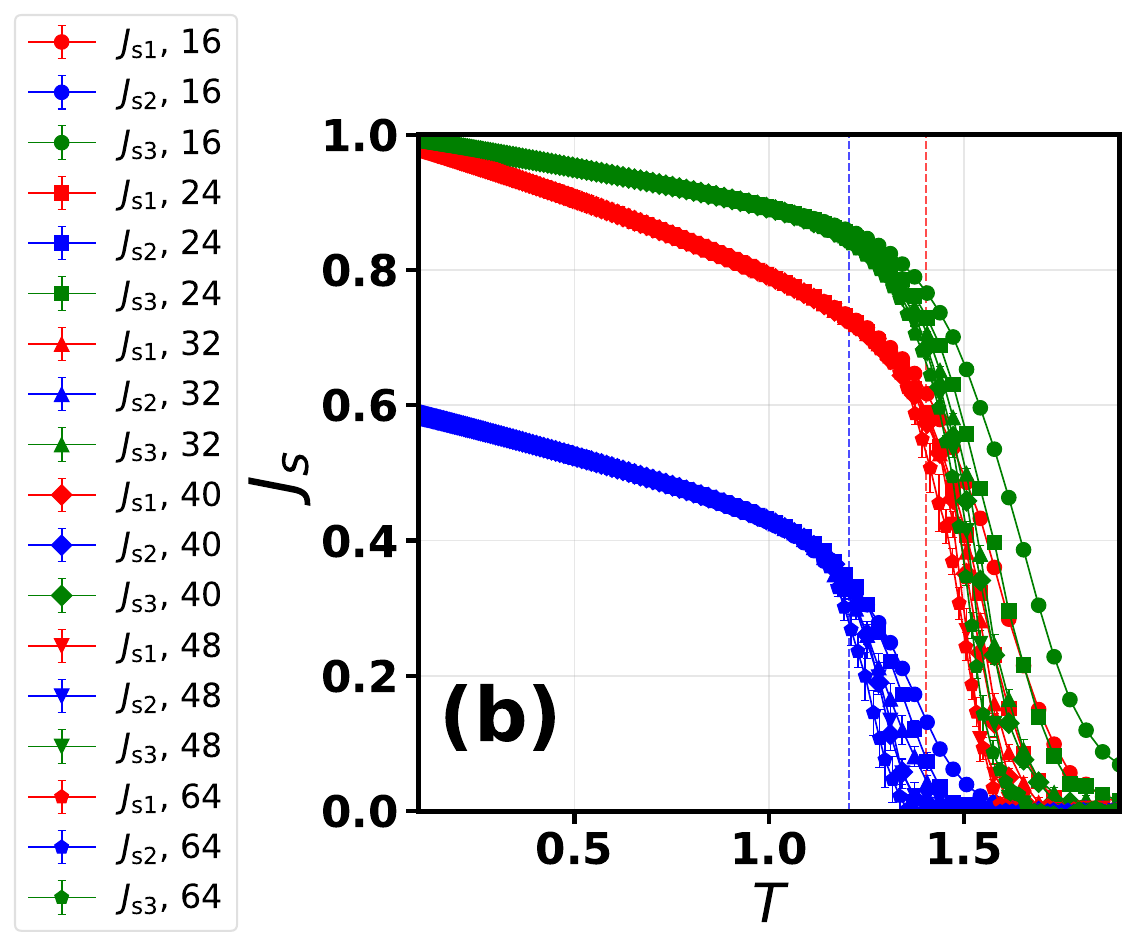}
    \end{minipage}

    \bigskip
    \begin{minipage}[b]{0.219\textwidth}
        \centering\includegraphics[width=\textwidth]{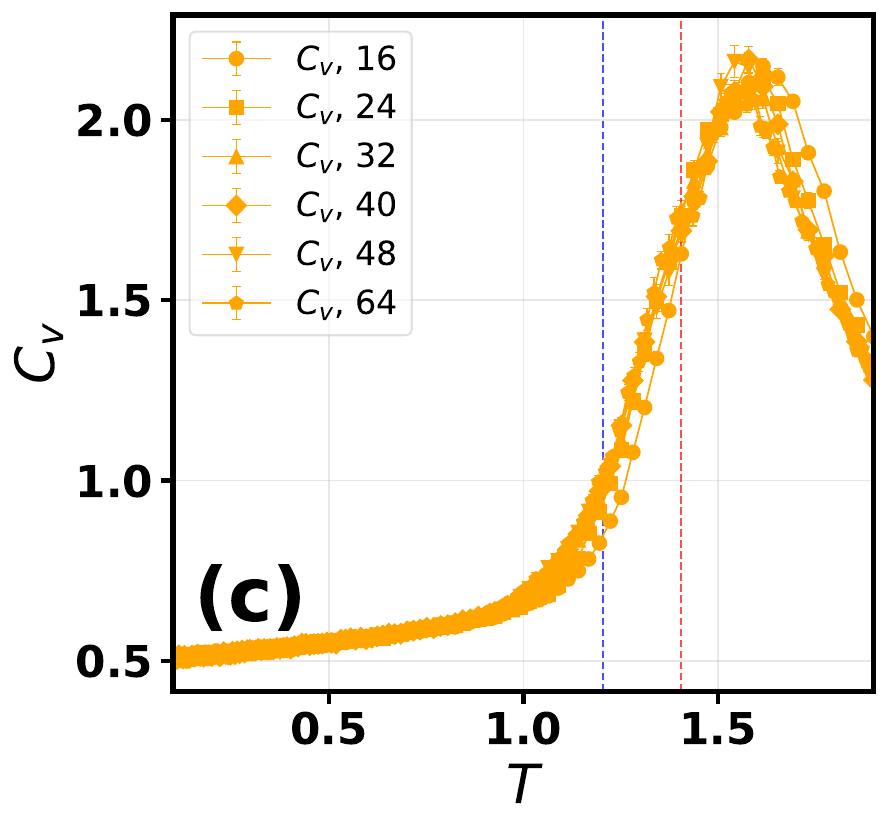}
    \end{minipage}
    \hfill
    \begin{minipage}[b]{0.229\textwidth}
        \centering\includegraphics[width=\textwidth]{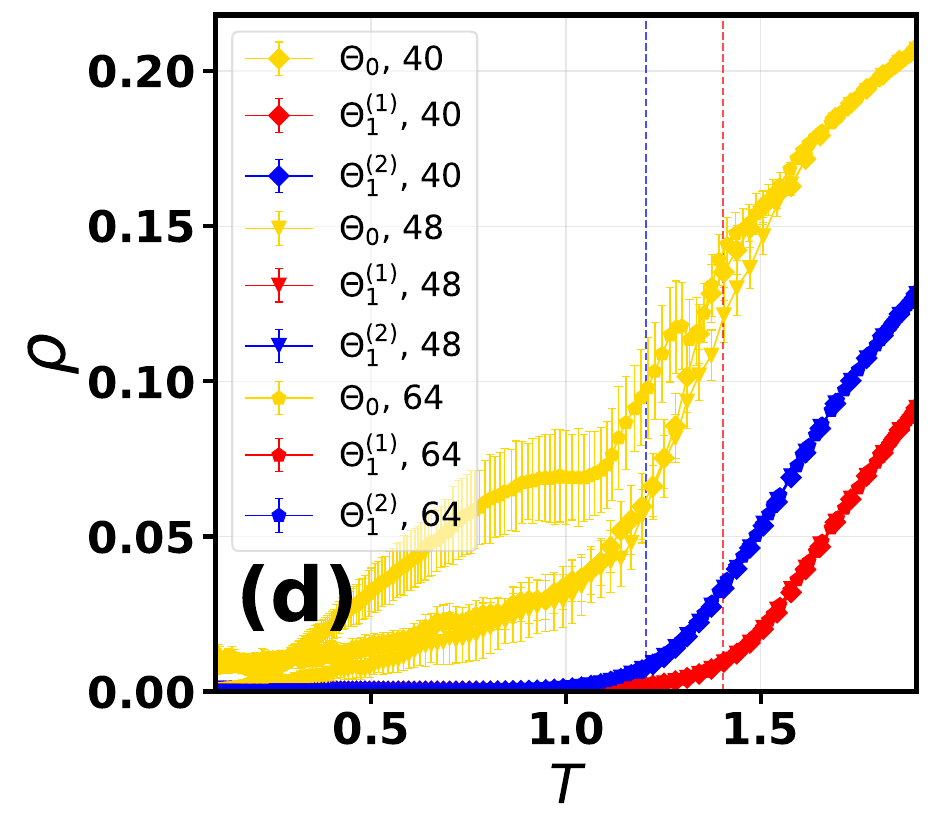}
    \end{minipage}
    \hfill
    \begin{minipage}[b]{0.216\textwidth}
        \centering\includegraphics[width=\textwidth]{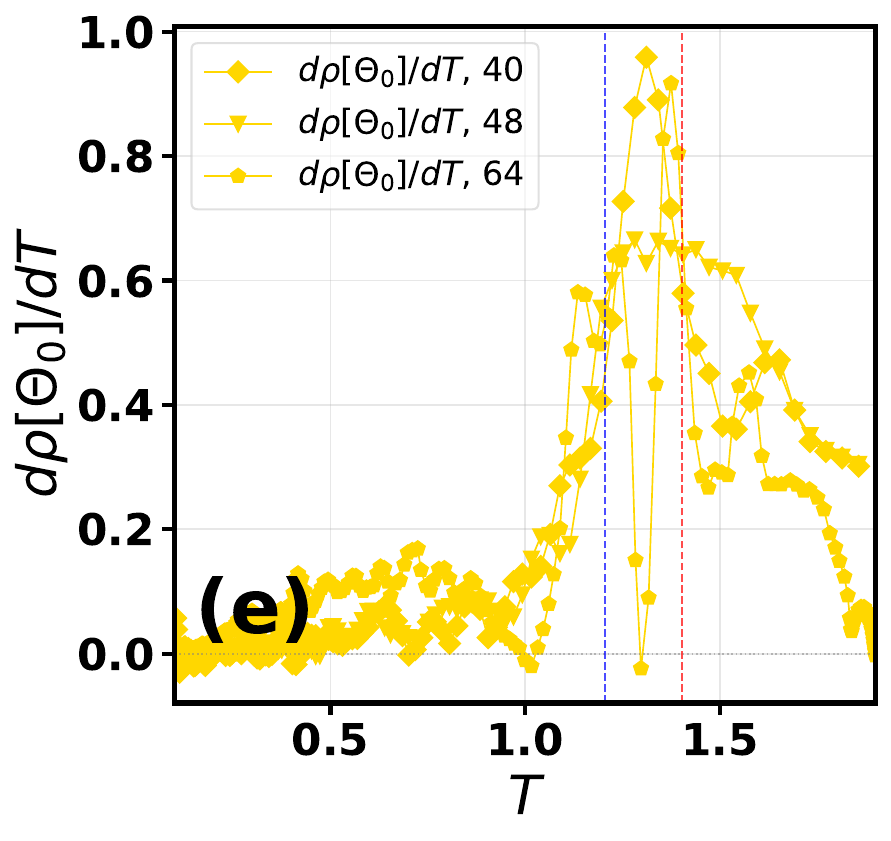}
    \end{minipage}
    \hfill
    \begin{minipage}[b]{0.3\textwidth}
        \centering\includegraphics[width=\textwidth]{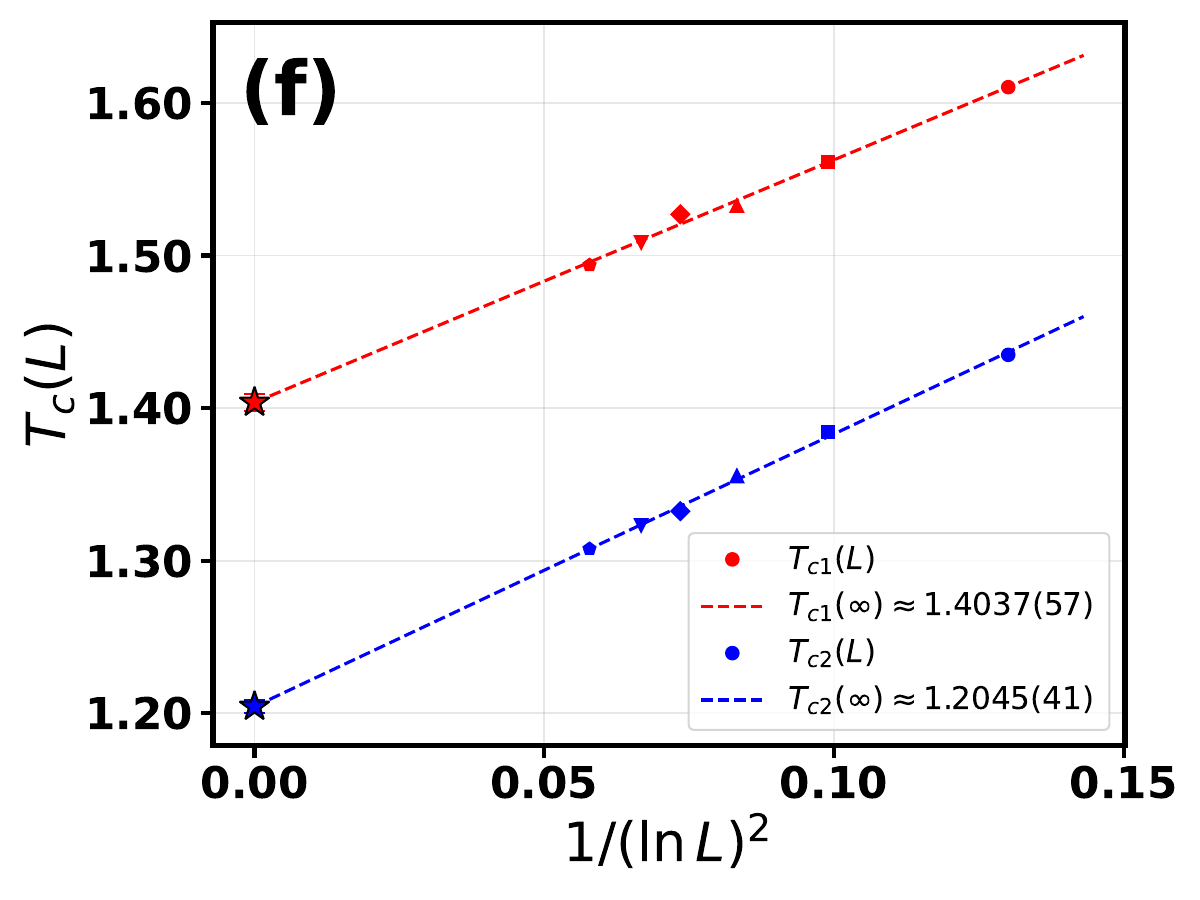}
    \end{minipage}
    \caption{Monte Carlo data for the anisotropic dipole-conserving XY model at $J_1=1$, $J_2=0.6$, $J_3=1$. (a) Correlation ratio $C(L/2)/C(L/4)$; the two crossing features correspond to the two critical temperatures $T_c^{(1)}$ and $T_c^{(2)}$, which are determined from the crossing points. (b) Generalized helicity moduli $J_{s1}$, $J_{s2}$, $J_{s3}$ showing two distinct drops at two separate temperatures, corroborating the splitting of the single KT transition into two. (c) Specific heat $C_v$ as a function of temperature. (d) Vortex densities $\rho(\Theta_0)$, $\rho(\Theta_1^{(1)})$, and $\rho(\Theta_1^{(2)})$ of all three species. The two PG-vortex species proliferate at different temperatures, providing defect-level evidence that anisotropy lifts the degeneracy between the two PG-vortex species. (e) Temperature derivative $\mathrm{d}\rho(\Theta_0)/\mathrm{d}T$, whose peak signals critical enhancement of phase-vortex generation. (f) Finite-size scaling analysis of the pseudocritical temperatures.}
    \label{fig:anisotropic_results_sm}
\end{figure}

We next consider $J_1=1$, $J_2=0.6$, $J_3=1$. Anisotropy lifts the degeneracy
between the two PG-vortex species, giving two distinct KT
transitions signaled by two scale-invariant crossings of the dipole-field
correlation ratio and separate drops in the generalized helicity moduli. The two
densities proliferate at different temperatures. Between the two transitions,
one channel retains quasi-long-range order while the other is
disordered---the intermediate phase. The full numerical
evolution is displayed in Fig.~\ref{fig:anisotropic_results_sm}, which
presents the complete set of observables for this regime, complementing the
overview in the main-text Figs.~\ref{fig:corr}, \ref{fig:stiffness}, and
\ref{fig:vortex}(b).

\subsection{Anisotropic case II: $J_3=0$ --- merging of the two KT transitions}

\begin{figure}[htbp]
    \centering
    \begin{minipage}[b]{0.35\textwidth}
        \centering\includegraphics[width=\textwidth]{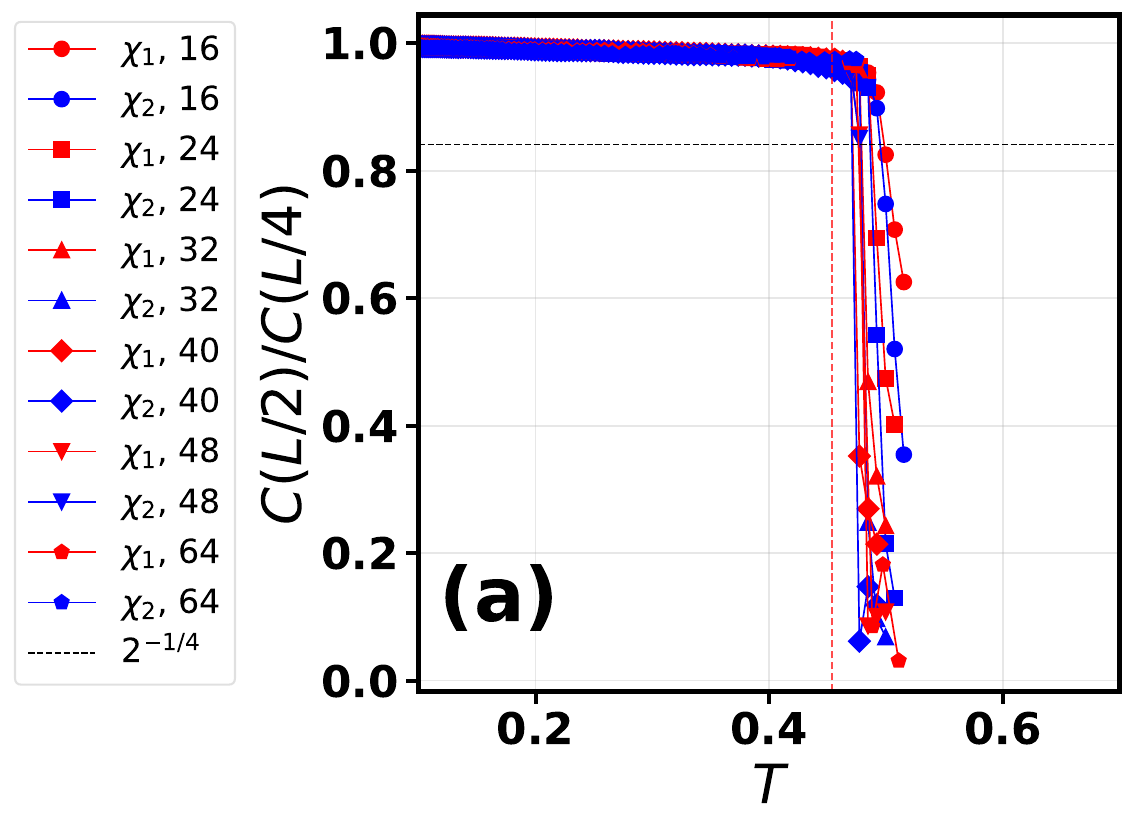}
    \end{minipage}
    \begin{minipage}[b]{0.35\textwidth}
        \centering\includegraphics[width=\textwidth]{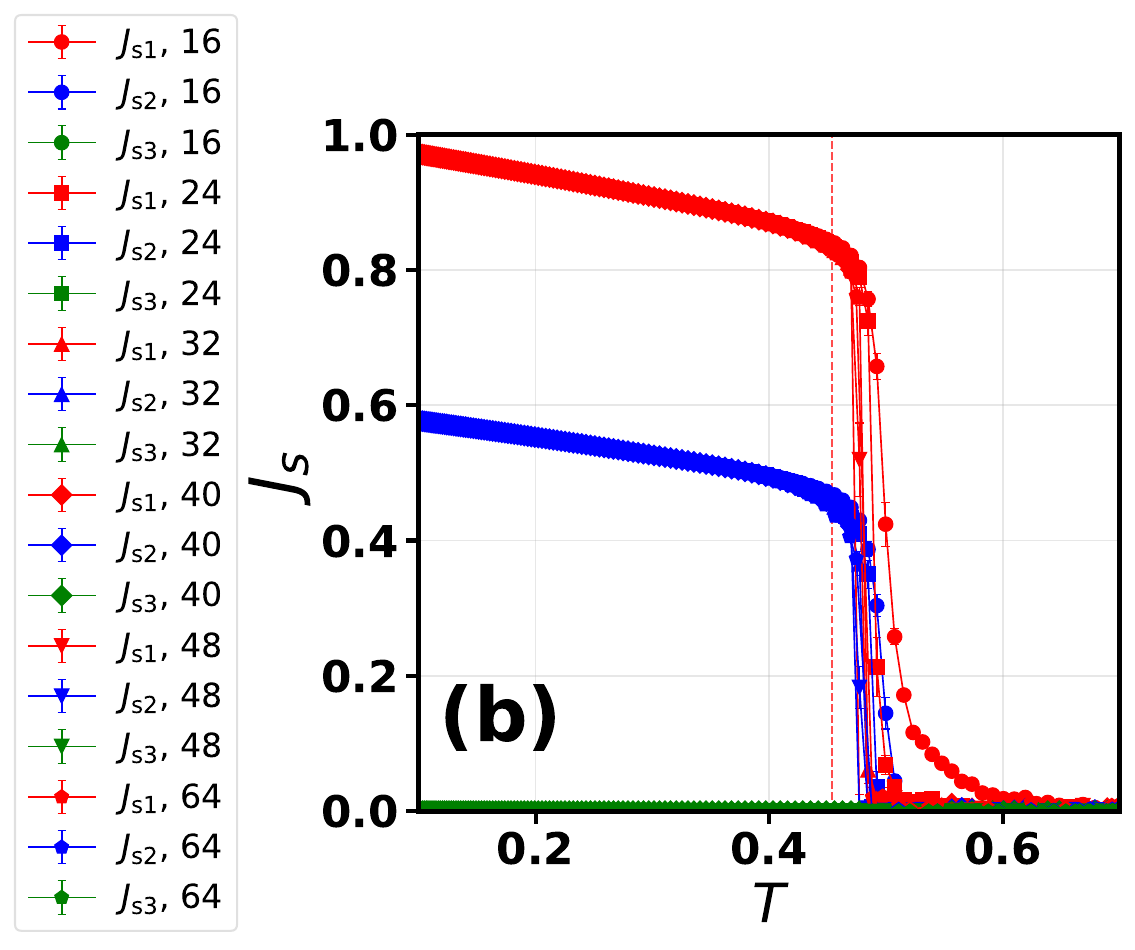}
    \end{minipage}

    \bigskip
    \begin{minipage}[b]{0.226\textwidth}
        \centering\includegraphics[width=\textwidth]{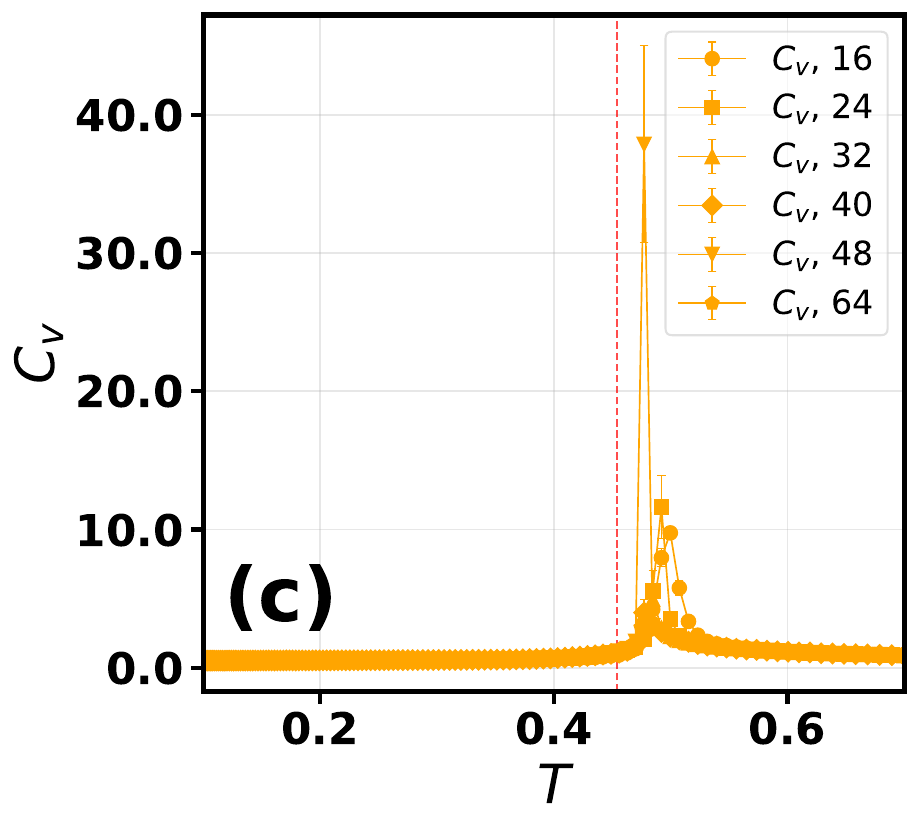}
    \end{minipage}
    \hfill
    \begin{minipage}[b]{0.229\textwidth}
        \centering\includegraphics[width=\textwidth]{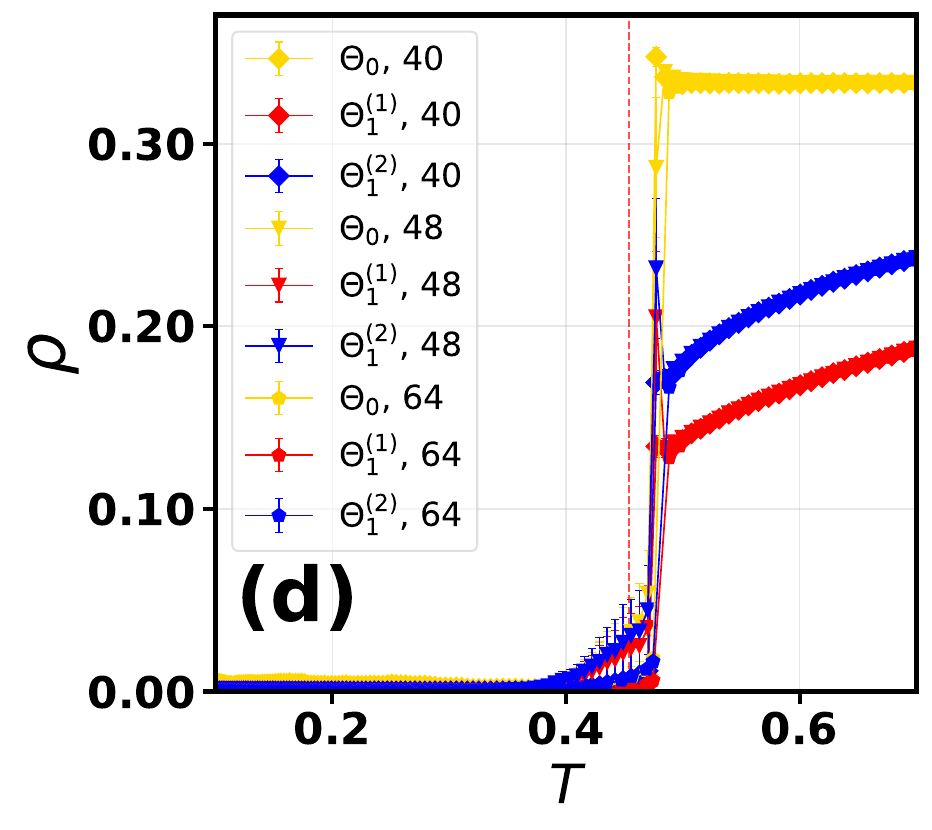}
    \end{minipage}
    \hfill
    \begin{minipage}[b]{0.226\textwidth}
        \centering\includegraphics[width=\textwidth]{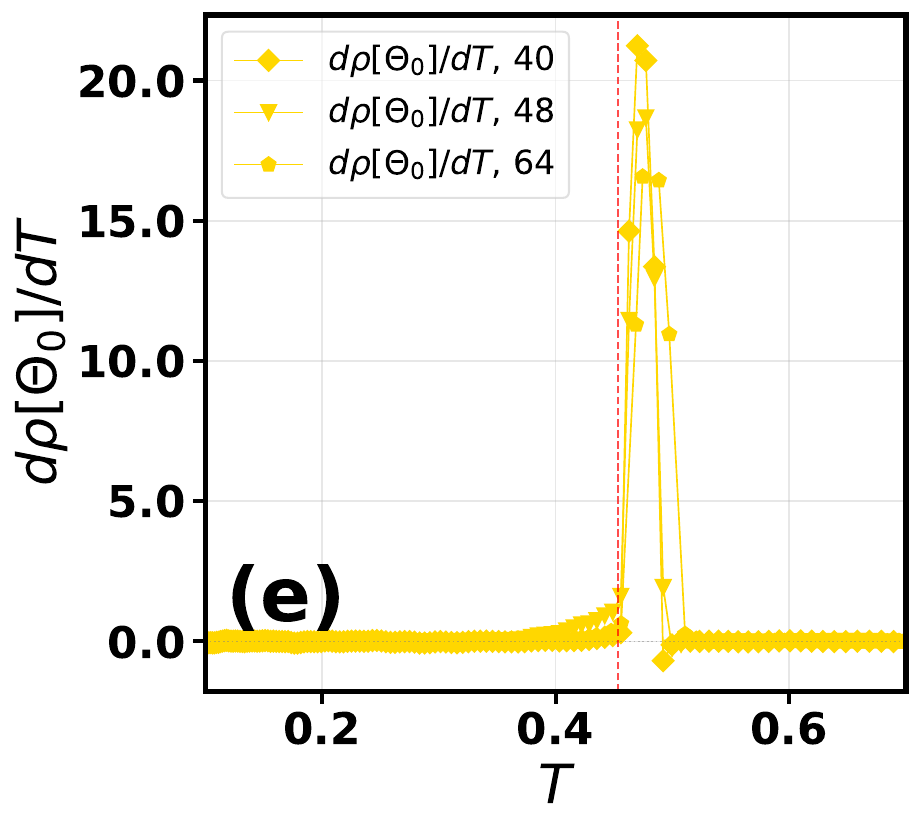}
    \end{minipage}
    \hfill
    \begin{minipage}[b]{0.3\textwidth}
        \centering\includegraphics[width=\textwidth]{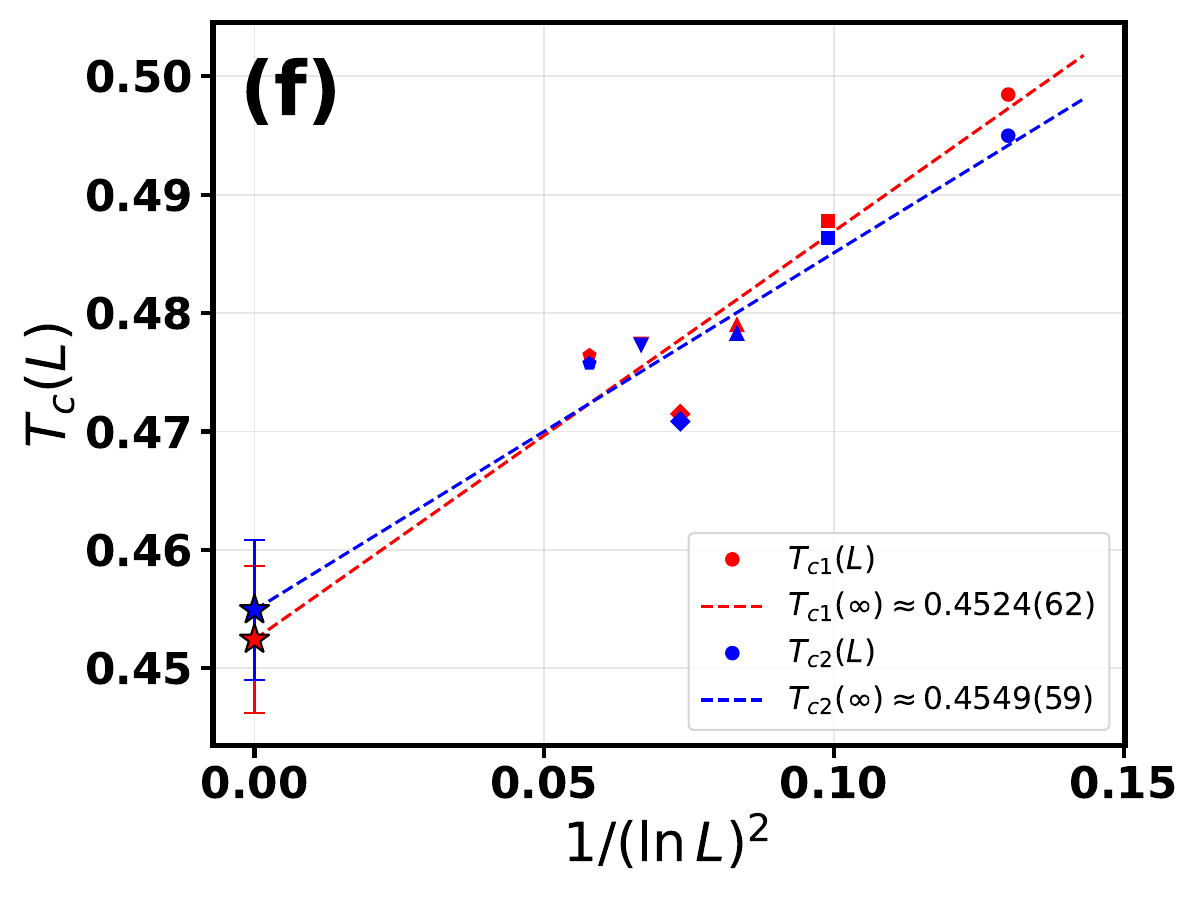}
    \end{minipage}
    \caption{Monte Carlo data for the anisotropic dipole-conserving XY model at $J_1=1$, $J_2=0.6$, $J_3=0$. (a) Correlation ratio $C(L/2)/C(L/4)$; a single crossing confirms one critical temperature. (b) Generalized helicity moduli $J_{s1}$, $J_{s2}$, $J_{s3}$; despite $J_1\neq J_2$, the helicity moduli $J_{s1}$ and $J_{s2}$ drop at the same temperature, corroborating the merging of the two KT transitions into a single transition when $J_3$ is set to zero. (c) Specific heat $C_v$. (d) Vortex densities $\rho(\Theta_0)$, $\rho(\Theta_1^{(1)})$, and $\rho(\Theta_1^{(2)})$ of all three species. The two PG-vortex species unbind at the same transition; $\rho(\Theta_0)$ (finite self-energy) is in principle nonzero but suppressed below numerical resolution at low temperatures in this regime. (e) Temperature derivative $\mathrm{d}\rho(\Theta_0)/\mathrm{d}T$, whose peak signals critical enhancement of phase-vortex generation. (f) Finite-size scaling analysis of the pseudocritical temperatures.}
    \label{fig:aniso2_results_sm}
\end{figure}

We next consider $J_3=0$ with $J_1=1$, $J_2=0.6$. The $J_3$
coupling vanishes, and the two dipole correlation exponents share a common
factor $\sqrt{J_{s1}J_{s2}}$:
\begin{equation}
    \eta_1 = \frac{1}{2\sqrt{2}\pi\beta\sqrt{J_{s1}\sqrt{J_{s1}J_{s2}}}},
    \qquad
    \eta_2 = \frac{1}{2\sqrt{2}\pi\beta\sqrt{J_{s2}\sqrt{J_{s1}J_{s2}}}}.
    \label{eq:eta_merged_sm}
\end{equation}
Both denominators contain $\sqrt{J_{s1}J_{s2}}$, so if either $J_{s1}$ or
$J_{s2}$ vanishes, both $\eta_1$ and $\eta_2$ diverge together, precluding an
intermediate phase. The two PG-vortex species therefore proliferate at the
same temperature despite the anisotropy, giving a single KT transition.
The data are shown in Fig.~\ref{fig:aniso2_results_sm}, complementing the End
Matter Fig.~\ref{fig:em_data}.

\begin{figure}[htbp]
    \centering
    \includegraphics[width=0.45\textwidth]{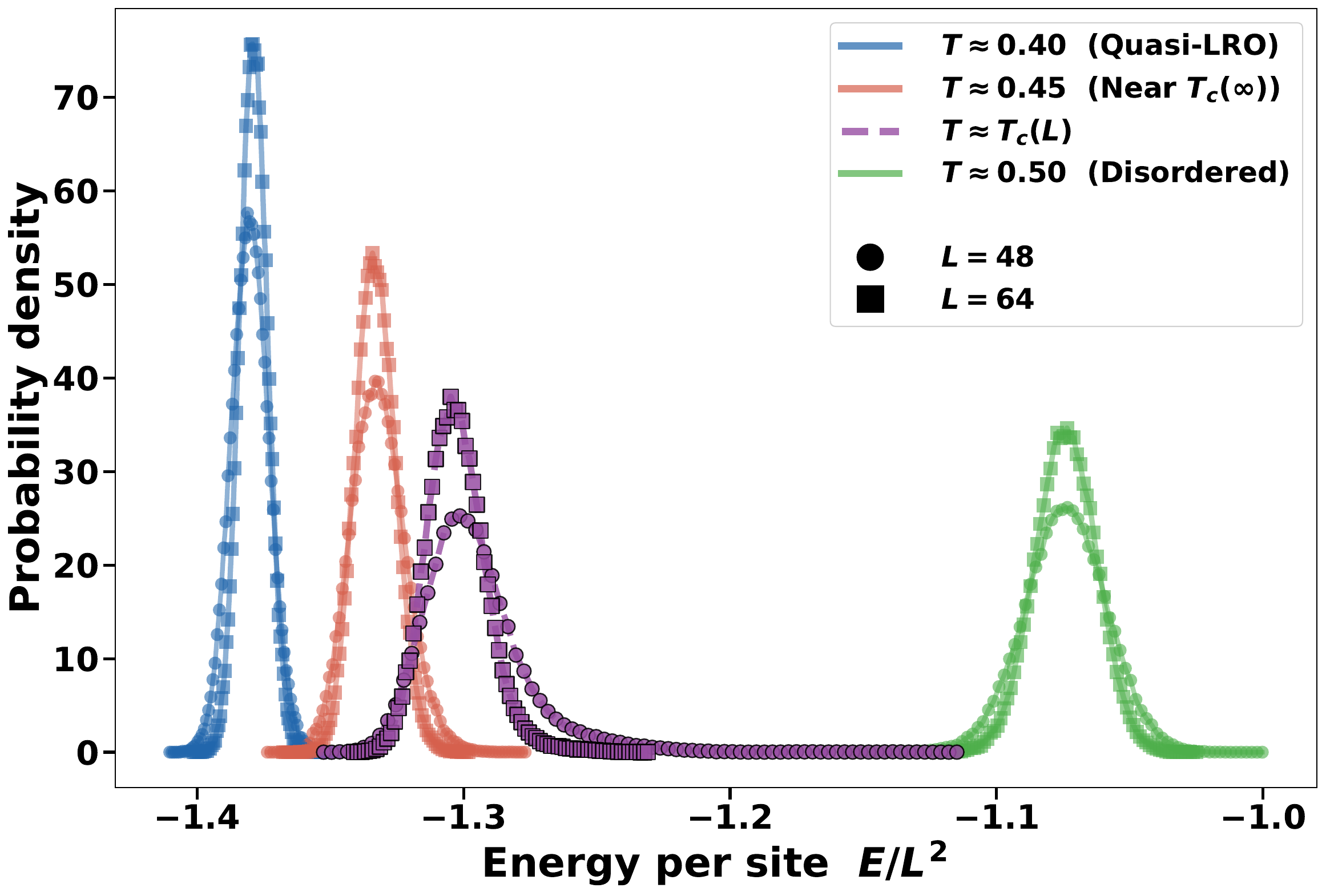}
    \caption{Energy per site histograms for the $J_3=0$ merging case ($J_1=1$, $J_2=0.6$, $J_3=0$) at four representative temperatures: the quasi-long-range ordered phase ($T\approx0.40$), near the thermodynamic critical temperature ($T\approx0.45$), at the finite-size pseudocritical temperature ($T\approx T_c(L)$), and in the disordered phase ($T\approx0.50$). The histograms remain unimodal across all temperatures, showing no evidence of phase coexistence at accessible system sizes and supporting the continuous KT interpretation.}
    \label{fig:energy_hist_J3=0_sm}
\end{figure}

The specific heat develops an unusually sharp peak at the merged transition. We
therefore examine the energy distribution directly, and the histograms in
Fig.~\ref{fig:energy_hist_J3=0_sm} confirm that this sharpening does not
originate from phase coexistence, supporting the continuous KT
interpretation.

\subsection{Anisotropic case III: $J_2=0$ --- a single PG-vortex KT transition}

\begin{figure}[htbp]
    \centering
    \begin{minipage}[b]{0.35\textwidth}
        \centering\includegraphics[width=\textwidth]{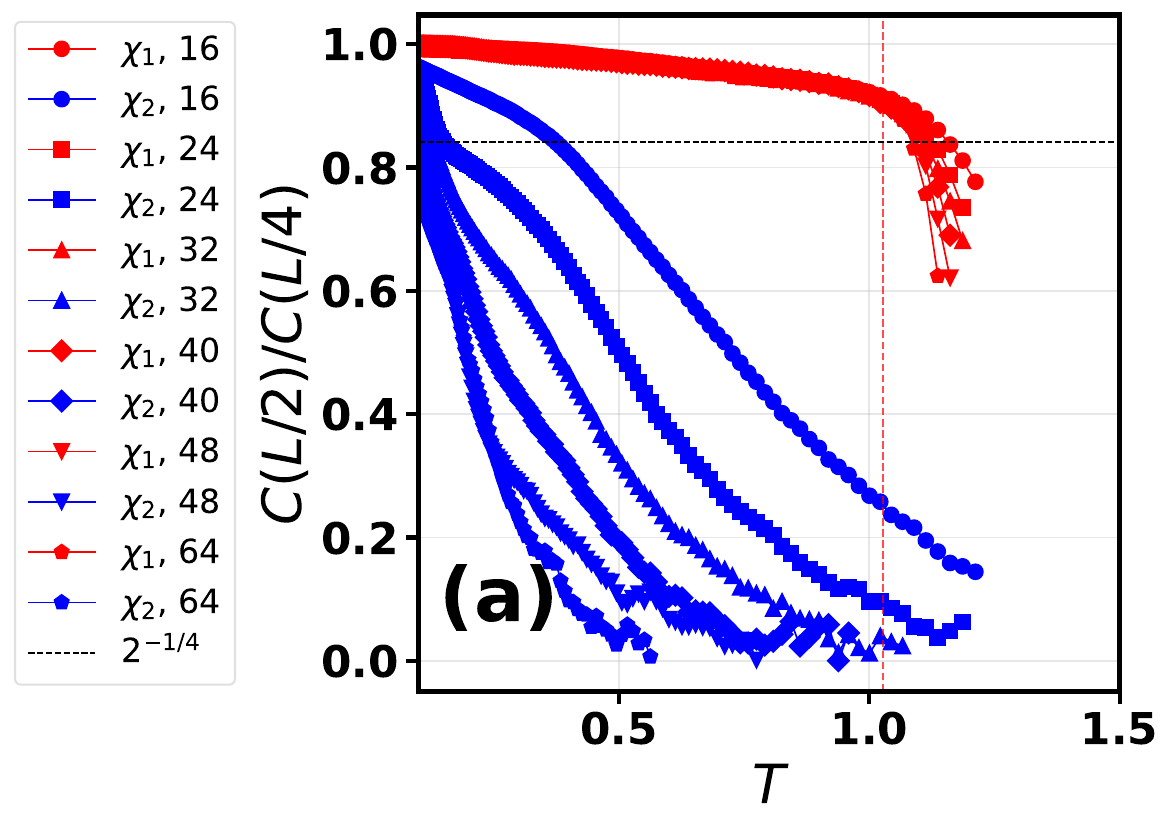}
    \end{minipage}
    \begin{minipage}[b]{0.35\textwidth}
        \centering\includegraphics[width=\textwidth]{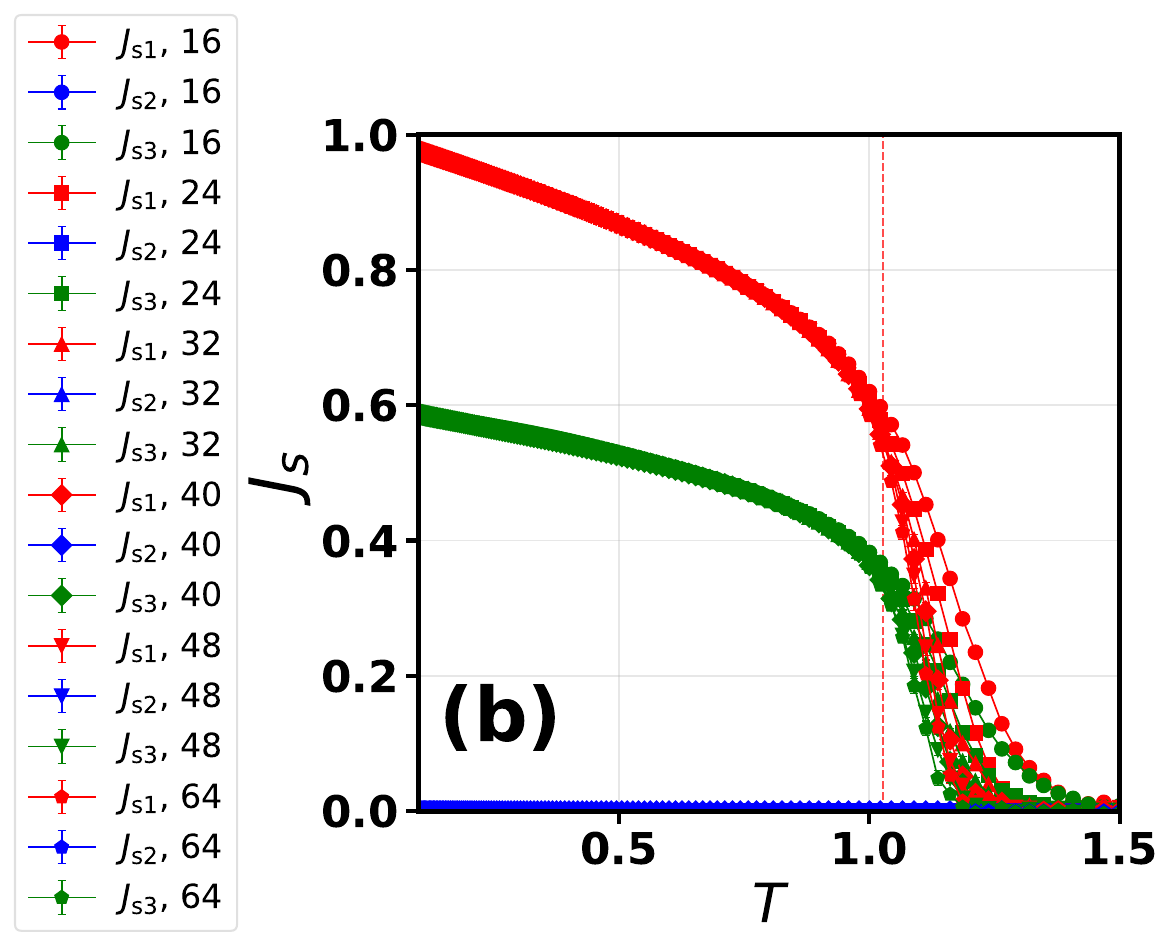}
    \end{minipage}

    \bigskip
    \begin{minipage}[b]{0.217\textwidth}
        \centering\includegraphics[width=\textwidth]{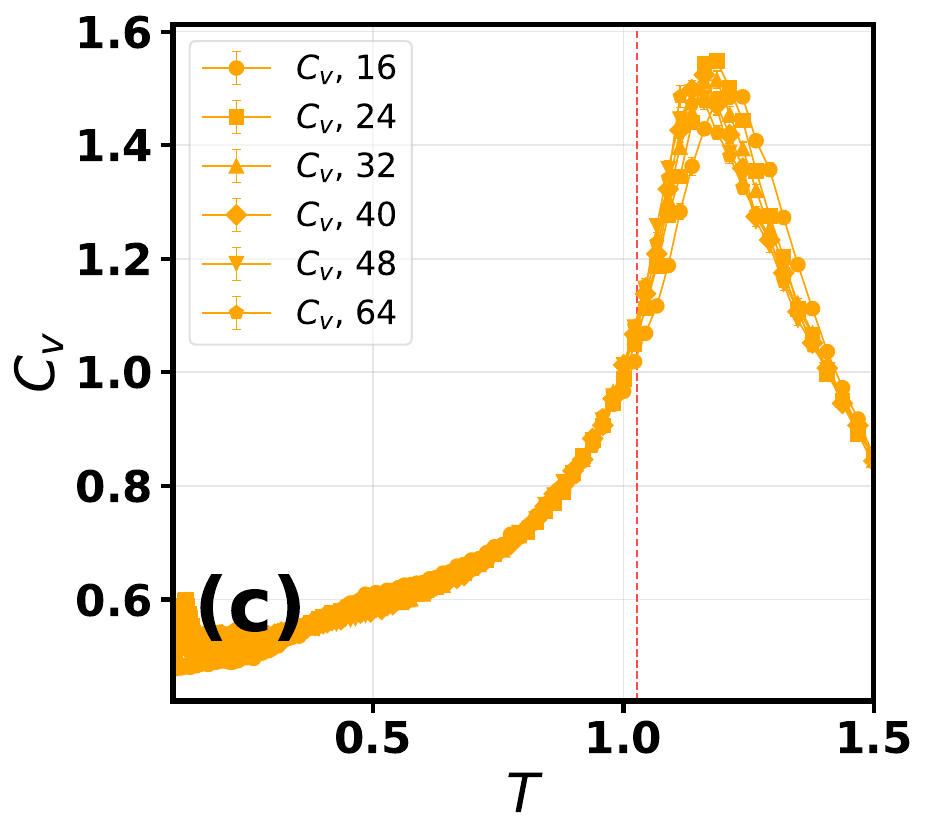}
    \end{minipage}
    \hfill
    \begin{minipage}[b]{0.229\textwidth}
        \centering\includegraphics[width=\textwidth]{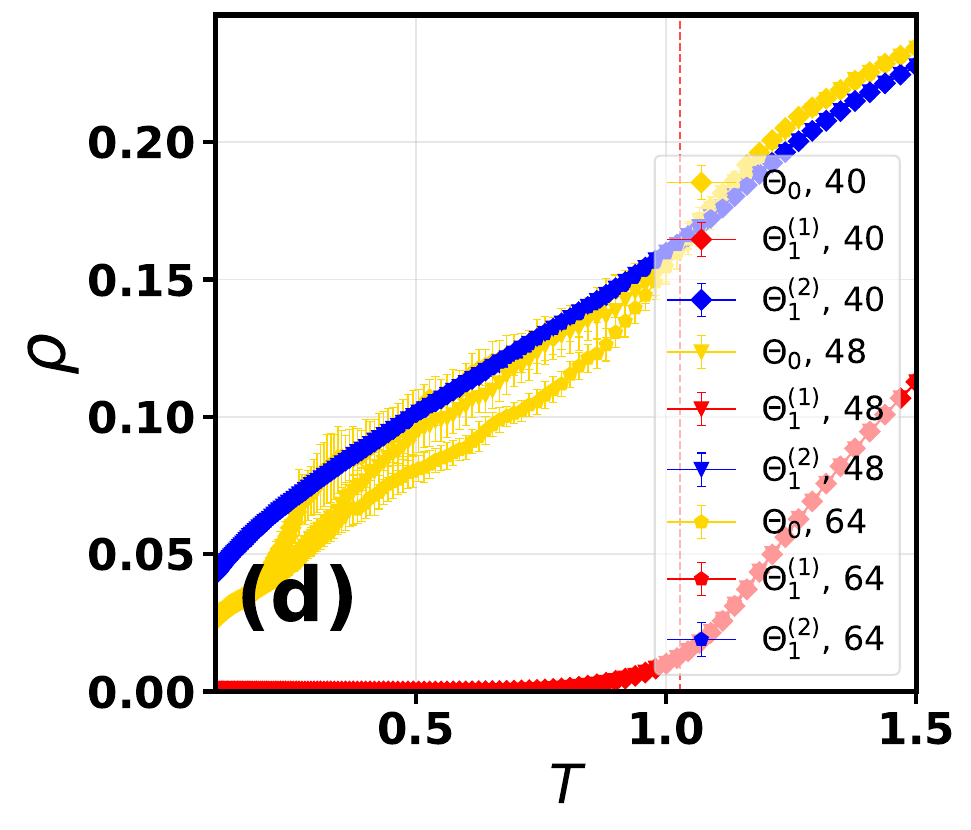}
    \end{minipage}
    \hfill
    \begin{minipage}[b]{0.219\textwidth}
        \centering\includegraphics[width=\textwidth]{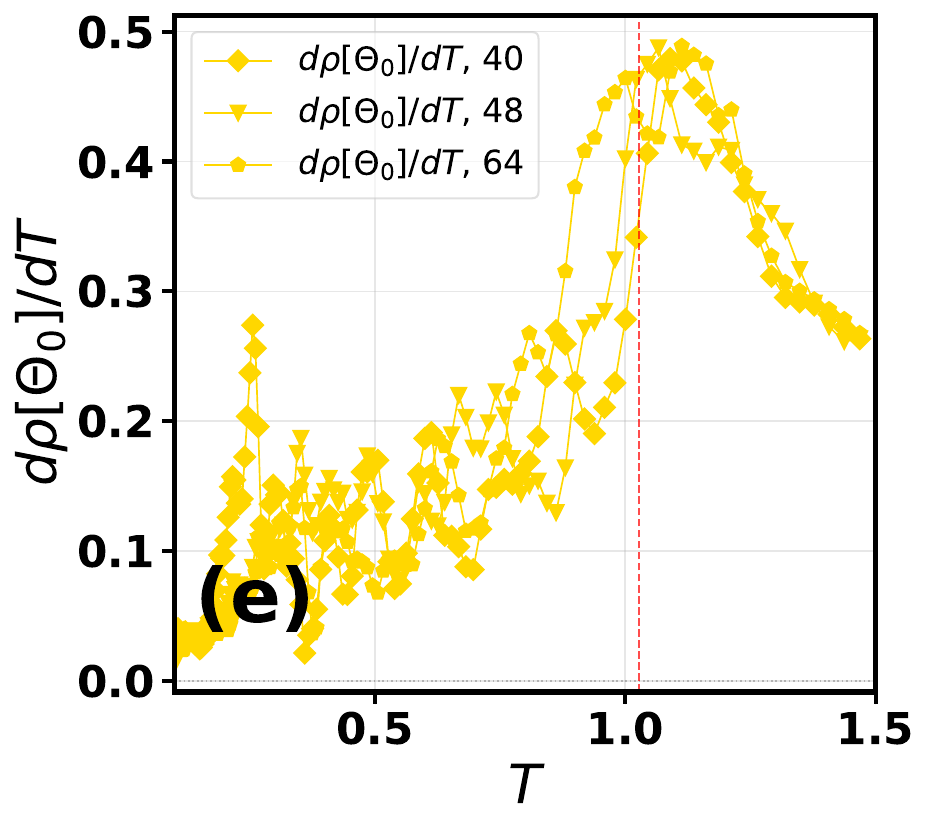}
    \end{minipage}
    \hfill
    \begin{minipage}[b]{0.3\textwidth}
        \centering\includegraphics[width=\textwidth]{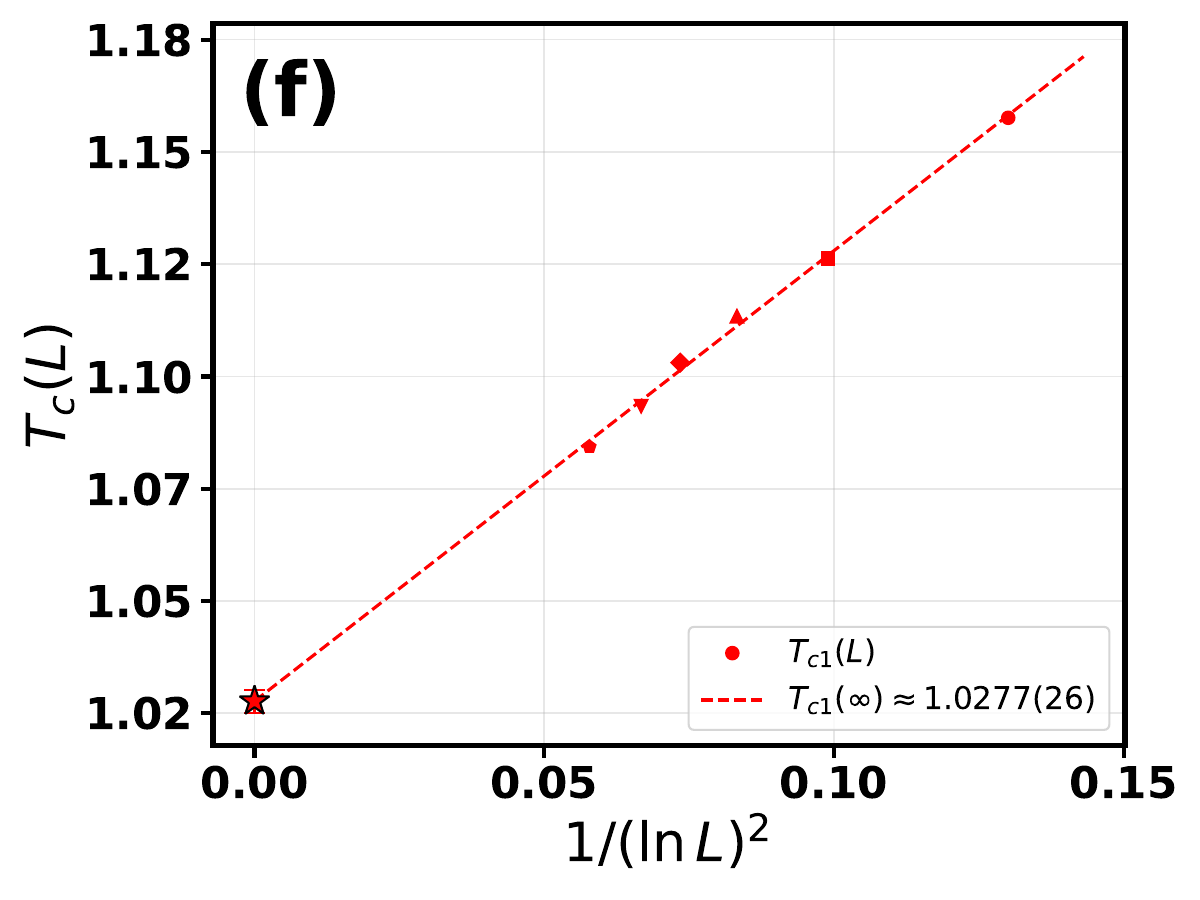}
    \end{minipage}
    \caption{Monte Carlo data for the anisotropic dipole-conserving XY model at $J_1=1$, $J_2=0$, $J_3=0.6$. (a) Correlation ratio $C(L/2)/C(L/4)$ showing a single crossing, determining the single critical temperature. (b) Generalized helicity moduli showing a single drop, corroborating that only one PG-vortex species drives a KT transition when $J_2=0$. (c) Specific heat $C_v$. (d) Vortex densities $\rho(\Theta_0)$, $\rho(\Theta_1^{(1)})$, and $\rho(\Theta_1^{(2)})$ of all three species. Only $\Theta_1^{(1)}$ proliferates at finite temperature, while $\rho(\Theta_1^{(2)})$ is nonzero at all temperatures and does not control criticality. (e) Temperature derivative $\mathrm{d}\rho(\Theta_0)/\mathrm{d}T$, whose peak signals critical enhancement of phase-vortex generation. (f) Finite-size scaling analysis of the pseudocritical temperatures.}
    \label{fig:aniso3_results_sm}
\end{figure}

Finally, we consider $J_1=1$, $J_2=0$, $J_3=0.6$. The
$(\partial_2^2\theta)^2$ term is absent, so the modulus $J_{s2}$ vanishes.
The exponent $\eta_2\to\infty$ from Eq.~\eqref{eq:eta}, meaning $\chi_2$ never
supports quasi-long-range order; consequently $\Theta_1^{(2)}$ cannot drive a
transition. Only $\Theta_1^{(1)}$, coupled to $\chi_1$, exhibits a single
KT transition, signaled by a single
correlation-ratio crossing and a drop in $J_{s1}$. The numerical data are collected in
Fig.~\ref{fig:aniso3_results_sm}, complementing the End Matter
Fig.~\ref{fig:em_data}.

\end{document}